\documentclass[12pt,a4paper,oneside]{book}

\usepackage{3Manuscript/codeformat}
\usepackage{3Manuscript/mathmacros}
\usepackage{caption}
\usepackage{subcaption}
\usepackage{graphicx}
\usepackage{tabularx}
\usepackage{multirow}
\usepackage{array}
\usepackage{indentfirst}
\usepackage{hyperref} 
\usepackage[backend=biber, style=numeric , 
citestyle=numeric]{biblatex}
\usepackage{algorithm}
\usepackage{algpseudocode}
\usepackage{float}
\usepackage{booktabs}
\usepackage{multirow}
\usepackage{longtable}  
\usepackage{pdflscape}  
\usepackage{siunitx}  
\usepackage[toc,page]{appendix}
\usepackage{comment} 

\usepackage{etex}
\usepackage{amsmath,amssymb,latexsym,mathtools}
\usepackage{amsthm}
\usepackage{setspace}
\usepackage{graphicx}
\usepackage[raggedright]{titlesec}
\usepackage{dsfont, enumerate}
\usepackage{color}
\usepackage{algorithm}
\usepackage{appendix}
\usepackage{bbding}
\usepackage{pifont}
\usepackage{wasysym}
\usepackage{amssymb}
\usepackage{pifont}
\newcommand{\cmark}{\ding{51}}%
\newcommand{\xmark}{\ding{55}}%

\makeatletter
\def\BState{\State\hskip-\ALG@thistlm}
\makeatother
\numberwithin{equation}{subsection}
\makeatletter
\def\@makechapterhead#1{%
  \vspace*{-40pt}%
  {\parindent \z@ \raggedright \normalfont \centering
    \ifnum \c@secnumdepth >\m@ne
      \if@mainmatter
        \Large\bfseries \@chapapp\space \thechapter
        \par\nobreak
        \vskip0truecm
      \fi
    \fi
    \interlinepenalty\@M
    \Large \bfseries #1\par\nobreak
    \vskip0.5truecm
  }}
\def\@makeschapterhead#1{%
  \vspace*{-20pt}%
  {\parindent \z@ \raggedright
    \normalfont \centering
    \interlinepenalty\@M
    \Large \bfseries  #1\par\nobreak
    \vskip0.5truecm
  }}
\makeatother

\titleformat{\section}{\large\bfseries}{\thesection}{1em}{}
\titlespacing*{\section}{0pt}{0.5cm}{0pt}
\titlespacing*{\subsection}{0pt}{0.3cm}{0pt}

\newtheorem{theorem}{Theorem}[section]

\theoremstyle{definition}
\newtheorem{definition}[theorem]{Definition}

\pdfoutput=1

\theoremstyle{plain}

\theoremstyle{remark}

\newcommand{\mytitle}{\textbf{\textit{The Families that Stay Together}: \\ A Network Analysis of Dynastic Power in Philippine Politics}}
\newcommand{\myauthor}{Rafael Acu\~{n}a, Aldie Alejandro, Robert Leung}

\newenvironment{conditions*}
  {\par\vspace{\abovedisplayskip}
   \tabularx{\columnwidth}{>{$}l<{$} @{${}={}$} >{\raggedright\arraybackslash}X}}
  {\endtabularx\par\vspace{\belowdisplayskip}}

\begin{document}

\thispagestyle{empty}

\begin{center}

\bfseries{\Large \mytitle}

\vfill
\bfseries{\itshape An Undergraduate Thesis Presented to the \\ Faculty of the Department of Mathematics \\ Ateneo de Manila University \\ in Partial Fulfillment of the Requirements for the Degree \\ Bachelor of Science in Applied Mathematics}

\vfill
\bfseries{\myauthor \\ April 10, 2025}
\end{center}
\newpage
\renewcommand{\contentsname}{Table of Contents}
\tableofcontents
\thispagestyle{empty}
\newpage
\pagestyle{plain}

\begin{center}
\bfseries{\Large Acceptance Page}
\end{center}
\addcontentsline{toc}{chapter}{Acceptance Page}

The Faculty of the Department of Mathematics of Ateneo de Manila University
accepts the undergraduate thesis entitled

\bigskip
\begin{center}
\bfseries{\itshape \mytitle}
\end{center}

\bigskip\noindent
submitted by {\myauthor} and orally presented on April 23, 2025, in partial fulfillment of the requirements
for the degree Bachelor of Science in Applied Mathematics.

\vskip1.1truecm

\begin{center}
\begin{tabular}{ccc}
\makebox[3.0in]{\hrulefill} & & \makebox[1in]{\hrulefill}\\
 Dr. Reginaldo M. Marcelo & & Date\\
Presentation Critic & & \\[5.4ex]
\makebox[3.0in]{\hrulefill} & & \makebox[1in]{\hrulefill}\\

 Dr. Philip Arnold P. Tua\~{n}o& & Date\\
Presentation Critic & & \\[5.4ex]
\makebox[3.0in]{\hrulefill} & & \makebox[1in]{\hrulefill}\\
Dr. Clark Kendrick C. Go & & Date\\
Adviser & & \\[5.4ex]

\makebox[3.0in]{\hrulefill} & & \makebox[1in]{\hrulefill}\\
 Dr. Jude C. Buot & & Date\\
Adviser & & \\[5.4ex]

\makebox[3.0in]{\hrulefill} & & \makebox[1in]{\hrulefill}\\
 Dr. Job A. Nable & & Date\\
Adviser & & \\[5.4ex]
\makebox[3.0in]{\hrulefill} & & \makebox[1in]{\hrulefill}\\
Dr. Romina Ann S. Yap & & Date\\
 Department Chair & &
\end{tabular}
\end{center}
\newpage
\begin{center}
\bfseries{\Large Acknowledgments}
\end{center}
\addcontentsline{toc}{chapter}{Acknowledgments}

\textit{First and foremost, we extend our deepest gratitude to Almighty God, whose grace and guidance have been our constant source of strength throughout the course of our thesis.}

\par\vspace{1em}

\textit{We would also like to thank our thesis advisers, Dr. Clark Go, Dr. Jude Buot, and Dr. Job Nable, for their unwavering support, insightful feedback, and invaluable expertise. Their patience and encouragement have shaped not only the outcome of this research but also our growth as scholars.}

\par\vspace{1em}

\textit{We would also like to acknowledge the Ateneo Policy Center, whose dataset started us in this topic in the first place. Their comments and suggestions were also valuable.}

\par\vspace{1em}

\textit{To our families, thank you for your unconditional love and understanding. Your encouragement has been our greatest motivation, especially during the challenging moments of this academic endeavor.}

\par\vspace{1em}

\textit{To our friends and classmates, especially the AMDSc Batch 2025, who shared our laughter and stress, offered help, and stood by us all the time---your presence has made this journey more meaningful.}

\par\vspace{1em}

\textit{Lastly, we thank each other---Rafael Acuna, Aldie Alejandro, and Robert Leung---for the shared determination, trust, and that despite our countless friendly arguments, we brought this research to completion.}

\par\vspace{1em}

\textit{To our panelist, apologies for the long manuscript and thank you for agreeing to critique our thesis.}

\newpage
\begin{center}
\bfseries{\Large Summary of the Thesis}
\end{center}
\addcontentsline{toc}{chapter}{Summary of the Thesis}

Dynasties have long dominated Philippine politics. Despite the theoretical consensus that dynastic rule erodes democratic accountability, there is limited empirical evidence establishing dynasties’ true impact on development. A key challenge has been developing robust metrics for characterizing dynasties that facilitate meaningful comparisons across geographies and election cycles. Using election data from $2004$ to $2022$, we leverage methods from graph theory to develop four indicators to investigate dynastic evolution: (i) Political Herfindahl-Hirschman Index (HHI), measuring the level of dynastic power concentration; (ii) Centrality Gini Coefficient (CGC), reflecting the inequality of influence between clan members; (iii) Connected Component Density (CCD), representing the degree of inter-clan connection; and (iv) Total Community Connectivity (TCC), quantifying intra-clan cohesion. Our analysis reveals three key findings. Firstly, dynasties have grown stronger and more interconnected, with dynasts occupying an increasing share of elected positions. Dominant clans have also remained tightly knit, but with great power imbalances between members. Secondly, we examine variations in party-hopping between dynastic and non-dynastic candidates. A Wilcoxon Signed Rank Test reveals that, across every election cycle, party hopping rates are significantly higher ($p < 0.01$) among dynastic candidates than non-dynasts---suggesting the dominance of dynasties may weaken institutional trust within parties. Finally, applying a Linear Mixed Model regression, which controls for geographic random-effects and time fixed-effects, we observe that provinces with high power asymmetries within clans (high CGCs) and with deeply interconnected clans (high CCDs) record significantly lower ($p < 0.05$) Human Development Index scores. These findings suggest that clan structure, rather than power concentration alone, may be the chief determinant of a ruling dynasty’s developmental impact. Our study is instrumental in refining the quantitative argument for Anti-Dynastic reform. In a legislative context where passing an Anti-Dynasty Law remains unlikely, our methods may illustrate alternative paths to reform for combating dynastic rule at its roots.

\newpage
\begin{center}
\bfseries{\Large Anti-Plagiarism Declaration}
\end{center}
\addcontentsline{toc}{chapter}{Anti-Plagiarism Declaration}

I declare that I have authored this thesis independently, that I have not used materials other than the declared sources or resources, and that I have explicitly marked all materials which have been quoted either literally or by content from the used sources.

\vskip2truecm

\begin{center}
\begin{tabular}{ccc}
\makebox[3.0in]{\hrulefill} & & \makebox[1in]{\hrulefill}\\
Rafael Acu\~{n}a & & Date\\
Student I.D. No. 210086 & & \\[5.4ex]
\makebox[3.0in]{\hrulefill} & & \makebox[1in]{\hrulefill}\\
Aldie Alejandro & & Date\\
Student I.D. No. 210233 & & \\[5.4ex]
\makebox[3.0in]{\hrulefill} & & \makebox[1in]{\hrulefill}\\
Robert Leung & & Date\\
Student I.D. No. 213402 & & \\[5.4ex]
\end{tabular}
\end{center}

\doublespacing

\chapter{Introduction}

\section{Background}
In the Philippines, politics is a family affair. The nation's political landscape has long been molded by the persistence of elite ruling clans. The monopolization of power in the hands of political dynasties has been theorized to undermine the development of democratic institutions---eroding accountability \cite{tadem2016}, impeding electoral competitiveness \cite{Teehankee2020}, and perpetuating elite capture \cite{querubin_political_2012}. Empirical evidence has emerged linking dynastic prevalence to a range of detrimental outcomes including heightened corruption risk \cite{davis_corruption_2024}, increased poverty incidence \cite{mendoza_political_2022-1}, and decreased Human Development investments \cite{villanueva_political_2022}. 

\par  The 1987 Constitution explicitly recognized the problem of dynasties. Section 26 of Article II provides that \textit{``the State shall...prohibit political dynasties"}, but leaves the task of defining what constitutes a dynasty to Congress \cite{philippine_constitution_1987}. Almost four decades later, there has been minimal progress toward passing a comprehensive and enforceable anti-dynasty law. The \textit{Sangguniang Kabataan} Reform Act of 2015 was hailed as a landmark in anti-dynasty legislation. Section 10 explicitly includes a clause disqualifying candidates related within the second degree of consanguinity to a local or national incumbent \cite{sk_reform_act_2015}. The Bangsamoro Electoral Code which was passed in 2022 adopts the same definition, prohibiting simultaneous candidates related within the second degree \cite{bangsamoro_autonomy_act_35_2022}. Such measures, however, have failed to gain traction at the broader national scale. Most recently, the Anti-Political Dynasty Act filed in the 19th Congress similarly defined dynasties as a \textit{"situation in which two incumbents who are related spousally or within the second degree of consanguinity run simultaneously or successively for public office within the same city and/or province"} \cite{dynastybill}. It is unlikely that this bill will pass when almost 80\% of incumbent district representatives are dynasts \cite{miranda_8_2024}. 

\par As attempts to legislatively combat the problem of dynasties have seen limited success, a natural question is whether it is possible to identify and address the roots of dynastic persistence. Scholars have theorized that the persistence of dynasties is both a cause and effect of the relative weakness of alternative forms of political organization \cite{querubin_political_2012, mendoza_dynastic_2020}. In contexts where party-based ties are relatively weak and unreliable, incumbents seeking to continue exerting political influence beyond their allowed terms turn to closer kinship networks \cite{querubin_political_2012}. As dynastic candidates occupy a larger share of seats, emerging parties are forced to curry favor with a ruling dynasty to secure resources and political clout \cite{ghosh_understanding_2023}. Thus begins a vicious cycle where strong dynasties hinder the emergence of strong parties, while the absence of strong parties fuels reliance on clan-based politics.

\par Thus far, however, there has been limited empirical research examining the interplay between dynastic networks and political parties \cite{pimentel_political_2024}. A key challenge in this regard has been developing robust indicators of dynastic concentration that capture the full scope and structure of dynastic networks. While second-degree consanguinity may serve as a practical legal heuristic for identifying dynastic candidates, in reality, political clans are often complex multi-generational networks extending far beyond second-degree kinship \cite{mendoza_political_2022, querubin_political_2012}. Political clans are often interlinked in a web of strategic intermarriages \cite{mendoza_dynastic_2020}. Further, political clans are not homogenous. A tight-knit dynasty composed of a single nuclear family may have very different motives, actions, and thus socioeconomic impact from a large multi-branch clan. Hence, investigations into both the roots and developmental consequences of dynastic concentration require a more comprehensive means of characterizing the structure of political clans.  

\par Network analysis applies methods from graph theory to the study of physical and social structures \cite{engel_network_2021}. Previous network analysis studies on political dynasties have largely focused on characterizing the influence of key individuals within the network of politicians in a given locale \cite{mendoza_dynastic_2020, cruzAER_2017} rather than drawing insights from their aggregate structure. Crucially, there is limited empirical literature examining both internal organization of political clans (e.g. whether they tend to form around a central matriarch/patriarch or disperse power more evenly), and describing the degree of interconnectedness between clans.

\section{Objectives of the Study}

\par This study aims to investigate the structure and evolution of dynastic networks in Philippine politics using local election data from 2004 to 2022. Specifically, we develop several graph-theory inspired indicators of dynastic prevalence at the provincial level. We then use these indicators to investigate the following questions:
\begin{enumerate}
    \item How has dynastic prevalence evolved across geographies and election cycles?
    \item How does partisan allegiance differ across dynastic and non-dynastic candidates?
    \item How does dynastic prevalence affect local development outcomes, specifically Poverty Incidence (POV) and Human Development Index (HDI)?
\end{enumerate}

\section{Scope and Limitations}
\par This study examines familial connections among locally-elected politicians who assumed office at least once between 2004 and 2022. The dataset comprises the following positions: Congressional District Representatives, Provincial Governor, Provincial Vice Governor, Municipal Mayor, Municipal Vice Mayor, Provincial Board Member (\textit{Sangguniang Panlalawigan}), and City Councilors (\textit{Sangguniang Panglungsod}). It is also important to mention that this study uses secondary data, mainly from the Commission on Elections (COMELEC) and Ateneo Policy Center (APC). As such, the information provided will no longer be fact-checked. 

A total of $80$ provinces (dropped from $87$ due to data availability) will be used for this study. Specifically, this study will not include the following provinces: Sulu, Tawi-Tawi, Compostela Valley, North Cotabato, Davao de Oro, Dinagat Islands, and Davao Occidental. Politicians within the same province are assumed to be related if the surname or middle name of one matches either the surname or middle name (or both) of the other based on their officially registered name at the time of election. While unrelated politicians may incidentally share the same surname or middle name yet no real familial linkages (though unlikely as will be discussed with the latter chapters), it is beyond the scope of this paper to identify such cases. Furthermore, this paper also does not consider inter-provinces linkages. Lastly, this study does not consider the national elections and will focus solely on local elections.

\section{Organization of the Thesis}
% ...
This thesis is organized as follows: Chapter 2 outlines key preliminary concepts from network analysis and graph theory that will be used throughout the study, Chapter 3 overviews the empirical literature on political dynasties in the Philippines, and in particular previous methods of quantifying dynastic power, Chapter 4 describes the methods and data used, Chapter 5 presents the paper's main results, and Chapter 6 concludes with directions for future research.

\chapter{Preliminaries}
% ...

\section{Basic Concepts and Terminology}
This chapter contains the basic concepts and terminology in graph theory that are relevant throughout this paper.

\begin{definition} [\cite{Zweig-2014}]
    A \textbf{graph} $G = (V, E)$ is a mathematical object consisting of:
    \begin{enumerate}
        \item[\textbullet] a finite non-empty set of objects called \textbf{vertices} $V(G)$ (the \textbf{vertex set}). In this paper, we sometimes refer to vertices as \textbf{nodes}.
        \item[\textbullet] a set of \textbf{edges} $E(G)$ (the \textbf{edge set}) such that $E(G) \subseteq \{ uv : u , v \in V, u \neq v \}$. An \textbf{edge} is a two-element subset typically written as \textit{uv} or \textit{vu}.
    \end{enumerate}
\end{definition}

\noindent A graph $G$ is synonymous to a \textbf{network} and will be used interchangeably in this paper. Additionally, if the edges of $G$ have no direction then we call $G$ an \textbf{undirected graph}.

\begin{definition} [\cite{Bickle-2020}]
    The \textbf{order} and \textbf{size} of a graph $G$ is the cardinality of the vertex set and the edge set of $G$, i.e. $|V(G)|$ and $|E(G)|$, respectively.
\end{definition}

\begin{definition} [\cite{Zweig-2014}]   Let $u,v,w \in V(G)$ and $e,f \in E(G)$.
    If $e = uv$, then we say that $u$ and $v$ are \textbf{adjacent}, or $u$ and $v$ are \textbf{neighbors}. In addition, we also say edge $e$ is \textbf{incident} with vertices $u$ and $v$. If $f = uw$, we also say that $e$ and $f$ are \textbf{adjacent} in $G$.
\end{definition}

\begin{definition} [\cite{Zweig-2014}]
    The \textbf{degree} of a vertex $v$, written as $d_G(v)$ (or $d(v)$ when the graph in question is clear), is the number of edges incident with $v$. A vertex with degree $0$
    is called an \textbf{isolated vertex}. The \textbf{neighborhood} of a vertex $v$, $N(V)$ is the set of vertices in $V(G)$ that are adjacent to $v$.
\end{definition}

\begin{figure}[H]
        \centerline{\includegraphics[width = 1\textwidth]{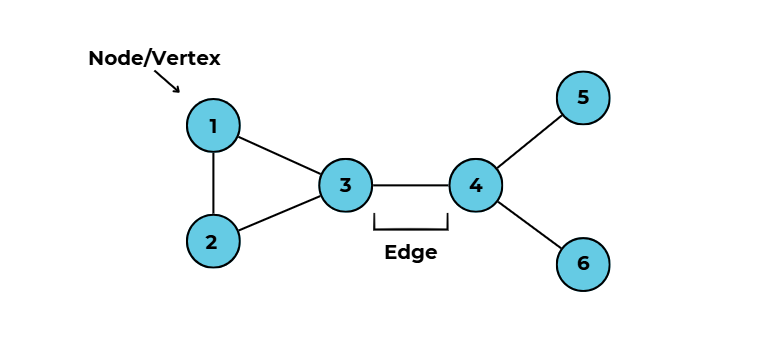}}
        \caption{An Example of a Graph}
        \label{Basic-Graph}
\end{figure}

\noindent Let $G$ be the graph illustrated in Figure \ref{Basic-Graph}. Then, the order and size of the graph $G$ is given by $|V(G)| = 6$ and $|E(G)| = 6$. On the other hand, the degree of Node 1 is $d(1) = 2$, while the degree of Node 5 is $d(5) = 1$. Furthermore, we say that Node $1$ is adjacent to Node $2$, Node $2$ is adjacent to Node $3$, and so on. There is no isolated vertex in $G$.

\section{Graph Classes}

\begin{definition} [\cite{Zweig-2014}]
    A \textbf{weighted graph} is a triple $G = (V,E, w)$ where $(V,E)$ forms a graph and $w: E \rightarrow \mathbb{R}$ is a function that assigns a real-valued weight to each edge.
\end{definition}

\noindent\textbf{Remarks:} It is possible to define a weighting system for nodes. In that case, $G'$ is now a tuple defined by $G' = (V, E, w_N, w_E)$ for which $w_N$ and $w_E$ are real-valued functions that assign weight to each node and edge, respectively.

\begin{definition} [\cite{Bickle-2020}]
    A \textbf{complete graph} of $n$ vertices, denoted by $K_n$, is a simple graph at which every pair of distinct vertices is adjacent.

\noindent\textbf{Remarks:} A complete graph of order $n$ has size equal to $\binom{n}{2} = \frac{n(n-1)}{2}$.

\end{definition}

\begin{definition} [\cite{Bickle-2020}]
    A graph $H$ is a \textbf{subgraph} of graph $G$ if $V(H) \subseteq V(E)$ and $E(H) \subseteq E(G)$. It is usually denoted as $H \subseteq G$.
    \begin{enumerate}
        \item[\textbullet] A subgraph $H$ is said to be \textbf{induced} if $E(H)$ contains all edges $uv \in E(G)$ for all $u,v \in V(H)$.
    \end{enumerate}
\end{definition}

\begin{definition} [\cite{Bickle-2020}]
    A path $P_n$ is a graph of order $n \geq 2$ for which $V(P_n) = \{v_1, v_2, \ldots,v_n \}$ and $E(P_n) = \{v_{k-1}v_k: 2 \leq k \leq n \}$.

\end{definition}

\begin{definition} [\cite{Zweig-2014}]
    The \textbf{length of a path} is defined as the sum of the weight of its edges if the graph is weighted, and the number of edges in it otherwise. From all the possible paths, the path with the minimal length is called the \textbf{shortest path}.

\end{definition}

\begin{definition} [\cite{Zweig-2014}]
    A \textbf{connected component} of a graph $G$ is an induced subgraph $H$ such that:
    \begin{enumerate}
        \item[\textbullet] For any two vertices $u, v \in H$, there exists a path in $G$ that connects $u$ and $v$.
        \item[\textbullet] $H$ is maximal, meaning that if $w \in V(G) \setminus V(H)$, then there is no edge in $E(H)$ that connects $w$ to any vertex in $H$.
    \end{enumerate}
\end{definition}

\section{Communities}

A \textbf{community} is a cluster of nodes that form relatively dense groups \cite{Traag2019}. Such modular structures are not usually known beforehand which makes community detection an important problem in network analysis.

\begin{figure}[H]
        \centerline{\includegraphics[width = 0.80\textwidth]{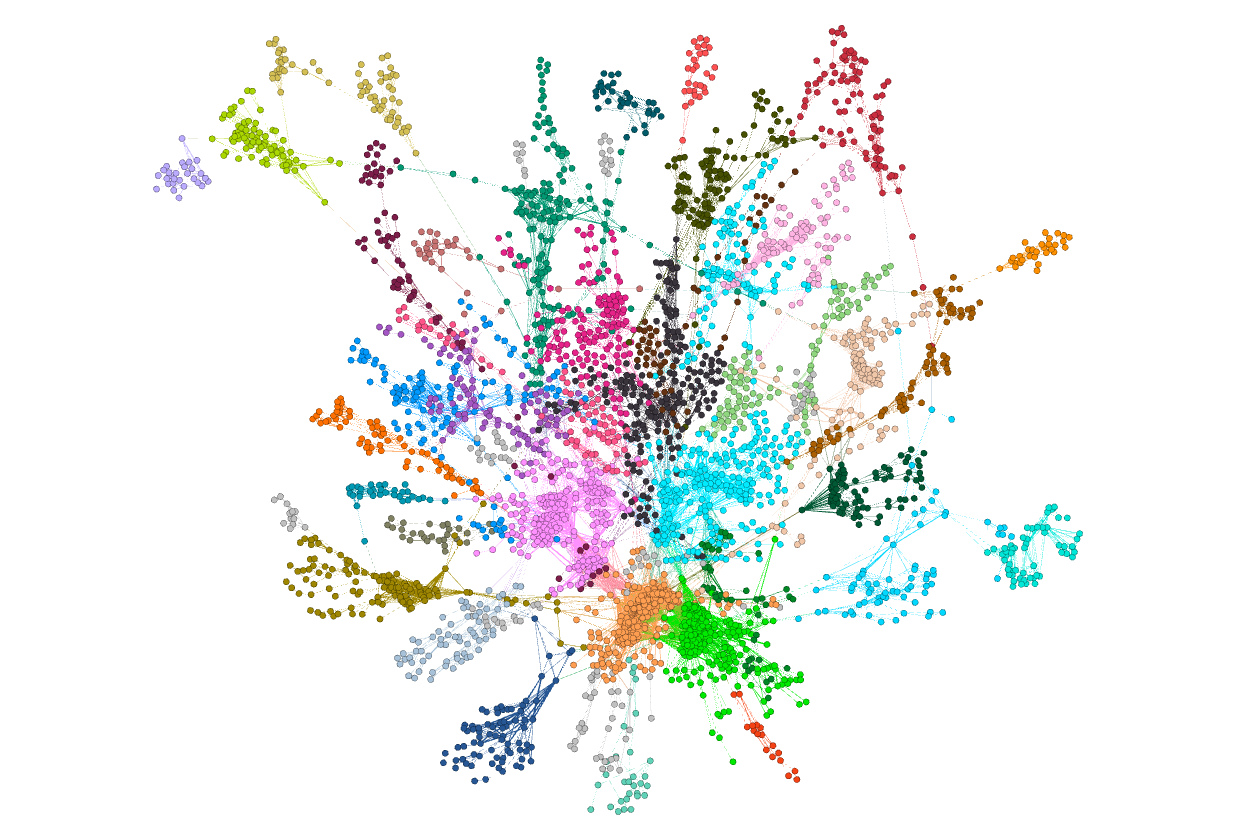}}
        \caption{An example of a network, color-coded by communities}
        \label{community-sample}
\end{figure}

\textbf{Modularity} is a measure used to quantify the quality of a division of a graph into communities \cite{Traag2019}. It evaluates how well a particular partition of the graph reflects the community structure, with a focus on maximizing the density of edges within communities while minimizing the density of edges between communities. Modularity is defined as follows:

\begin{equation*}
\mathcal{H} = \frac{1}{2m} \sum_{c} \left( e_{c} - \gamma \frac{K_{c}^{2}}{2m} \right)
\end{equation*}
where, 
\begin{enumerate}
    \item[\textbullet] $e_c$ is the actual number of edges in community $c$;
    \item[\textbullet] $K_c$ is the sum of the degrees of the nodes in the community;
    \item[\textbullet] $m$ is the size of the graph network; and
    \item[\textbullet] $\gamma$ is the resolution, which is a constant parameter that controls how fine-grained the community structure revealed.
\end{enumerate}

\noindent\textbf{Remark:} Higher resolutions lead to more communities, while lower resolutions lead to fewer communities.

\subsection{Leiden Algorithm} \label{Leiden-Definition}

The \textbf{Leiden Algorithm} is a heuristic algorithm that optimizes modularity \cite{Traag2019}. Unlike algorithms that preceded it, Leiden algorithm guarantees that communities are well-connected. The pseudocode of the algorithm ~\ref{alg:leiden} is given below.

\begin{algorithm} [H]
    \caption{Leiden Algorithm for Community Detection} \label{alg:leiden}
\begin{algorithmic}
\Require Graph \( G = (V, E) \)
\Ensure Partition of nodes into communities

\State \textbf{Initialize:} Assign each node to its own community.
\Repeat
    \State \textbf{Phase 1: Local Moving}
    \Repeat
        \For{each node \( v \in V \)}
            \State Move \( v \) to the community that maximizes modularity gain.
        \EndFor
    \Until{no more improvement}
    \State \textbf{Phase 2: Refinement}
    \State Identify sub-communities within each existing community.
    \State Merge poorly connected sub-communities.

    \State \textbf{Phase 3: Aggregation}
    \State Construct a new graph where nodes represent communities.
    \State Recompute edge weights based on inter-community connections.

\Until{no significant improvement in modularity}
\State \textbf{Return} final community structure.

\end{algorithmic}
\end{algorithm}

\begin{figure}[H]
        \centerline{\includegraphics[width = 0.8\textwidth]{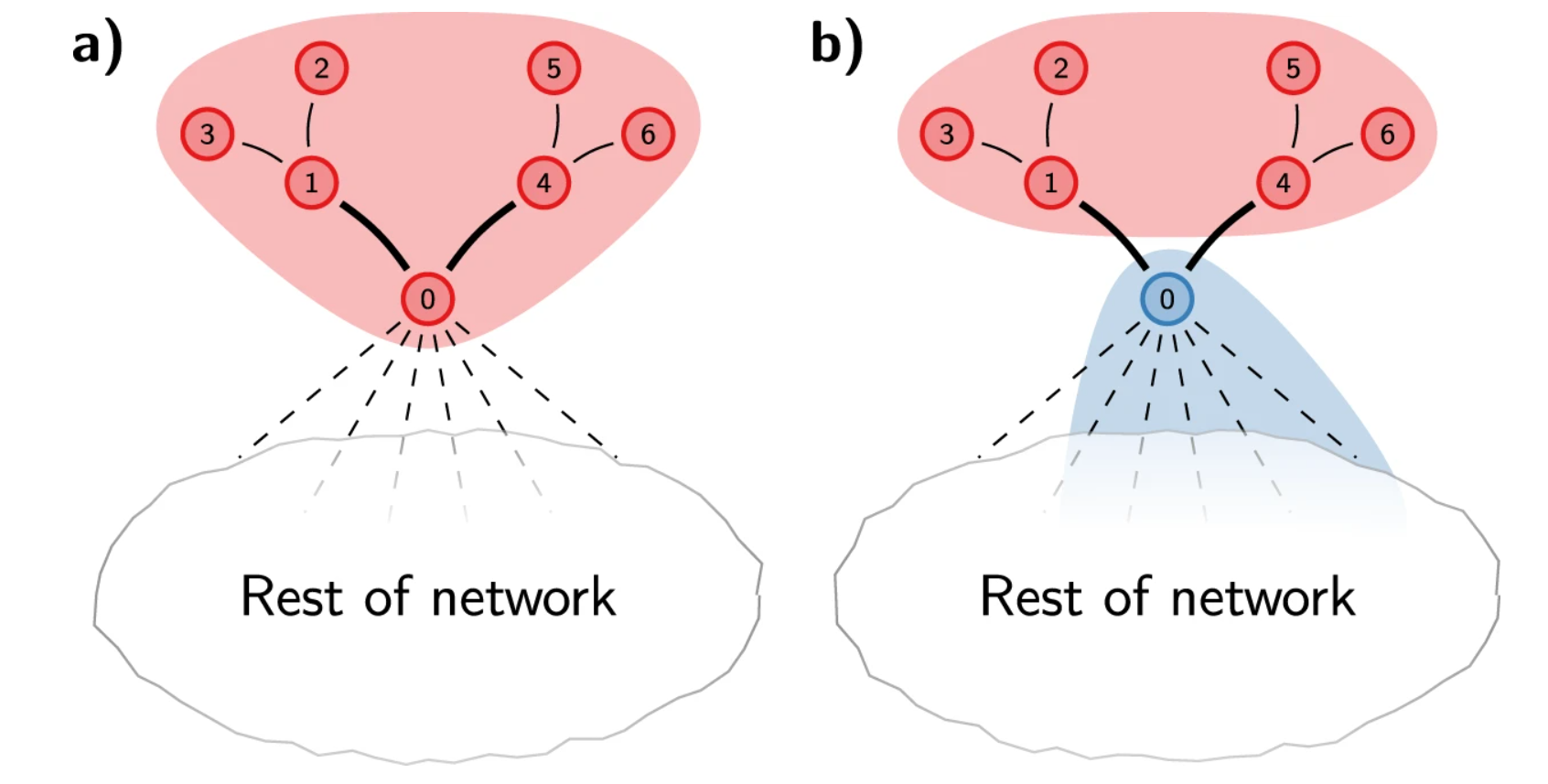}}
        \caption{In some algorithms, it is possible that the node $0$ in (a) will be transferred to a different community that will cause an internal disconnectedness such as in (b). No such cases will occur while using the Leiden Algorithm.}
        \label{Leiden-Advantage}
\end{figure}

The Leiden Algorithm is suitable for this study because, unlike other algorithms, it ensures that there will be no disconnected communities \cite{Traag2019}. As illustrated in Figure \ref{Leiden-Advantage}, it is possible that the node $0$ in (a) will be transferred to a different community that will cause an internal disconnectedness such as in (b).

\section{Network Metrics} 
In this section, we will be defining some key definitions and formulations for the graph metrics used in the paper. 

\subsection{Network Community Concentration}

\begin{definition} \label{HHI} [\cite{HHI-Paper}]
    In economics, the \textbf{Herfindahl-Hirschman Index (HHI)} measures the aggregate concentration present in a certain market, signifying the potential competition to a shared commodity. It is usually given by the following formula:
    \begin{equation*}
        HHI = \sum_{i} \left(\frac{s_i}{s_t} \times 100  \right )^2,
    \end{equation*}
where $s_i$ denotes the market share of $i$th firm in the market, while $s_t$ is the aggregate shares of all the firms in the market. 
\end{definition}

\begin{definition}\label{PHHI}
    Let $G = (V, E, w_N)$ be a weighted network, with each node $v$ having an associated weight $w_N(v)$. Let $J$ be a community existing within $G$. Then we define what we will call as the \textbf{Political Network HHI} of community $J$ as the value given by the equation:
    \begin{equation}
        HHI = \displaystyle\sum_{j \in J} \left(\frac{W_j}{W_t} \times 100  \right )^2,
        \label{PNHHI}
    \end{equation}
where 
\begin{enumerate}
    \item[\textbullet] $W_j$ is the sum of all the node weights belonging to community $J$; and
    \item[\textbullet] $W_t$ is the total node weights across all nodes in the network.
\end{enumerate}

\end{definition}

Inspired by the economic HHI, the Political Network HHI measures the political concentration in an area with respect to the percentage share of the communities that are present. If no confusion arises, sometimes we refer to the Political Network HHI simply as Political HHI (or just HHI). 

Equation \ref{PNHHI} differs from that in other studies, which considers each unique surname in a network as a different clan. Surname-based Political HHI tends to understate the level of dynastic concentration compared to the Political HHI we describe here. To illustrate this, consider a hypothetical provincial election with 10 seats controlled by three political clans. Suppose politicians with surname A control 5 seats, those with surname B control 3 seats, and those with surname C control the remaining 2. Suppose further that clans A and B are connected by an inter-marriage alliance (which usually would only be indicated via middle name matches). Therefore, the Political HHI for this province, calculated using surname matching, would be $HHI = 10,000\times \left[(\frac{5}{10})^2 + (\frac{3}{10})^2 + (\frac{2}{10})^2\right] = 3,800$. However, taking into account middle name matches, would indicate that clans A and B are really part of the same larger family. So, the Political HHI would now become $HHI = 10,000\times \left[(\frac{8}{10})^2  + (\frac{2}{10})^2\right] = 6,800$, an increase of almost double.

\subsection{Network Sparsity}

\begin{definition} \label{Gini} [\cite{GINI-Paper}] In economics, the \textbf{Gini Coefficient} measures the inequality of resource distribution in a certain population or area. Its index value is normalized and lies between $0$ and $1$, with increasing inequality present for values closer to $1$.
\end{definition}

\begin{definition} [\cite{sparsityGINI}]
Let $G = (V(G), E(G))$ be a simple, undirected network with $|V(G)| = n$. Let $A = [a_{ij}]_{nxn}$ be the corresponding adjacency matrix of $G$, where $x_{ij}$ is the \textit{weighted} connection between nodes $i$ and $j$. Define the following:
\begin{enumerate}
    \item[\textbullet] $x_i$ is the number of connections made by the $i$th individual, i.e. \[
x_i = \sum_{j=1}^{n} x_{ij}, \quad \forall i = 1, 2, \ldots, n.\]
    \item[\textbullet] $\mathbf{x} = [x_1', x_2',\ldots,x_n']$ is the list of connections for all $n$ nodes, arranged in \textit{ascending} order. 
\end{enumerate}
Then the \textbf{Centrality Gini Coefficient (CGC)} is defined by
\begin{equation}
    CGC = 1 - 2\sum_{i=1}^{n}  \left[\frac{x_i}{\sum_{i=1}^{n}x_i} \cdot\frac{n - i + \frac{1}{2}}{n} \right], x_i \in \mathbf{x}.
    \label{Orig-CGCformula}
\end{equation} 
\end{definition}

For coding convenience, in this paper, we used our own formulation of the Centrality Gini Coefficient (CGC).

\begin{definition} \label{CCD}
Given a network $G = (V, E)$, the \textbf{CGC} measures the Gini coefficient of \textit{weighted centrality} of each $v \in V(G)$ and is defined by:
\begin{equation}
    CGC = \dfrac{\displaystyle\sum_{i=1}^{n}\left(2i - n -1 \right)x_i}{n \displaystyle\sum_{i=1}^{n}x_i}, x_i \in \mathbf{x}
    \label{Actual-CGCFormula},
\end{equation}
where:
\begin{enumerate}
    \item[\textbullet] $n$ is the total number of politicians in the network; 
    \item[\textbullet] $x_i$ is the number of connections made by the $i$th individual, i.e. \[
x_i = \sum_{j=1}^{n} x_{ij}, \quad \forall i = 1, 2, \ldots, n; \text{ and}\]
    \item[\textbullet] $\mathbf{x} = [x_1', x_2',\ldots,x_n']$ is the list of connections for all $n$ nodes, arranged in \textit{ascending} order.
    
\end{enumerate}

Equations \ref{Orig-CGCformula} and \ref{Actual-CGCFormula} can be shown to be equal as follows:
\begin{align*}
    CGC &= 1 - 2\displaystyle\sum_{i=1}^{n}  \left[\dfrac{x_i}{\displaystyle\sum_{i=1}^{n}x_i} \cdot\dfrac{n - i + \frac{1}{2}}{n} \right] 
    = 1 - 2\left[\dfrac{\displaystyle\sum_{i=1}^{n}x_i\left(n-i+\dfrac{1}{2}\right)}{n\displaystyle\sum_{i=1}^{n}x_i}\right] \nonumber \\
    &= \dfrac{n\displaystyle\sum_{i=1}^{n}x_i - \displaystyle\sum_{i=1}^{n}x_i\left(2n - 2i + 1\right)}{n\displaystyle\sum_{i=1}^{n}x_i} 
    = \dfrac{\displaystyle\sum_{i=1}^{n}\left(2i - n -1\right)x_i}{n\displaystyle\sum_{i=1}^{n}x_i}. 
\end{align*}

\noindent\textbf{Remarks:}
\begin{enumerate}
    \item $CGC$ measures the \textbf{compactness} and \textbf{sparsity} of a network as it both considers the \textit{weights} and \textit{connections} present.
    \item $CGC$ value is normalized, i.e. $CGC \in [0,1]$. A $CGC$ value close to $1$ suggests that the network $G$ is centered around highly-connected nodes while closer to $0$ indicates homogeneous connection distribution in $G$.
\end{enumerate}
\end{definition}

\subsection{Inter-Community Connectedness} \label{CCD}
\begin{definition}
    Let $G = (V, E)$ be a network and $C(H)$ be the set of all connected components $H$ present in $G$.
    Then the \textbf{Connected Component Density} $(CDC)$ is defined as:
    \begin{equation}
        CCD = 1 - \frac{|C(H)|}{|V(G)|}.
        \label{CDC-Formula}
    \end{equation}

\noindent\textbf{Remarks:}
$CCD$ reflects the \textit{distribution} of all nodes among the \textit{connected components} present in the network $G$. That is, if $CCD \rightarrow 0$, almost all nodes are its own component. Meanwhile, $CCD \rightarrow 1$ implies that almost all nodes belong to a single connected component, i.e. $CCD = 1 - \frac{1}{|V(G)|} \rightarrow 1$ for a large network.
\end{definition}

\subsection{Intra-Community Connectivity} \label{ACC}

For this section, we define $G_{P_{i}} = (V(G_{P_{i}}), E(G_{P_{i}}))$ as a specific connected \textbf{community} $i$ that persists in Province $P$ for a specific year, comprising of $k$ families, namely, $F_1, F_2,\ldots, F_k$.

\begin{definition}
The \textbf{vertex community connectivity} for a specific community $G_{P_i}$, denoted by $\gamma(G_{P_{i}})$, is the minimum number of vertices that needs to be removed to dismantle a connected community into disconnected, individual subunits. 
\end{definition}

The vertex community connectivity measures the \textbf{intra-community strength} that persists between the community extracted from Leiden Algorithm as the said metric directly evaluates the number of connections present in the community as well as the number of dynastic units in the collection.

\begin{definition}
The \textbf{average community connectivity (ACC)} is defined as the sum of all the ratios between the vertex community connectivity $\gamma(G_{P_{i}})$ to the total number of political units present in the community, $|V(G_{P_{i}})|$, for all the extracted Leiden communities present in the province. Therefore, the \textbf{ACC} is defined mathematically as:
    
    \begin{equation*}
        ACC=\sum_{i \in P} \frac{\gamma(G_{P_{i}})}{|V(G_{P_{i}})|}.
    \end{equation*}
\end{definition}

\section{Regression}
\begin{definition}
    Let $\{(x_{ij}, y_{j})\}_{i=0}^N{}_{j=1}^D$ be $D$ total number of data points comprising of $x_1, x_2,\ldots,x_N$ independent features and a dependent label $y$. A \textbf{simple regression} characterizes the latent relationship and its strength between $N$ number of features $x_i$ and a label $y$ by establishing the following regression equation: 
    \begin{equation}
        \label{eq:OLS}
        \hat{y} = x_0 + \beta_1x_1 + \beta_2x_2 + \cdots + \beta_Nx_N + \epsilon,
    \end{equation}     
    where $\beta_i$ is the intercept of variable $x_i$ and $\epsilon$ is a random error for which $\epsilon \sim \mathcal{N}(0, \sigma^2)$.
\end{definition}

\begin{definition}
    Let $y$ and $\hat{y}$ be the actual and predicted values of the labels using Equation \ref{eq:OLS}, respectively. The \textbf{ordinary Least Squares (OLS) regression} minimizes the squared differences between $y$ and $\hat{y}$ across all data points, i.e. $\sum_{i = 1}^{D} (y_i - \hat{y}_i)^2$. 
\end{definition}

\subsection{Regression Modeling Techniques}
In this section, a panel data is the set $\{(x_{1it},x_{2it},\ldots,x_{Dit}, y_{it})\}_{i=1}^G{}_{t=1}^T$, with $D$ total number of features present in the data across $N$ entities over $D$ time periods. 

\begin{definition} The \textbf{fixed effects modeling} is a more robust simple regression model that controls for all time-invariant differences that persist between entity $G$ (known as fixed effects) or analyzes the impact of features that vary over time $D$ (time effects). Mathematically, it is given by the following regression formula:
    \begin{equation}
            \label{eq:FE}
            \hat{y_{it}} = \beta_0 + \beta_1X_{1_{it}} + \beta_2X_{2_{it}} + \cdots + \beta_DX_{D_{it}} + \alpha_i + \epsilon_{it},    
    \end{equation}
    where $\alpha_i = \beta_0 + \beta_PZ_i$ are the fixed-effects of entity $i$ caused by the time-invariant heterogeneity across entity $i$ and $\epsilon_{it}$ is the random error present in the regression model.
\end{definition}

\begin{definition} The \textbf{random effects modeling} measures a believed difference or variability between entities that have an expected influence on the dependent variable. Mathematically, it is given by:
    \begin{equation}
            \label{eq:RE}
            \hat{y_{it}} = \alpha + \beta_1X_{1_{it}} + \beta_2X_{2_{it}} + \cdots + \beta_DX_{D_{it}} + U_i + \epsilon_{it},    
    \end{equation}
    where $U_i$ is the random effect for entity $i$ and $\epsilon_{it}$ is the residual, with $U_i$ and $\epsilon_{it}$ being positively correlated under the same entity.
\end{definition}

\begin{definition}
    Consider $(K+1)$ features $\{x_{\phi _1it}, x_{\phi_2{it}}, \ldots, x_{\phi _K{it}}\}$ from the panel data as fixed variables and the $(J+1)$ features $\{x_{\psi_1it}, x_{\psi_2{it}}, \ldots, x_{\psi_J{it}}\}$ as the random variables, not necessarily distinct from features considered as fixed. The \textbf{linear mixed modeling (LMM)} takes into account a repeated measurement of features present in hierarchical level of data defined in the model. The LMM is defined 
    mathematically as the following:
    \begin{align}\label{eq:LMM}
            \hat{y_{it}} &= \beta_0 + \beta_{\phi_1}X_{\phi _1{it}} + \cdots + \beta_{\phi_K}X_{\phi_K{it}} + \upsilon_{\psi_1} \\
            &+\eta_{\psi_1}X_{\psi _1{it}} + \cdots + \upsilon_{\psi_J} + \eta_{\psi_J}X_{\psi_J{it}} + \epsilon_{ij},
    \end{align}
    where $\beta_0$ is the model's overall intercept, $(\beta_{\phi_k'}X_{\phi_k'{it}})$ is the fixed effects of feature $k'$ $X_{\phi_k'{it}}$ in the model with coefficient $\beta_{\phi_k'}$, $ (\upsilon_{\psi_j'} + \eta_{\psi_j'}X_{\psi_j'{it}})$ is the random effect of feature $j'$ with feature-specific slope and intercept equal to $\eta_{\psi_j'}$ and $\upsilon_{\psi_j'}$, respectively, and residual $\epsilon_{ij}$.
    
\end{definition}

\subsection{Regression Metrics}
\begin{definition}
    Let $y_i$, $\hat{y_i}$, and $\bar{y}$ be the response value, predicted value from the regression model, and the mean of all response values, respectively. The \textbf{coefficient of determination} $R^2$ measures the proportion of variation in the dependent variable that is explained by the features. Mathematically, it is given by the equation:

    \begin{equation}
        \label{eq:R2-formula}
        R^2 = 1-\frac{\displaystyle\sum_i(\hat{y_i} - y)^2}{\displaystyle\sum_i(y_i - \bar{y})^2}.
    \end{equation}
\end{definition}

\begin{definition}
    Let $\sigma^2_f$ and $\sigma^2_e$ be the fixed-effects and residual variance in a typical LMM model. Define $\sigma^2_l$ as the random variance component for each level $l$ assumed as the random factor. The \textbf{conditional $R^2$} measures the proportion of variance explained by both fixed and random factors assumed in the LMM model. Mathematically, it is formulated as:

    \begin{equation}
        \label{eq:R2cond-Formula}
        R^2_{cond} = \frac{\sigma^2_f + \displaystyle\sum_l\sigma^2_l}{\sigma^2_f + \displaystyle\sum_l\sigma^2_l + \sigma^2_e} .
    \end{equation}
\end{definition}

\begin{definition}

    Let $\widehat{L(\theta)}$ be the maximum value of the likelihood function derived from the regression model. Additionally, let $k$ be the number of estimated parameters in the model. The \textbf{Akaike Information Criterion (AIC)} is defined as an estimator of prediction error which provides a relative quality of fit of the statistical model for the given set of data. The formula for AIC is given by:

    \begin{equation}
        \label{eq:AIC-formula}
        AIC = 2k - 2\ln(\widehat{L(\theta)}).
    \end{equation}
\end{definition}

\chapter{Review of Related Literature}
% ...

\section{On the Origins and Persistence of Dynastic Rule}

Though many dynastic clans have colonial roots, scholars have argued that the modern system of dynastic rule largely originates from the introduction of term limits with the adoption of the 1987 Constitution \cite{querubin_political_2012, beja_jr_inequality_2012}. Before the declaration of Martial Law in 1972, local officials were elected for four-year terms with no term limits. The 1987 Constitution introduced the current system whereby local officials are allowed to hold office for no more than three consecutive three-year terms to foster political competition by curbing incumbent advantage.

A seminal study by Querubin \cite{querubin_political_2012} argues that term limits had the perverse effect of strengthening political dynasties. Term limits forced officials to turn to alternative means of retaining their grasp on power-–often fielding their relatives to replace them. Eventually, members of the same family alternate positions, entirely circumventing term limits. Querubin also suggests that the introduction of term limits had a secondary effect of discouraging would-be challengers. Rather than run against a dynastic incumbent, challengers prefer to wait till their term limit approaches–making elections \textit{less competitive}. Analyzing the results of congressional and gubernatorial races from 1947 to 2010, the study concluded that not only did term limits have no significant effect on curbing political clans' hold on power, the incumbency advantage of \textit{families} significantly increased after 1987.

A crucial question the term limit hypothesis leaves unanswered is why incumbents turned to their families rather than political parties to circumvent term limits. Querubin hypothesizes that dynasties initially emerge as a natural alternative in instances where political trust is low. Dynasties take on the role of being ``long-lived" units of political organization that can promise policy continuation. Over time, they build up ``brand name" recognition in the same way that parties do. Dynasties remain powerful, and no parties can consolidate to challenge them. 

This, however, does not explain why dynasties have remained powerful even in contexts where strong political parties \textit{have} emerged (i.e. the Liberal Party under the Aquino Administration, \textit{Partido Demokratiko ng Pilipinas (PDP)-Laban} under the Duterte Administration). Why do candidates return to relying primarily on kinship networks for political longevity when wider-reaching, formally organized parties exist as an alternative? Much remains to be established regarding how dynasties interact with political parties, and how different interactions translate to policy.

Labonne et al. \cite{labonne_political_2021} revisit the term limit hypothesis with a focus on the role the introduction of term limits played in the rise of female candidates post-1987.  They found that women were both more likely to run and more likely to win in forced-open races (races where the previous incumbent was term-limited). However, they noted that the increase in female candidates was almost entirely caused by a surge in female relatives of dynastic incumbents. They suggested that the significant disparity in electoral success between dynastic and non-dynastic female candidates was only even further indicative of the heightened role of dynastic power upon the introduction of term limits. 

Mendoza et al. \cite{mendoza_term_2019} challenged the view that term limits were the primary factor that lead to the surge in dynastic politics. They claimed it was the failure to introduce key ancillary reforms accompanying the introduction of term limits that created the conditions fostering fat dynasties. Examining evidence from dominant political clans in Samar and Lanao Del Norte, they further argued that removing term limits in the current political landscape would only allow dynastic incumbents to secure their political foothold, while failing to introduce meaningful political competition. Though term limits may have catalyzed the rise of dynasties, their introduction alone does not explain the extent of dynastic prevalence in Philippine politics today. Much remains to be investigated surrounding how dynasties have consolidated power over the years and why they continue to maintain their grasp on power. 

\section{Quantifying Dynastic Power} \label{sec:Quantifying Dynastic Power}

Previous studies have devised a variety of indicators characterizing the power and scope of dynasties. These indicators vary in complexity and scale. The most basic indicators categorize dynasties along binary characteristics. One of the most commonly used binary indicators is the notion of \textit{fatness} and \textit{thinness}. A dynasty is ``fat" if multiple members of the same clan occupy different elected positions at the same time (\textit{sabay-sabay}) and ``thin" if members of the same clan successively occupy the same position (\textit{sunud-sunod}). This indicator is inadequate, however, as dynasties in practice are often both \textit{fat} and \textit{thin} at the same time. Further, the notion of fatness and thinness alone does not distinguish between clans of different sizes and degrees of influence. These limitations are highlighted in a study by Mendoza, Jaminola and Yap (2019) \cite{mendoza_fat_2019}. Using a local election dataset spanning 1988 to 2019, they find that the share of fat dynasties grew from 19\% in 1988 to 27\% in 2017. However, they note that the ``fat-share" alone does not capture the fact that ``fat" dynasties have also been growing more expansive. In 2001, for instance, there were only 157 dynasties with 4 or more active members. That number rose to 217 by 2019. 

Another ubiquitous binary indicator is the concept of the \textit{Governor-Congressman-Mayor (GCM) Link}. In a given municipality, there is a GCM Link if the Governor, Congressman, and Mayor are blood-related. GCM Links are an indicator of dynastic capture at the highest local level positions. As these positions often have outsized influence on local policy, a single family controlling all three seats creates a powerful political monopoly. 

An alternative approach has been to analyze dynastic power using indicators typically applied to political parties. These include measures such as \textit{dynastic membership} (the number of active incumbents belonging to a particular clan), \textit{dynastic seat share} (the proportion of available positions in a given race that are occupied by a given family), and \textit{size of largest dynasty} (the number of politicians belonging to the biggest clan in a given province). Though such indicators are useful for comparing clans within the same constituency, they do not allow for reliable cross-province comparisons.

To meaningfully compare dynastic presence across provinces, Mendoza, Beja, Venida, \& Yap (2013) \cite{mendoza_political_2013} adopt the \textit{Political Herfindahl-Hirschman Index (Political HHI)}. In microeconomics, HHI is a standard indicator of market concentration (see \hyperref[HHI]{Definition 2.5.1}). Political HHI extends this concept to dynasties, treating each clan as a separate ``firm" and taking the sum of squared dynastic seat share across all clans in a given province. The key advantage of Political HHI is that it allows for robust comparisons of electoral competitiveness both across provinces and election cycles. It provides a meaningful sense in which dynastic concentration has ``increased" or ``decreased" over time. The same study, using data on provincial elections between 2001 and 2010, found that dynastic influence as measured by Political HHI and dynastic seat share increased on average. 

This construction of Political HHI, which is adopted in other studies \cite{mendoza_political_2022, mendoza_political_2022-1, davis_corruption_2024, mendoza_political_2016}, makes two key assumptions. First, each distinct surname is treated as a separate political clan (i.e. even if two families are related through intermarriage, they are considered two separate clans). Secondly, all positions are given the same weight when computing dynastic seat share (e.g. mayoral seats are given the same weight as municipal councilors). This simplification fails to reflect the actual disparities in the distribution of power between local positions. 

 Crucially, no single dynastic indicator should be taken as ``standard" as they reflect different features of dynastic power. In practice, multiple indicators are often used in conjunction to capture a variety of dynastic characteristics most relevant to the research problem. 

\section{Identifying Dynastic Ties}

Dynasties are often interconnected in a complex web of marriage alliances. The challenge for empirical research on dynastic power concentration is on determining whether two related families should be treated as separate entities or as part of a larger clan. The earliest studies \cite{balisacan_going_2004} relied on manual identification strategies to identify the different clans within a province. As manual identification methods are infeasible for larger election datasets, most studies \cite{mendoza_dynastic_2020, mendoza_political_2022-1, mendoza_political_2022} rely on within-province surname matching to identify members of the same clan. Two politicians within the same province (either within the same or successive election cycles) are considered related if they share a common surname \cite{mendoza_political_2022}. Restricting the search for political kin to only those within the same province minimizes the misidentification of politicians who incidentally happen to share the same last name. While the Philippines has a fair share of common surnames (e.g. Dela Cruz, Santos, Abad, etc.), Querubin \cite{querubin_political_2012} argued that historical factors make it unlikely for rival political clans within the same province to share a coincidental surname. 

A more significant limitation of surname matching is that it ignores the spousal connections indicated by a politician's middle name. Beja, Mendoza, Venida, and Yap (2012) \cite{beja_jr_inequality_2012} assert that for Philippine election data, matching based on surname alone yields similar results as using both surname and middle name. This assertion, however, does not hold when introducing more complex measures of dynastic power (see the discussion after Definition \ref{PHHI}). 

Relying solely on surname identification tends to understate the true level of dynastic concentration. On the other hand, broadening the definition of clans to include relatives up to any degree would be equally impractical. Thus, there is a need for a robust means of identifying and distinguishing between political families that more accurately reflects true clan dynamics.

\section{Network Analyses of Political Dynasties}

 Network analysis is a broad field that applies graph-theoretic concepts and techniques to the study of social and physical networks \cite{engel_network_2021}. Viewing dynasties from the perspective of network science paves the way for novel methods of characterizing not only their degree of political influence but also their structure and evolution \cite{mendoza_dynastic_2020}. A close-knit dynasty composed of a single nuclear family may have very different incentives, and thus contribute to different development outcomes, than a large extended dynasty with many expansive branches.

While several studies have applied network analysis to the study of Philippine political dynasties, these have largely focused on characterizing the influence of individual politicians within a given network rather than extracting insights from the aggregate structure of the network. Using survey data taken during the 2013 and 2016 local elections, Cruz et al. \cite{cruzAER_2017} reconstructed the familial networks of provincial election candidates from over 15,000 villages. Within each village, they identified the eigenvector centrality of each candidate and major clan. They found that candidates disproportionately originate from more central clans. Moreover, within familial networks, candidates of higher centrality received a higher share of votes. They found that the advantage of candidates with higher centrality is significant even when controlling for factors such as wealth, historical elite status, and previous electoral success. Their study highlighted how, even within dynasties, the distribution of power is often highly unequal.

Mendoza et al. \cite{mendoza_dynastic_2020} similarly adopted network analysis to analyze the evolution of political clans in Western Samar from 1988 to 2016. They found that the two major clans in the province, the Uy and Tan dynasties, were organized in a similar hub-and-spoke structure wherein certain prominent ``heads" of both families had significantly higher centralities than the rest of their kin. They also found that, when these two clans became interlinked via a marriage alliance, the connecting individual (in this case Stephany Uy-Tan) became one of the most central politicians in the clan and subsequently assumed higher ranked positions. They also found that instances of political violence tend to be targeted against individuals who were less central to the network (although the evidence presented is largely anectodal). 

In both of the above studies, the emphasis in applying network analysis is to draw novel insight about the individual politicians within the network rather than examining the network's aggregate structure. Balanquit et al. \cite{balanquit_measuring_2017} used a combination of Political HHI and a novel graph-theory inspired \textit{dynastic index} to explore the evolution of dynastic concentration within the major municipal governments of Metro Manila. The dynastic index of a clan takes the sum of weighted degree centralities of all members of that clan holding office. The weight of the edge connecting two related politicians is based on both of their elected positions. A connection between two ``high-ranking" politicians (defined as Mayor, Vice Mayor, or Congressman) is assigned weight 3. A connection between two ``low-ranking" politicians (all other positions) is given weight 1. An edge between a low- and high-ranking politician is given weight 2. In order to account for both the ``fatness" and ``thinness" of dynasties, the dynastic index considers connections between politicians in successive election cycles. The authors found that between 1988 and 2013, dynastic power concentration in all cities within Metro Manila has increased, with Las Pinas having the greatest level of dynastic concentration. 

Network analysis offers a promising new lens for understanding the power and persistence of dynastic rule. In this study, we expand on this approach both by developing novel metrics for characterizing the network graph of dynasties on an inter-provincial scale, but also by demonstrating that these network-derived metrics are significant predictors of socioeconomic outcomes–-thus providing an empirical method of verifying the effect dynasties have on local development.

\section{Dynasties and Local Development Outcomes}

Dynasties have long been theorized to cause the deterioration of local development outcomes. Once a local dynasty has accumulated a critical mass of influence, its members are able to evade checks and balances. The presence of influential dynasties further deters non-dynastic challengers from running against them, eroding democratic accountability. This creates a vicious cycle by which dynasts are able to remain in power while only minimally catering to voter interests. 

Balisacan and Fuwa  \cite{balisacan_going_2004} were the first to incorporate dynastic concentration in a formal macroeconomic model of sub-national growth. Using provincial-level economic data from 1988 to 1997, they found that dynastic concentration (as measured by the share of provincial officials who are blood-related to at least one other official concurrently in office) has a significant negative relationship with per capita income and mean expenditure growth rate, but no significant association with poverty reduction. The last finding is notable as it runs contrary to the theoretical consensus that dynasties exacerbate poverty. The authors themselves include the caveat that this may have resulted from an error in their manual identification strategy. 

The aforementioned study by Mendoza, et al. (2013) revisits the relationship between dynasties and poverty. While Balisacan and Fuwa \cite{balisacan_going_2004} used only dynastic share as an indicator of dynastic power, Mendoza et al.  \cite{mendoza_political_2013} also included the Political HHI and the size of the largest dynasty. They similarly established that the relationship between dynastic share and poverty is insignificant. However, they noted that both poverty incidence and income per capita \textit{were} related to both Political HHI and the size of the largest dynasty. They suggest this is indicative of a more nuanced relationship between dynasties and poverty; that is, while poverty does not induce dynasties to emerge, it may contribute to the expansion and concentration of power in established clans. 

Expanding on this link, Mendoza et al. \cite{mendoza_political_2022-1} adopted a more complex mixed-effects model of provincial poverty that accounts for both the dynastic share and the local economic competitiveness using data from 2009 to 2018. Two proxy variables are used to gauge local competitiveness: \textit{business dynamism} represented as the annual business tax collection of the region, and \textit{ownership} which is a discrete variable representing the perceived extent of local business ownership by politicians. Ownership was derived from subjective expert survey data. Dynastic share had a significant positive relationship with poverty incidence ($\beta \approx 0.214, \,p < 0.01$). When considering only provinces outside of Luzon, this effect was even stronger at $\beta \approx 0.706, \,p < 0.01$. \textit{Ownership} had a weak significant relationship with poverty, and business dynamism was insignificant. 

Other studies have found more direct links between dynastic concentration and a variety of developmental outcomes. A recent study by Davis et al.  \cite{davis_corruption_2024} investigates the link between dynasties and corruption. Using a novel Corruption Risk Indicator (CRI) dataset assembled using public procurement data on government infrastructure contracts, they developed a fixed-effects model of provincial corruption risk as a function of Political HHI, the size of largest dynasty, GCM Links, population, and Internal Revenue Allotment (IRA). They found a weak positive ($\beta \approx 0.002, \, p < 0.05)$ but significant relationship between CRI and HHI, and a moderate positive ($\beta \approx 0.02, \, p < 0.05)$ but significant relationship between CRI and the size of the largest dynasty. These results cohere to the theoretical consensus that dynastic rule undermines systems of checks and balances and increases the risk of corruption. 

Scholars have more recently examined alternative measures of development such as the provincial Human Development Index (HDI). Villanueva (2022) \cite{villanueva_political_2022} investigated the relationship between dynastic share and HDI among 13 municipalities within Rizal province between 2001 and 2017. They found a significant negative relationship between HDI and dynastic share---supporting the hypothesis that prolonged dynastic rule deteriorates holistic standards of living. 

Much remains to be investigated in establishing the precise line between dynastic rule and local development. The literature thus far has almost exclusively used dynastic indicators related to power but not clan structure. Moreover, there is a need to incorporate the effect of inter-dynastic dynamics. A province with multiple competing dynasties may enjoy more positive development outcomes than one where all the ruling clans are inter-allied. Understanding how specific clan characteristics shape their effect on development may serve as a crucial guide to crafting anti-dynastic policy.


\chapter{Methodology}

The methodology comprises three major sections. We first develop a series of robust indicators of local dynastic power leveraging concepts from graph theory. From these metrics, we draw insights regarding the evolution of political dynasties over time and across geographies. In the second section, we explore the relationship between dynastic membership and partisan allegiance as indicated by variations in party-hopping rate. We further investigate whether there are significant differences in the extent to which dynastic versus non-dynastic candidates tend to bandwagon with the most dominant national party at the time. Finally, we investigate the relationship between the dynastic indicators developed and local development outcomes. We develop a series of Linear Mixed-Models (LMMs) to test the relationship between our dynastic indicators and poverty incidence and human development index (HDI) respectively. This chapter also provides important details on how the methods stated above were implemented, such as the pre-processing of the dataset, the formulation of the proposed metrics, as well as the tests and verification of results.

\section{Election Dataset}

\subsection{Data Sourcing}

This study analyzes the results of seven local election cycles from 2004 to 2022. The following positions were considered: Governor, Congressional District Representatives, Provincial Governor, Provincial Vice Governor, Municipal Mayor, Municipal Vice Mayor, Provincial Board Member (\textit{Sangguniang Panlalawigan}), and City Councilors (\textit{Sangguniang Panglungsod}). The main election dataset was compiled from a variety of publicly available sources. Data was initially sourced from the Ateneo Policy Center's (APC) Political Dynasties Dataset, which contains a record of all locally elected Philippine officials from $2004$ to $2016$ \cite{APC-2016}. Table \ref{tab:meta1} shows the metadata of the APC Political Dynasties Dataset that will be used as a base template for this study, including their respective data types.

\begin{table}[H]
   \centering
       \caption{Metadata of APC Political Dynasties Dataset}
    \renewcommand{\arraystretch}{1.2} 
    \begin{tabular}{>{\centering\arraybackslash}m{3cm} | >{\centering\arraybackslash}m{6cm} | >{\centering\arraybackslash}m{3cm} }
        \hline
        \textbf{Variable} & \textbf{Description} & \textbf{Data Type} \\
        \hline
        Last Name & Indicated Last Name in COMELEC Election Returns & String \\
        \hline
        First Name & Indicated First Name in COMELEC Election Returns & String \\
        \hline
        Position & Indicated Position in COMELEC Election Returns & String \\
        \hline
        Party & Indicated Party Affiliation in COMELEC Election Returns & String \\
        \hline
        Region & Indicated Regional Designation in COMELEC Election Returns & String \\
        \hline
        Province & Indicated Provincial LGU in COMELEC Election Returns & String \\
        \hline
        Year & Election Year & Integer \\
        \hline
        Municipality.City & Indicated City or Municipality in COMELEC Election Returns & String \\
        \hline
        Fat & Marker that flags a fat dynasty & Binary \\
        \hline
    \end{tabular}
    \label{tab:meta1}
\end{table}

However, there were a few notable deficiencies in the APC dataset. The dataset contained no record of politicians' middle names which was crucial to reconstructing the provincial dynastic networks. The dataset also lacked results for the 2019 and 2022 local elections. 

To address these issues, election results were extracted from a variety of other sources. A full list of candidates for the years $2010$, $2016$, and $2019$ was obtained from the Commission on Election's (COMELEC) - National Election Results website. These lists contained both the candidates middle names and their official party list. Since the COMELEC data contained candidates (including non-winners), a cross-checking between the other dataset was performed to extract only the middle names. Furthermore, PDF copies of the official certified election results from $2004$ to $2019$ were sourced from the Bureau of Local Government Supervision (BLGS). These documents were made available in response to an online Freedom of Information (FOI) request. However, the $2004$ results only contained candidates at the provincial level and did not contain their political party affiliation. Another issue that arose was the fact that the PDF documents themselves were inconsistently formatted, requiring manual spot checking. For $2022$, data was gathered from the Department of Interior and Local Government's (DILG) Official Masterlist of local officials, which contained the full names (including middle names) of all local incumbents elected during the 2022 elections \textit{except} for those in the provinces of Tawi-Tawi, Sulu, Dinagat Islands, and Compostela Valley. The DILG Masterlist however, did not include candidate's party affiliations. The election data gathered for $2010$, $2016$, $2019$, and $2022$ were cross-referenced with records from National Citizens' Movement for Free Elections (NAMFREL), which contained a complete record of all candidates party affiliations even for 2022. Data availability is summarized in Table \ref{tab:data-sources}.

\begin{table}[H]
    \caption{Data Availability on different sources}
    \centering
    \renewcommand{\arraystretch}{1.2}
    \setlength{\tabcolsep}{8pt}  
     \begin{tabular}{>{\centering\arraybackslash}m{2cm} | >{\centering\arraybackslash}m{1.5cm} | >
     {\centering\arraybackslash}m{1.5cm} | >
     {\centering\arraybackslash}m{1.5cm} | >
     {\centering\arraybackslash}m{3cm} }
        \hline
        \textbf{Source} & \textbf{Years} & \textbf{Party?} & \textbf{Middle Name?} & \textbf{Remarks} \\
        \hline
        APC & 2004 - 2016 & Yes & No & \\  
        \hline
        COMELEC & 2010, 2016, 2019 & Yes & Yes & Candidates only \\  
        \hline
        BLGS & 2004 - 2019 & Yes & Yes & 2004 has no party affiliations \\ 
        \hline
        NAMFREL & 2016 - 2022 & Yes & No & Candidates only \\  
        \hline
        DILG & 2022 & No & Yes & \\  
        \hline
    \end{tabular}

    \label{tab:data-sources}
\end{table}

\subsection{Preprocessing}

Records obtained in PDF format were first processed via Optical Character Recognition (OCR) through Python, which extracted the data into machine-readable text. The relevant fields (e.g. name, party affiliation, province, etc.) were then extracted from each file. This was then converted into a comma separated values (csv) file format and then appended onto a single dataset alongside containing the information for all years. Data from the DILG Masterlist was directly extracted and appended onto the dataset.

An issue encountered while matching candidates' middle names and party affiliations was that the naming schemes used by the various datasets were often inconsistent. This was solved using Record Linkages under the \textit{Record Linkage} library in Python, which allows merging various datasets while taking into account that the data fields are prone to variations such as due to spelling errors or inconsistent formatting. Minor inconsistencies in naming scheme were handled by enforcing a standardized handling rule (e.g. replacing all ``ñ" characters with ``n" to avoid UTF encoding errors, representing all independents as ``IND"). In cases of missing fields where no practical method of accurately determining the appropriate identifier (e.g. candidates with missing middle names or party affiliations), the fields are left blank.

\subsection{Final Dataset}

Due to incomplete data that could not be imputed from the available sources, seven provinces were dropped from the final dataset: Sulu, Tawi-Tawi, Compostela Valley, North Cotabato, Davao de Oro, Dinagat Islands, and Davao Occidental. The final dataset thus spans a total of $80$ provinces (counting the districts of NCR) across $7$ electoral years, covering over 122,000 (non-unique) local officials. Of these, middle names were found for 108,787 or 89.14\% of all candidates in the dataset, and only 1,170 candidates (less than 1\% of the dataset) have unidentified party affiliations.

To facilitate analysis in later sections, the dataset includes three additional features. ``Community" is an integer identifier denoting the clan a politician belongs to within their province and electoral year as determined by the Leiden algorithm (discussed in the next section). ``Dynastic" is a binary identifier equal to $1$ if a candidate is dynastic i.e. if they enter office simultaneously with at least one  other person belonging to the same community in the same province during the same election year. Finally, ``Hopper" is a binary identifier equal to $1$ if a candidate changed party affiliations from the last election cycle. Since the hopper identifier is dependent on the previous election year, only the years $2007$ until $2022$ were considered. Furthermore, it should be noted that due to the limitations of the dataset, only candidates who occupy successive terms in office can be identified as party hoppers (e.g. a candidate who runs under Party A but loses, then wins the next term under Party B would not be considered a party hopper).

 Table \ref{tab:meta2} summarizes the variables present in the final dataset. 

\begin{table}[H]
\caption{Metadata of the Final Dataset used in this Study}
   \centering
    \renewcommand{\arraystretch}{1.2} 
    \begin{tabular}{>{\centering\arraybackslash}m{3cm} | >{\centering\arraybackslash}m{6cm} | >{\centering\arraybackslash}m{3cm} }
        \hline
        \textbf{Variable} & \textbf{Description} & \textbf{Data Type} \\
        \hline
        Last Name & Indicated Last Name in COMELEC Election Returns & String \\
        \hline
        First Name & Indicated First Name in COMELEC Election Returns & String \\
        \hline
        Position & Indicated Position in COMELEC Election Returns & String \\
        \hline
        Party & Indicated Party Affiliation in COMELEC Election Returns & String \\
        \hline
        Region & Indicated Regional Designation in COMELEC Election Returns & String \\
        \hline
        Province & Indicated Provincial LGU in COMELEC Election Returns & String \\
        \hline
        Year & Election Year & Integer \\
        \hline
        Municipality.City & Indicated City or Municipality in COMELEC Election Returns & String \\
        \hline
        Middle Name & Indicated Middle Name in COMELEC Election Returns & String \\
        \hline
        Community & Inferred Community as detected by Leiden Algorithm& Integer \\
        \hline
        Dynastic & Marker that flags if politician belongs to a dynasty & Binary\\
        \hline
        Hopper & Marker that flags politican changed parties & Binary \\
        \hline
    \end{tabular}
    
\label{tab:meta2}
\end{table}

\section{Dynasties Network Construction}

To construct the political network graph for a given province and year, each politician is first assigned to a node $i$, with an assigned node weight based on their current elected position. Though the specific weighting scheme used in this study is arbitrary, the weights are assigned to capture the relative difference in influence held by officials in each position. The weight assignments are shown in table \ref{tab:political_weights}. 

\begin{table}[h]
\caption{Node Weights Assigned to Political Positions}
    \centering
    \begin{tabular}{|l|c|}
        \hline
        \textbf{Position} & \textbf{Weight} \\
        \hline
        Councilor & 2 \\
        Provincial Board Member & 2 \\
        Vice Mayor & 3 \\
        Vice Governor & 3 \\
        Mayor & 5 \\
        Member, House of Representatives & 5 \\
        Governor & 5 \\
        \hline
    \end{tabular}
    \label{tab:political_weights}
\end{table}

Two nodes $i$ and $j$ are connected by an edge denoted by $ij$, if and only if they \textbf{share a common last name or middle name}. The weight of $ij$ is the product of the node weights of $i$ and $j$ multiplied by a scalar factor to account for the approximate degree of consanguinity between politicians. Approximate consanguinity is used as name data alone does not reflect the actual relationship between politicians. The approximate degree of consanguinity is shown in Table \ref{tab:scalar_factors}.

\begin{table}[H]
\caption{Scalar Factors for Name-Based Connections}
    \centering
    \begin{tabular}{|l|c|}
        \hline
        \textbf{Connection} & \textbf{Scalar Factor} \\
        \hline
        Same Middle Name and Last Name & 1.00 \\
        Same Last Name Only & 0.75 \\
        Matching Middle Name and Last Name & 0.50 \\
        Same Middle Name Only & 0.25 \\
        \hline
    \end{tabular}
    \label{tab:scalar_factors}
\end{table}

\begin{figure}[H]
        \centerline{\includegraphics[width = 1\textwidth]{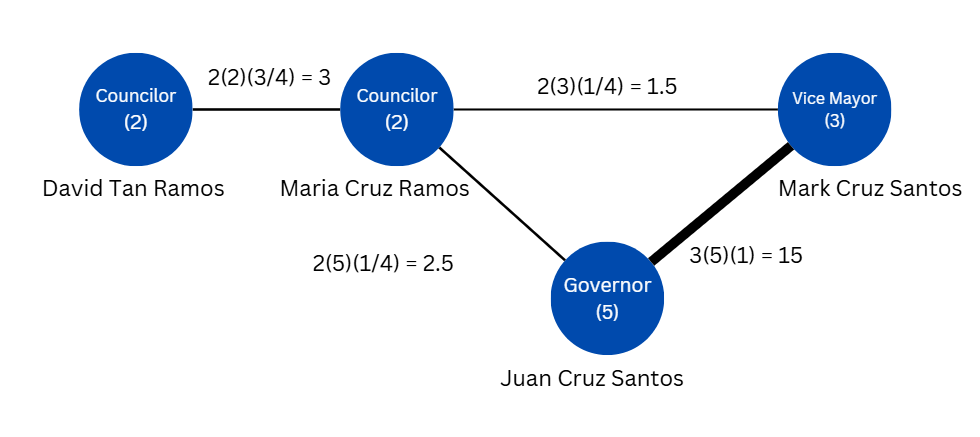}}
        \caption{Example of the Graph Weighting Scheme}
        \label{GRAPH-WEIGHT}
\end{figure}

Figure \ref{GRAPH-WEIGHT} shows an example of a dynastic network, with their corresponding node weights and edge weights. For instance, Juan and Mark are both ``Cruz" and ``Santos", so the scalar factor of their approximate degree of consanguinity is $1$, and since Juan is a Governor and Mark is a Vice-Mayor, the weight between them is computed as $3 \cdot 5 \cdot 1 = 15$.

Finally, the network is generated using the \textit{networkx} library from Python, which was constructed as an undirected graph. To facilitate visualization, \textit{Gephi}---which is an open graph viz platform---was used for clarity and ease of analysis. 

\subsection{Community Detection}

A common problem faced in dynastic literature is how to determine the boundaries of a given clan. When two clans are thoroughly interwoven, where does one dynasty stop and another begin? Rather than rely on subjective boundaries or defaulting to simple but inaccurate heuristics such as assuming each unique last name is a separate clan, we develop a novel community-detection based approach.

\begin{figure}[H]
        \centerline{\includegraphics[width = 1\textwidth]{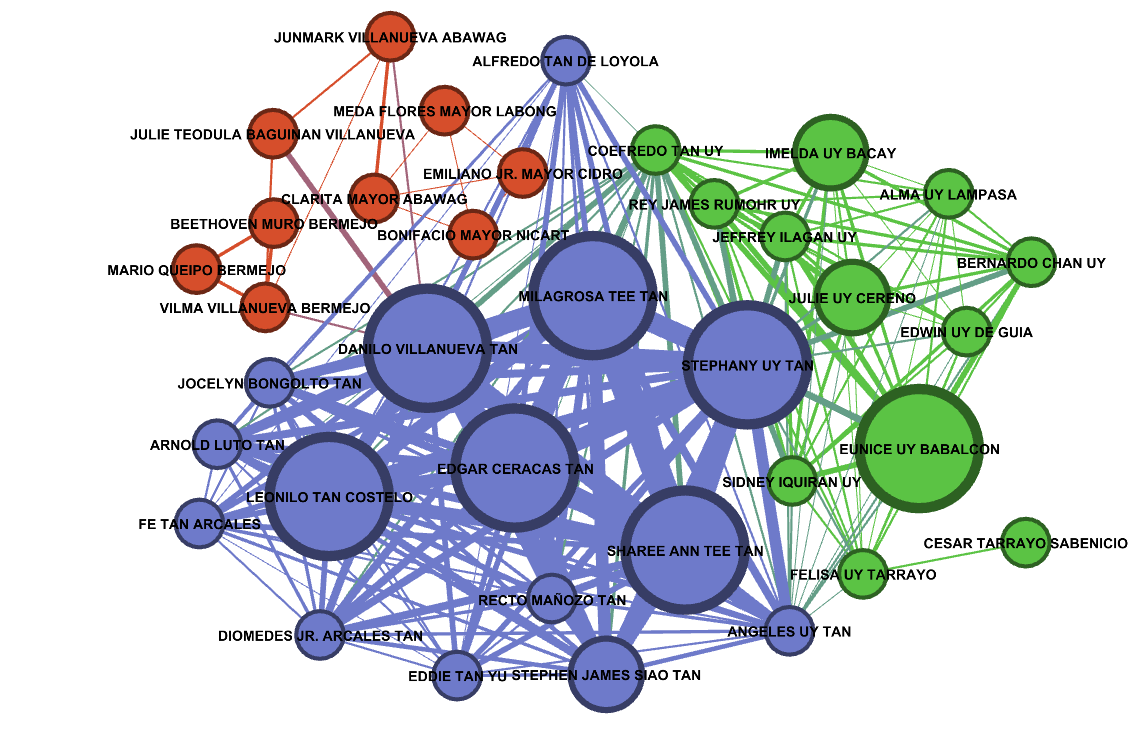}}
        \caption{The Largest Connected Component of Political Network in Samar  (2016)}
        \label{SAMAR-GEPHI}
\end{figure}

The Leiden Algorithm (see Definition \ref{Leiden-Definition}) is an efficient large-network community-detection algorithm that divides nodes in a network in a way that maximizes the modularity of each community. Upon constructing the political network graphs of each province for each election year in the dataset, we then run the Leiden Algorithm with Resolution = $1$ (as implemented in the \textit{networkx} package for Python) to separate the nodes into communities that each politician most likely belongs to. Here, communities correspond to a ``family" of politicians at a specific electoral year. Consequently, this study will use   ``communities", ``dynasties", ``families" and ``clans" interchangeably but are essentially equivalent. An example of a network, color-coded by their respective communities, is shown in Figure \ref{SAMAR-GEPHI}. 

This community-detection based approach to identifying dynasties has two key advantages over previous methods. Firstly, it provides a standard, mathematical criterion for identifying clans–eliminating the need for either the researchers' (or their survey respondents) to make subjective judgment calls as to which dynasty a politician belongs to. Secondly, it allows for the efficient macro-scale analysis of even large collections of political networks. Instead of heuristics such as a sole surname-based identification, we believe that clans identified through community-detection represent dynastic ties more accurately and realistically.

\section{Dynastic Indicators}

Viewing each dynasty as a community of nodes embedded in their province's political network allows us to understand these networks from a graph-theoretic frame. We develop four (4) quantitative indicators to capture the degree and characteristics of dynastic prevalence in political networks: Political Herfindahl-Hirschman Index (HHI), Centrality Gini Coefficient (CGC), Community Connectivity Density (CCD), and Average Community Connectivity (ACC). At a high level, each of these indicators captures a different facet of how power and influence are structured and maintained within political networks. They differ in the specific aspect of the network they represent, and thus each offer a different view of dynastic influence. In conjunction, these indicators provide richer, novel insights into the evolution of dynastic prevalence.

\subsection{Political HHI}

Political HHI (see Definition \ref{HHI}) has previously been used to capture the extent to which political influence, measured by the share of dynastic seats, is concentrated among the ruling clans in a given province (\cite{mendoza_political_2022, mendoza_political_2022-1, davis_corruption_2024, mendoza_political_2016}). The indicator is a reflection of how ``competitive" (in terms of how power is distributed) the political landscape of a given province is. 

In this study, we compute the Political HHI of a province in a given election year as:
    \begin{equation*}
        HHI = \sum_{j \in J} \left(\frac{W_j}{W_t} \times 100  \right )^2,
    \end{equation*}
where $W_i$ represents the sum of all node weights of members of community i, and $W_t$ represents the sum of all node weights in the province's network. 

Our formulation of Political HHI improves upon previous methods in two key ways. Firstly, we define ``dynasties" as communities detected by the Leiden Algorithm which allows us to account for connections between politicians who may not share the same last name. Formulations that assume each unique surname is a separate dynasty will tend to understate the true degree of power concentration. Using communities as the basis avoids that. Aside from more accurately reflecting the true degree of political concentration in a given province, this also means that key phenomenon such as political inter-marriages that result in large clan mergers have a more drastic numerical effect on Political HHI. Secondly, where previous formulations consider each position as having an equal contribution to seat share, our formulation of Political HHI uses weighted seat share. The weighted seat share of a clan is the sum of the node weights of all its members divided by the total weight of all nodes (seats) in the province's network. This way, we incorporate not just how many seats a clan's members occupy, but also how powerful their seats are relative to others in the province.

Political HHI offers a straightforward numerical representation of the degree of dynastic concentration in a political landscape. If one or two dynasties control a large portion of the network, the system is highly concentrated and consequently, the HHI will be higher (up to a maximum of 10,000) suggesting the existence of dynasties that monopolize power. This scenario suggests that a few families not only dominate political decision-making, but also likely influence policies and resource allocation to a significant degree. Furthermore, a high concentration also often signals a potential lack of competition. This can mean that the established families have already built a system that reinforces their positions over generations.

While the HHI lays the groundwork by a offering a macro-level view of concentration, it is primarily focused on the distribution of total influence across communities. Political HHI alone does not, however, capture the nuances of how that influence is distributed among individuals within those dynasties. 

\subsection{Centrality Gini Coefficient}

While HHI measures the overall level dynastic power concentration, the Centrality Gini Coefficient (CGC) provides insight into how evenly or unevenly political influence (as represented by weighted node centralities) are distributed. The mathematical formulation of CGC is discussed in-detail in \ref{Gini}. Conceptually, CGC takes each candidate's (node) weighted degree centrality as an approximate measure of how ``influential" they are relative to others in the network. It then accounts for the distribution of weighted centralities across all the politicians in the network and distills the dispersion of this distribution in a single indicator (a Gini coefficient). A CGC close to 1 indicates a high degree of inequality; that is, the network is dominated by a handful of powerful, well-connected central figures and their families. On the other hand, a value near 0 signifies a more equitable distribution of connections and influences. 

CGC allows for finer-grained analysis of the dispersion of influence within political networks. Two provinces may have the exact same Political HHI, and thus the same degree of dynastic power concentration, but have very different CGC's. In the province with a high CGC, dynasties could tend to be structured around a single influential matriarch or patriarch (thus one individuals has a much higher weighted centrality than all others). On the other hand, provinces with a low CGC, even though dynasties may still control a majority of seats, the different members of the dynasty have a more equal share of power. This could have strong implications for how each of the dynasties governs. Even if dynasties are rampant in a province, a low CGC means that dynasties can still actively compete with each other, and in this case dynastic rule \textit{could} still result in developmentally beneficial policies. In contrast, if there is a strong dynasty and high CGC, the central figure could become virtually unaccountable and thus could still reign unchecked. 

Moreover, in a political system dominated by a few central families, the removal or downfall of one could lead to a sudden and significant shift in the balance of power. This fragility, can often be masked by the apparent strength of established dynasties. Yet in actuality, dynasties structured around singular dominant individuals may be more vulnerable to crises of succession. In contrast, a more balanced network, where influence is distributed more uniformly, tends to be more robust and adaptive in the face of change. 

\subsection{Connected Component Density}

Another natural question to ask about political networks is how we might quantify the degree to which clans are interconnected. Even though two clans may be separate communities within a network, they may nevertheless share ``weak" connections making them part of the same \textit{connected component} (see Definition \ref{CCD}). Connected Component Density (CCD) measures the extent of cohesion between clans–shifting the focus from the distribution of influence to the overall connectivity and fragmentation of the political network. 

CCD broadly reflects the degree inter-clan cohesion in a network. CCD is given as one minus the ratio of connected components in a network to the total number of vertices. In a ``perfectly" non-dynastic network where none of the politicians are related, each node is its connected component; thus, CCD equals 0. A low CCD implies the political network is largely decentralized, suggesting a healthy amount of pluralism and competition. On the other hand, in a network where all politicians are related in some way and part of a single larger ``mega-clan", CCD is equal to 1. A high CCD suggests that the various clans in a network may be highly integrated, though through weak ties. These weak ties could nevertheless facilitate coordinated political action and more inter-clan collusion.

\subsection{Average Community Connectivity}

Where CCD describes inter-clan cohesion, the Average Community Connectivity (ACC) describes the intra-cohesion and robustness of the dynasties in the network. Community connectivity essentially asks the question: ``How many key nodes must be disrupted to break apart this political community?". A higher value suggests the community is highly resilient, implying that power is deeply embedded in multiple, redundant connections. We then take the Average Community Connectivity across each of the clans in a given province's political network. This provides a sense of how ``tight-knit" or ``compact" the dynasties in a given province are. 

\section{Dynasties and Party Loyalty}

In this second section, we examine how partisan allegiance varies among dynastic versus non-dynastic incumbents. The aim here is to establish empirical evidence to evaluate the hypothesis that dynasties persist as the preferred unit of political organization because parties remain unreliable and unable to guarantee continued electoral success. This hypothesis would predict that dynastic candidates, who can rely on their families for resources and political clout, have less incentive to stay loyal to their chosen parties compared to their non-dynastic counterparts.

We approach this by considering three phenomena that serve as close proxy indicators of party loyalty: incumbent party hopping, dynasty-party overlap, and bandwagoning. In the first approach, we examine whether there is a significant difference in the rate of party-hopping between dynastic and non-dynastic incumbents of the same province and election cycle. Though our method does not establish a causal link between dynastic membership and party hopping, by controlling for both province and election year we compare the party-hopping behavior of candidates under the same prevailing political conditions. 

In the second approach, we ask to what extent members of the same clan tend to affiliate with the same political party by examining the average party homogeneity in each province across each election year. For each community (clan) within a province, we measure dynasty-party overlap as the proportion of community's members that belong to the most common party within that community (i.e. the proportion of dynasty members that share the same party affiliation). We then take the average across all communities within the province, excluding non-dynastic candidates. In a province where all dynasts belong to the same party as their kin, we expect dynasty-party overlap to equal $1$, while in a province where dynasty have completely fragmented party affiliations, it would equal $0$.

In the third approach, we inspect differences in bandwagoning between dynastic and non-dynastic candidates. Bandwagoning is a well-documented phenomenon in Philippine politics wherein politicians switch party allegiances to that of the majority ruling party at the time (typically the party of the President). We inspect differences in the rate of bandwagoning between dynastic and non-dynastic politicians. We identify the ``major party" as the party of the sitting president at the time of the election (e.g. in 2016, the year Duterte assumed office, the ``major party" was still Aquino's Liberal Party). In the next Chapter, we show that this closely aligns with the party with a numerical majority of seats each year. We take the number of dynastic (non-dynastic) incumbents who hopped to the majority party in that year and divide by the total number of dynastic (non-dynastic) candidates who won a seat in that election cycle. In essence, we ask ``out of all the dynastic (non-dynastic) candidates who won office that year, how many of them were bandwagon-ed to the leading party?"

\section{Socioeconomic Data}

\subsection{Socioeconomic Indicators}

We incorporated two socioeconomic variables into our analysis---Poverty Incidence and the Human Development Index (HDI) which serves as vital lenses for understanding the broader implications (if any) of the persistence of political dynasties. 

Poverty Incidence is a well-established metric in recent literature that explores the extent of political dynasties in local regions such as in the case of Mendoza et.al.,(2016) \cite{mendoza_political_2016}. Essentially, Poverty Incidence captures the percentage of population living below a set poverty threshold.

Human Development Index (HDI), on the other hand, is a composite measures that integrates indicators of life expectancy, educational attainment, and per capita income, offering a holisistic assessment of human-well being and development \cite{beja_jr_inequality_2012}. Unlike the other two socioeconomic variables, this metric has not been used much in terms of measuring the socioeconomic ramifications of dynasties in local provinces. 

\subsection{Socioeconomic Data Gathering and Processing}

Data for Poverty Incidence were obtained from the Family Income and Expenditure (FIES), a nationally representative survey conducted triennially by the Philippine Statistics Authority (PSA). To ensure temporal alignment with the electoral cycle, we adopted a two-year lag approach; for example, poverty data from $2006$ were paired with the $2004$ dynastic metrics, $2009$ data with the $2007$ elections, and so on. The HDI data, on the other hand, was acquired from the Human Development Network Philippines for the period spanning $1997$ to $2021$ \cite{HDNPH-Data}. Since the HDI survey was also triennial, a two-year lag was also used to remain consistent with the poverty data. 

\section{Dynastics and Development}
In this final section, we examine the potential bidirectional relationships that exist between the dynastic metrics devised in the previous sections and key development indicators for each province. In this way, we can scrutinize how the strength and persistence of dynastic clans interfere with provincial socioeconomic development, and how such indicators can also provide insights to the dynasty's continuance across electoral periods. 

One usual approach in establishing causal relationships between factors of interest is Simple OLS Regression \ref{eq:OLS}. However, using this baseline regression is deemed ineffective in capturing time-dependent associations, which is a direct violation of \textit{independence} assumption imposed to this model. And given the longitudinal nature of our dataset, with repeated measures being conducted for each provincial dynastic portrait across $7$ electoral years from $2004$ to $2022$, it becomes imperative for us to utilize a more robust methodology that could capture such trends. 

A formal way to bypass this limitation of OLS is to perform averaging in our variables, which in our case, is to take the average of all provincial dynastic metrics across the $7$ electoral cycles. However, performing such operation leads to a reduction in possible insights that can be generated. However, our study not only wants to establish association between dynasty and development. Rather, we also want to capture the potential influence of evolving dynamics in this association in relation to the changing political landscape in Philippine elections. 

To account for this, we employ a more robust regression technique called Linear Mixed Modeling (LMM) \ref{eq:LMM} to model the association between dynastic metrics and socioeconomic variables.

\subsection{Linear Mixed Modeling}

Table \ref{tab:preprocessing} provides an overview of the data preprocessing for all variables that are used in the LMM regression. First, due to limited data availability, we only considered the poverty incidence (POV) and human development index (HDI) for each province from $2004$ to $2019$ (a total of $6$ electoral periods). This is $1$ less electoral period than what is available from dynastic indicators whose data until $2022$ are known. Thus, to match these, we had to drop all $2022$ metrics and only consider the metrics extracted from $2004$ to $2019$.

\begin{table}[H]
    \centering
    \setlength{\tabcolsep}{6pt}      
    \renewcommand{\arraystretch}{1.2} 
    \caption{Summary of Data Preprocessing done for Linear Mixed Modeling}
    \begin{tabular}{
        l 
        S[table-format=3.2] S[table-format=3.2] S[table-format=3.2] 
        S[table-format=3.2] S[table-format=3.2]}
        \toprule
        {Variables} & {Log?} & {Lagged Values?} & {Years with Complete Data} \\
        \midrule
        \multicolumn{4}{l}{\textbf{Dynastic Indicators}} \\
        ACC & \xmark & \xmark & \texttt{2004 - 2022} \\
        CCD & \xmark & \xmark & \texttt{2004 - 2022} \\
        GINI & \xmark & \xmark & \texttt{2004 - 2022} \\
        HHI & \cmark & \xmark & \texttt{2004 - 2022} \\
        \midrule
        \multicolumn{4}{l}{\textbf{Socio-Economic Indicators}} \\
        POV & \xmark & \cmark & \texttt{2004 - 2019} \\
        HDI & \xmark & \cmark & \texttt{2004 - 2019} \\
        \bottomrule
    \label{tab:preprocessing}
    \end{tabular}
\end{table}

Additionally, we established another variable that considers the 3-year lag value of each socioeconomic indicator. We argue that there might be lagging effects of the local development measured in Electoral Year $X$ towards the same metric indicated for Electoral Year $Y + 3$, but such values may be unaccounted for in the survey. To establish the lagged variables for poverty (POV) and development index (HDI), we considered $2004$ as our base year. Lastly, we \textbf{log-scaled} the values for the $HHI$ metric to match with the range of values for the other metrics. In the future discusions of LMM, we interpret the results for $HHI$ in terms of its \textit{logged} counterpart.

After these preprocessing techniques, we are now left with $8$ variables across the two main categories, $6$ of which are originally extracted while the remaining $2$ variables represent the lagged values for development indicators tagged as \texttt{<Indicator>\_lag\_3year}. Likewise, we only considered the time period $2007$ to $2019$ as the final periods of consideration. Our final dataset spans for $5$ electoral years and across $80$ provinces.

Afterwards, we implemented a linear mixed modeling to establish correlation between dynastic indicators and socioeconomic variables, and vice versa which we would tag as \texttt{Direction 1} and \texttt{Direction 2}, respectively. The assumptions for linear mixed modeling are presented in the next section which we are also able to check for our dataset. 

Overall, we performed a total of $10$ LMM regressions, $2$ of which are for \texttt{Direction 1} while the rest are for \texttt{Direction 2}. We also conducted \texttt{OLS Regression} and \texttt{Fixed-Effects Modeling} on the same dataset to establish significance of using LMM to capture a more complex and dynamic co-influences of our variables of interest. We constructed these regressions as defined in \ref{eq:OLS} and \ref{eq:FE}, respectively. 

\subsection{Assumptions}
As linear mixed models fall under the category of general linear models, using this model constitutes satisfying a set of assumptions for a more appropriate fit to the data. These are summarized below:
\begin{itemize}
    \item The explanatory variables are related \textbf{linearly} to the response.
    \item The residuals are independent and identically distributed, i.e. $\epsilon_{it} \sim \mathcal{N} (0, \sigma)$.
    \item Predictors are linearly independent, meaning there is no perfect multicollinearity.
\end{itemize}

The first and second assumptions can be easily verified by plotting Normal Q-Q plot of residuals. This is to ensure that our residuals can be assumed to be normally distributed with mean 0 and variance $\sigma^2$, i.e. $\epsilon_{ij} \sim \mathcal{N}(0,\sigma^2)$. Contrariwise, the last assumption can be checked by ensuring that the Variance-Inflator Factor (VIF) of each variables are below the tolerable range of $VIF < 5$. 

Relevant regressions are performed in \texttt{Python programming language} using \texttt{statsmodel} package and are cross-compared using \texttt{R Programming Language}. Meanwhile, the final LMM formulation for all direction similar to \ref{eq:LMM} is derived using \texttt{equatiomatic} package in \texttt{R}.

\subsection{Direction 1}
The first direction of LMMs establishes an association between provincial dynastic metrics and development indicators. More specifically, it tries to address the question: \textit{Can the persistence of clan structure and political power concentration in the province explain development status in the mentioned locality?} By exploring this direction, we may be able to establish significant correlations between concepts \textbf{dynasty} and \textbf{development} which are essentially useful in providing the portrait of dynastic influences. The formulas for the regression models are presented below:

Although our main result will revolve around \textit{linear mixed models,} we still performed \textit{OLS Regression} and \textit{Fixed-Effects Modeling}. This was done for us to establish the need to analyze longitudinal dynastic panel data in mixed models versus limiting them to simpler regression models. In these three regressions, we assumed the following regression formulation presented in the following:
\vspace{-20pt}

\begin{equation}
    \label{eq: D1-OLS}
        \mathtt{OLS_{D1}}:\mathtt{SOCIO_P = \beta_0 + \beta_1DYN_{1_P} + \beta_2DYN_{2_P} + \beta_3DYN_{3_P} + \beta_4DYN_{4_P}}
\end{equation}

\vspace{-20pt}
\begin{equation}
    \label{eq: D1-FIXED}
    \begin{aligned}
            \mathtt{FIXED_{D1}: SOCIO_{T,P}} &= \mathtt{\alpha + \beta_1DYN_{1_{T,P}} + \beta_2DYN_{2_{T,P}} + \beta_3DYN_{3_{T,P}}}\\ 
            &+ \mathtt{\beta_4DYN_{4_{P,T}} + \beta_5YEAR_{5_{T,P}} + \epsilon_{T,P}}
    \end{aligned}
\end{equation}

\vspace{-20pt}

\begin{equation}
    \label{eq: D1-LMM}
    \begin{aligned}
           \mathtt{LMM_{D1}}: \mathtt{SOCIO_{i}} &\sim \mathcal{N}(\mu, \sigma^2), \\
            &\mu = \mathtt{\alpha_{j[i]} + \beta_1DYN_1 + \beta_2DYN_2 + \beta_3DYN_3 + \beta_4DYN_4 + \beta_5YEAR_5} \\
            & \alpha_j \sim \mathcal{N}(\mu_{\alpha_j}, \sigma_{\alpha_j}^2), \text{for Province } j = 1, \ldots, P
    \end{aligned}
\end{equation}

For the OLS regression model, we took the average of all numerical values across each province to satisfy the independence between data measurement. Meanwhile, for fixed-effects (FE) modeling, we assumed that the time horizon (years) in conducting the repeated measures may have an overall effect with the metric values. Thus, we applied time-fixed effects only for our second regression model.

Lastly, for our LMM regression model, we considered \texttt{Province} as a potential source of variation among the dynastic measures. We argue that the overall topology of dynastic clan present across all electoral cycles may not necessarily be the same across all provinces. Thus, to say that dynastic structure of Province $A$ is similar to Province $B$ could provide a rather skewed results in our regression. Therefore, hierarchical grouping was considered for \texttt{Province} by incorporating \textit{varying intercepts} for each provincial data to take into account a potential variations that are induced by these geographical differences. However, we also argued that incorporating \textit{varying slopes} into provincial regression lines may induce overfitting to our model, and thus varying only the \textit{intercepts} would suffice. Meanwhile, we considered only time-fixed effects in our LMM model, and such is generated by considering the $YEAR$ in the regression. This is equivalent in producing dummy variables that constitute the value of the dependent variable in a specific electoral year. In total, we performed $2$ sets of each of the regression models devised for this direction. 

\subsection{Direction 2}
The second direction explores the reverse-causality of what we have established previously. That is, we want to investigate if the provincial development indicators can explain its prevailing dynastic indicators. Simply put, we aim to address the following question: \textit{Can the level of provincial development provide significant insights to understand evolution of dynastic structures and power dynamics in the area?} 

Similar to the first direction, we considered both \texttt{OLS Regression} and \texttt{Time Fixed-Effects Modeling} as comparison model for the results provided in \texttt{LMM}, with the \textbf{YEAR} assumed as the source of time-fixed effects while we consider again \textbf{Province} as source of random effects. In these three regressions, we assumed the following regression formulas defined at the next page. Overall, we conducted $8$ sets of regressions, $2$ for each dynastic metric which we considered here as a dependent variable.

\vspace{-20pt}

\begin{equation}
    \label{eq: D2-OLS}
    \begin{aligned}
        \mathtt{OLS_{D2}}:\mathtt{DYN' = \beta_0 + \beta_1SOCIO_1 + \beta_2SOCIO_1^{lag}} \\
    \end{aligned}
\end{equation}

\vspace{-20pt}
\begin{equation}
    \label{eq: D2-FIXED}
    \begin{aligned}
            \mathtt{FIXED_{D2}}: \mathtt{DYN'_{P,T}} = \mathtt{\alpha + \beta_1SOCIO_{1_{P,T}} +  
            + \mathtt{\beta_2SOCIO_{1_{P,T}}^{lag} + \beta_3YEAR_P + \epsilon_{P,T}}}
    \end{aligned}
\end{equation}

\vspace{-20pt}
\begin{equation}
    \label{eq: D2-LMM}
    \begin{aligned}
           \mathtt{LMM_{D2}}: \mathtt{DYN'_{i}} &\sim \mathcal{N}(\mu, \sigma^2), \\
            &\mu = \mathtt{\alpha_{j[i]} + \beta_1SOCIO_1 + \beta_2SOCIO_1^{lag} + \beta_3YEAR_j} \\
            & \alpha_j \sim \mathcal{N}(\mu_{\alpha_j}, \sigma_{\alpha_j}^2), \text{for Province } j = 1, \ldots, P
    \end{aligned}
\end{equation}

\subsection{Regression Metrics}
The \texttt{LMM} regression model for each direction was compared to its \texttt{FE} and \texttt{OLS} counterpart. This is for us to establish the significance in using a more complex regression model like \texttt{LMM} in constructing a more robust association between the variables. To do this, we used known regression metrics like $R^2$ and $AIC$ defined in \ref{eq:R2-formula} and \ref{eq:AIC-formula}, respectively. In justifying the need to consider provinces as a source of within-group differences in our regression, we want the AIC metric from \texttt{LMM} be the lowest among \texttt{OLS} and \texttt{FE} regressions.

\chapter{Results and Discussion}

This chapter presents the study's primary findings, and a discussion on their implications for dynastic research and policy. Results are presented in the same order of sections as in the Methodology.

\section{The Evolution of Dynastic Power}

We begin by investigating each of the indicators developed in the previous chapter to examine how dynasties have evolved in various senses from $2004$ to $2022$. Table \ref{tab:Dynasty Stats} shows a descriptive summary of the values for HHI, CGC, CCD, and ACC. The comprehensive results for each province and metrics can be found in the Appendix.

\begin{table}[H]
    \centering
    \setlength{\tabcolsep}{6pt}      
    \renewcommand{\arraystretch}{1.2} 
    \caption{Summary statistics over the years for HHI, CGC, CCD, and ACC}
    \begin{tabular}{
        l 
        S[table-format=3.3] S[table-format=3.3] S[table-format=3.3] 
        S[table-format=3.3] S[table-format=3.3] S[table-format=3.3]
        S[table-format=3.3]
    }
        \toprule
        {Years} & {2004} & {2007} & {2010} & {2013} & {2016} & {2019} & {2022} \\
        \midrule
        \multicolumn{8}{l}{\textbf{HHI}} \\
        Mean & 147.7126 & 177.4465 & 185.3237 & 187.3095 & 195.8673 & 192.4335  & 216.4767 \\
        SD   & 72.9320  & 87.4919  & 95.8931 & 85.7573  & 97.1249  & 100.7711 & 116.1469 \\
        \midrule
        \multicolumn{8}{l}{\textbf{CGC}} \\
        Mean & 0.7981   & 0.7693   & 0.7595   & 0.7602   & 0.7555   & 0.7567   & 0.7521 \\
        SD   & 0.0639   & 0.0743   & 0.0753   & 0.0714   & 0.0737   & 0.0726   & 0.0720 \\
        \midrule
        \multicolumn{8}{l}{\textbf{CCD}} \\
        Mean & 0.2481   & 0.3042   & 0.3281   & 0.3300   & 0.3428   & 0.3194   & 0.3611 \\
        SD   & 0.1104   & 0.1324   & 0.1357  & 0.1336   & 0.1378   & 0.1248   & 0.1313 \\
        \midrule
        \multicolumn{8}{l}{\textbf{ACC}} \\
        Mean & 7.0904   & 6.1828   & 6.8809   & 7.1091   & 6.9123   & 7.9136   & 6.9631 \\
        SD   & 4.6837   & 3.8256   & 3.8984   & 3.7803   & 3.6974   & 4.7925   & 3.7862 \\
        \bottomrule
    \end{tabular}

    \label{tab:Dynasty Stats}
\end{table}

At an initial glance, the results depicted in Table 5.1 reveal a few notable trends. Political HHI rose steadily from 147.71 in 2004 to 216.47 in 2022, with its standard deviation growing from 72.93 to 116.15. These suggest that dynastic concentration has been increasing over the years, but with large deviations between provinces. In contrast, the Centrality Gini Coefficient (CGC) has remained high throughout the years with a slight steady decline. Its low, stable standard deviation (0.06-0.08) suggests CGC is relatively stable across provinces. Connected Component Density (CCD) increased from 0.25 to 0.36, and although its variability is higher than CGC’s, it too remains relatively stable across years. Finally, Average Community Connectivity (ACC)---which quantifies intra-clan cohesion—--displayed a non‑monotonic mean, fluctuating between 6.18 and 7.91. Its standard deviation similarly rose and fell, indicating moderate but variable cohesion within clans over time.

Noting these overarching trends, we investigated each metric in depth, closely examining the provinces that attain the highest and lowest values, and interpret how the results yielded by our metrics align with historical expectations and intuition surrounding the political landscape in those cases.

\subsection{Political Herfindahl-Hirschman Index (HHI)}

\begin{figure}[H]
        \centerline{\includegraphics[width = 0.80\textwidth]{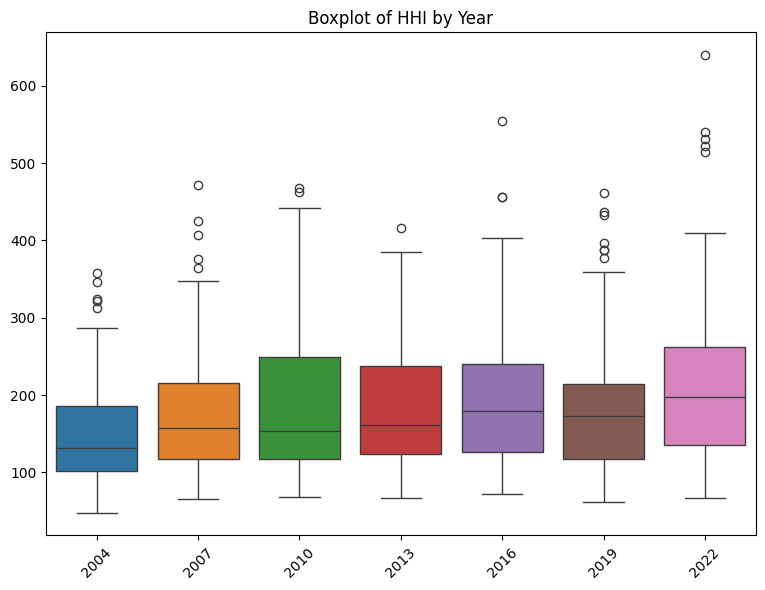}}
        \caption{Boxplot of Political Herfindahl-Hirschman Index (HHI)}
        \label{HHI-boxplot}
\end{figure}

Figure \ref{HHI-boxplot} shows the distribution of Political HHIs across provinces over the period of study. We confirmed the increasing trend by performing a Linear Trend Analysis (see Appendix \ref{Trend-HHI})  yields a slope of $2.9382$ with a p-value of $0.0041 < 0.05$, confirming that the upward slope is significant. This implies that on average, provincial Political HHI has increased by roughly $2.9382 \times 3 = 8.8146$ units per electoral year. This implies that political concentration is increasing on average nationwide.

The upward trend in Political HHI has troubling implications. It suggests that not only do dynasties remain a dominant force in modern Philippine politics, but that they are also occupying an increasing share of seats. If the theoretical consensus that dynastic concentration degrades checks and balances holds, then this consistent rise in Political HHI represents a dangerous threat to the integrity of local democratic institutions. Further, this finding may support Querubin's \cite{querubin_political_2012} hypothesis that the increase in dynastic candidates has a ``chilling effect"–discouraging non-dynastic candidates from seeking election to begin with. 

Our findings on the rising trend in Political HHI concur with those of previous studies that noted similar trends using the surname-based formulation of Political HHI \cite{mendoza_political_2022, mendoza_political_2016, beja_jr_inequality_2012, balanquit_measuring_2017}. We note however, that as expected, the particular values of HHI generated by our community-based method are higher than those reported by previous studies that use surname-based HHI. Mendoza (2022) \cite{mendoza_political_2022} for instance reports a mean provincial HHI of just 34.046 in the period spanning 2004 to 2018, implying a much lower degree of political concentration than we find here. This demonstrates that surname-based methods for identifying dynasties do in fact tend to understate the true level of dynastic concentration, which has significant implications when investigating the link between dynastic concentration and development outcomes.

\begin{figure}[H]
    \centering
    \begin{subfigure}{0.49\textwidth} 
        \includegraphics[width=\linewidth]{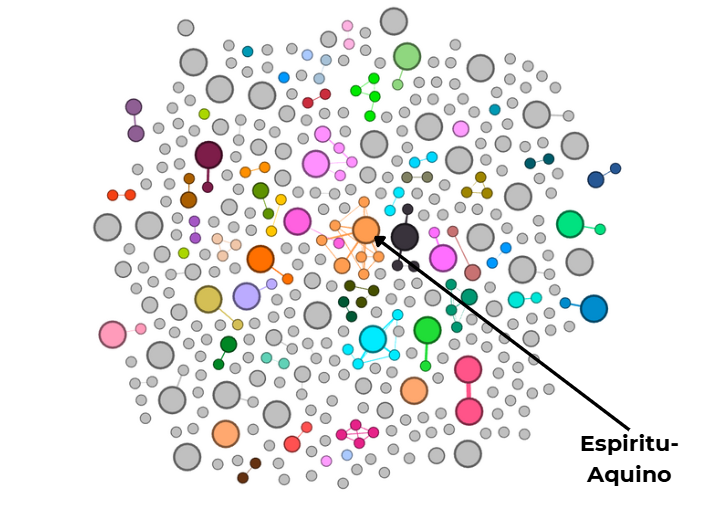}
        \caption{Camarines Sur in 2004}
        \label{fig:first}
    \end{subfigure}
    \hfill
    \begin{subfigure}{0.49\textwidth}
        \includegraphics[width=\linewidth]{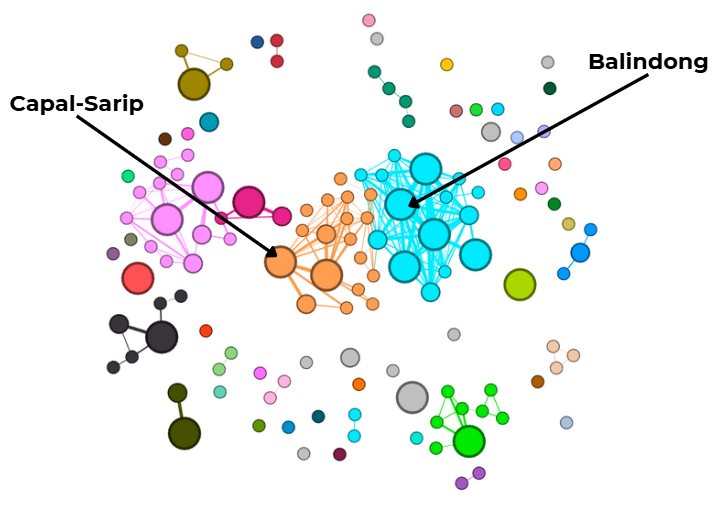}
        \caption{Lanao del Sur in 2022}
        \label{fig:second}
    \end{subfigure}
    \caption{Networks with the Lowest and Highest Political HHI}
    \label{fig:network-HHI}
\end{figure}

Figure \ref{fig:network-HHI} illustrates the dynastic networks in Camarines Sur (2004) and Lanao del Sur (2022), with node size proportional to the positions each politician held. Camarines Sur registered the lowest HHI at a value of $47.89$, while the highest recorded HHI was in Lanao del Sur (2022), with a value of $630.04$.  The complete ranking of all provinces is shown in Appendix \ref{HHI-RANK}.

In provinces with high HHI, a major share of the seats are controlled by a few dynasties, as they capture a larger share of the overall political influence. For example in Lanao del Sur (2022), the Capal-Sarip clan (orange) held 18 seats (2 mayors, 2 vice-mayors, 14 councilors), and the Balindong clan (blue) held 15 seats (1 congressional representative, 3 mayors, 2 vice-mayors, and 9 councilors) out of 131 winning politicians. Notably, this is supported by the fact that the enduring influence of the Balindong family  traces back to the 17th century Lanao Sultanate \cite{balindong-sultanate}. Furthermore, while the Sarips are also commonly identified as a dynasty in the province, there were no mentions of the Capal family nor any literature relating the two. This suggests that our method of using middle-name linkages alongside community detection techniques may uncover dynastic connections that would be obscured to traditional surname-based methods.

As seen in Appendix \ref{HHI-RANK}, other notable provinces with consistently high HHI include Batanes, Maguindanao, Guimaras, and Siquijor. Though Maguindanao has a notable history of dynastic rule (i.e. Ampatuans, Sinsuats, Midtimbangs) \cite{Lucero_2016}, the high HHIs recorded in the other three provinces is likely explainable by their size. These are small provinces with relatively few open positions. Thus, even a handful of dynastic clans occupying a share of seats results in high value of HHI. This is in contrast with low HHI-provinces such as Camarines Sur in 2004 which is characterized by fragmented dynasties, with the largest family being the Espiritu-Aquino family (orange), consisting only of 8 politicians simultaneously running (1 mayor, 1 vice-mayor, and 6 councilors) out of the 391 winning politicians. 

It is also worth noting that a number of large provinces, including Cebu, Iloilo and Leyte, which are historically known for having powerful dynastic clans, recorded unexpectedly low HHIs. In these cases, the HHI is “diluted” as although dynastic clans do indeed occupy a large share of seats, there is either such a large pool of open positions or seats are dispersed across multiple clans that the HHI remains low. These cases underscore the complementing concentration measures that would avoid this issue and adjust for local scale and topology.

\subsection{Centrality Gini Coefficient (CGC)}

\begin{figure}[H]
        \centerline{\includegraphics[width = 0.75\textwidth]{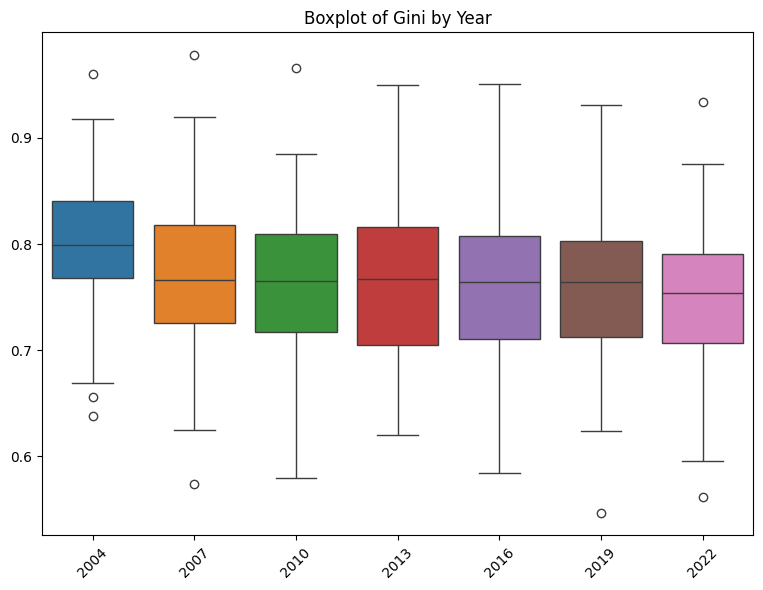}}
        \caption{Boxplot of Centrality Gini Coefficient (CGC)}
        \label{CGC-boxplot}
\end{figure}

Figure \ref{CGC-boxplot} depicts the distribution of provincial CGCs from $2004$ to $2022$. Based on the mean in Table \ref{tab:Dynasty Stats}, CGC appears fairly consistent over the years. However, performing Linear Regression Trend Analysis (see \ref{Trend-CGC}) yields a slope of $-0.002$, with a p-value of $0.0243$, confirming that the downward trend is significant, though minor. This means that the mean provincial CGC nationwide has decreased $0.006$ units per electoral year.  

Recall in Definition \ref{Gini} that a value closer to $0$ implies a more equal distribution of influence, while a value closer to $1$ suggests a more asymmetric distribution. Across all electoral years surveyed, the national average was well above 0.5, suggesting that there tends to be a high degree of inequality in their degree centrality (and thus political influence) within dynastic networks, suggesting that these networks tend to revolve around central dynasties.

\begin{figure}[H]
    \centering
    \begin{subfigure}{0.49\textwidth} 
        \includegraphics[width=\linewidth]{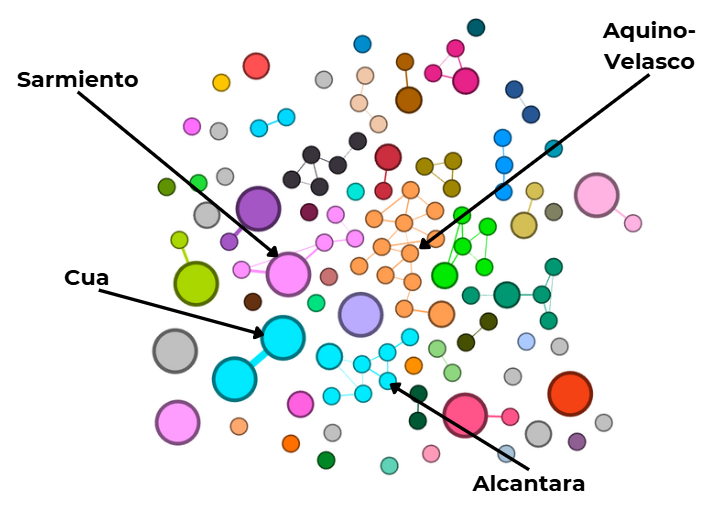}
        \caption{Catanduanes in 2019}
        \label{fig:first}
    \end{subfigure}
    \hfill
    \begin{subfigure}{0.49\textwidth}
        \includegraphics[width=\linewidth]{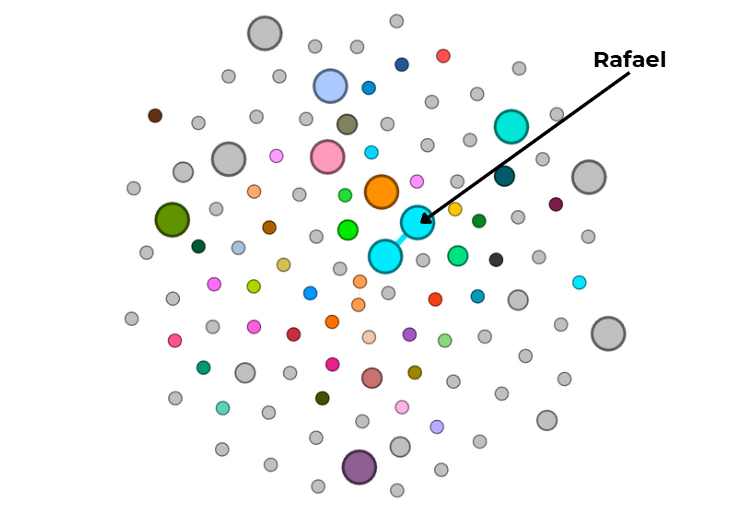}
        \caption{Mountain Province in 2007}
        \label{fig:second}
    \end{subfigure}
    \caption{Networks with the Lowest and Highest Centrality Gini Coefficient}
    \label{fig:network-CGC}
\end{figure}

Figure \ref{fig:network-CGC} illustrates the dynastic networks in Catanduanes (2019) and Mountain Province (2022), with node size proportional to the positions each politician held. Catanduanes recorded the lowest CGC with a value of $0.5469$, while the highest was in Mountain Province (2007) with a value of $0.9780$. As shown in Figure \ref{fig:network-CGC} (a), the political network in Catanduanes, which had the lowest CGC, is highly fragmented, with no clear powerhouse dynasty. On one side, there is the Aquino-Velasco Family (Orange) consisting of 12 politicians out of 114, yet consists only of councilors. On the other side there is also former Catanduanes Governor Joseph Chua Cua running simultaneously alongside former San Andres Mayor Peter Cua which holds considerably powerful positions. There are also other minor dynasties such as the Rosa-Alcantaras (Light Blue), Magallanes-Sarmiento (Pink), Vargas-Zafe (Green), etc. This is also the case in provinces such as Aklan, Nueva Ecija, Pampanga, Batanes, etc., which registered a low CCD due to the lack of presence of a dominant dynasty.

In contrast, consider the case of Mountain Province in 2007, which had the highest CGC. The province’s high CGC stems from the fact that only two politicians in the network are linked: Cesar Balacanao Rafael and his daughter Ana Maria Paz Rafael, mayors of Paracelis and Natonin in 2007, respectively. All other politicians in the network are non-dynastic and thus share no kinship ties. This is an example of a province with a very high CGC, but a relatively low HHI. Notably, Mayor Cesar Balacanao Rafael was very dominant in the province, often being portrayed by his detractors as a “warlord”, in the 1980s. In $2007$, however, Mayor Rafael was assassinated in a believed politically-motivated ambush-slay \cite{Felipe_2011}, which further proves that he held considerable power in the province. The scenario is similar to other provinces who continuously logged a high value of CCD. The presence of Romualdos in Camiguin, Espinas in Biliran, and Balagso-Dasayon in Kalinga, often unopposed, are the main driver of the high CGC in their respective provinces.

\subsection{Connected Component Density (CCD)}

\begin{figure}[H]
        \centerline{\includegraphics[width = 0.80\textwidth]{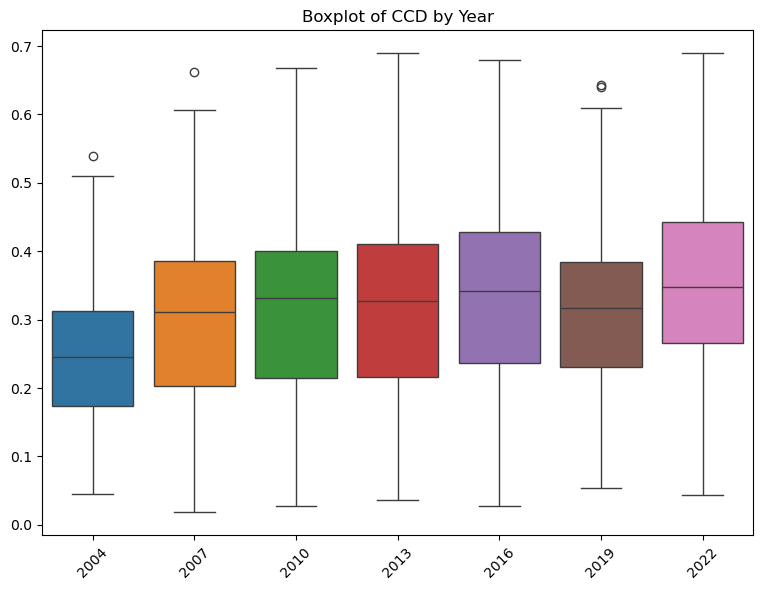}}
        \caption{Boxplot of Connected Component Density (CCD)}
        \label{CCD-boxplot}
\end{figure}

Figure \ref{CCD-boxplot} shows the distribution of CCD from $2004$ to $2022$. We can confirm the upward trend again by performing Linear Regression Trend Analysis (see Appendix \ref{Trend-CCD}) which yields  a slope of $0.0046$ with a p-value of $0.0231 < 0.05$, confirming that the trend is significant. This means that on average, provincial CCD has increased by $0.014$ units per electoral year.

Recall that CCD gauges inter-clan connectivity (see Definition \ref{CCD}). A higher CCD  indicates that a large portion of nodes are part of one or a few large connected clusters, meaning that the network is highly integrated. The upward trend in CCD suggests that, on average, political clans have grown more interconnected over time. This could mean either that more members of dominant dynastic clans are entering politics while displacing their non-dynastic rivals (thus the network becomes more connected), or, as Mendoza et al. \cite{mendoza_dynastic_2020} suggest, that dynasties have increasingly been forming strategic alliances through political inter-marriage. In other words, previously isolated or fragmented political groups are now more likely to interact, possibly promoting political collusion. 

\begin{figure}[H]
    \centering
    \begin{subfigure}{0.5\textwidth} 
        \includegraphics[width=\linewidth]{3Manuscript/images/MTPROVINCE-2007-GEPHI.png}
        \caption{Mountain Province in 2007}
        \label{fig:first}
    \end{subfigure}
    \hfill
    \begin{subfigure}{0.5\textwidth}
        \includegraphics[width=\linewidth]{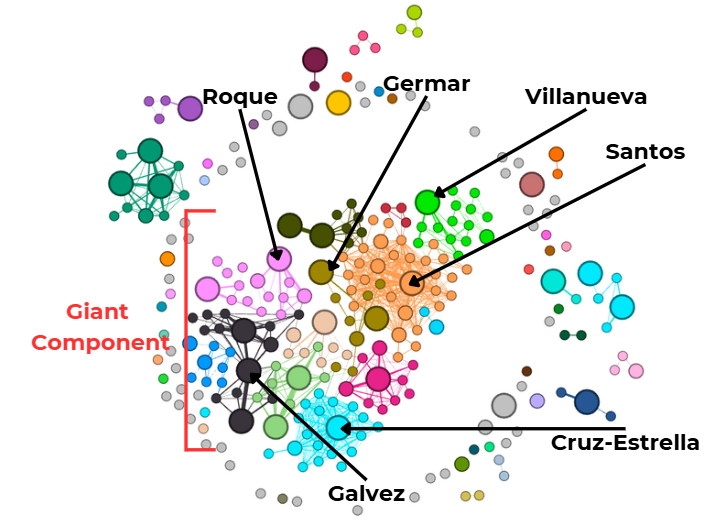}
        \caption{Bulacan in 2013}
        \label{fig:second}
    \end{subfigure}
    \caption{Networks with the Lowest and Highest Connected Component Density (CCD)}
    \label{fig:network-CCD}
\end{figure}

Figure \ref{fig:network-CCD} shows the networks with the lowest and highest CCD. The lowest recorded value was found in Mountain Province ($2007$) with a value of $0.018349$, while the highest was in Bulacan ($2014$) with a value of $0.68394$. As discussed in the previous subsection, Mountain Province in 2007 only had two dynastic politicians. As nearly every node is disconnected from all the others, this naturally yields a very low CCD. This is in contrast with Bulacan in 2013, where 148 out of 262 politicians ($56.49$\%) belong to a ``Giant Component", as seen in the middle. The ``Giant Component" is a large cluster of different clans that all share at least one inter-linkage. Major families found in the Giant Component includes Fernando and Sy-Alvarado (for Provincial Level), Galvez's in San Ildefonso, Villanueva's in Bocaue, Cruz-Estrella's in Guiguinto, Roque's in Pandi, Germar's in Norzagaray, and Estrella's in Baliwag City. While the Leiden Algorithm detects them as separate communities, CCD captures the inter-community relationships, which Querubin \cite{querubin_political_2012} claims are a key way in which dynasties are able to consolidate power and manipulate political outcomes. Even if no single dynasty wins a majority of seats, if dynasties are interlinked as in Bulacan's case, they may form a "majority block".  Other provinces with similarly high CCD include Pampanga, Pangasinan, and Batangas, each with values reaching more than 0.6 by 2022, reflecting extensive inter-clan linkages that likewise warrant a deeper, province-level analysis.

A peculiar case worth mentioning is the province of Maguindanao, which ranked 25th in CCD in 2007 (with a value of $0.3776$) but rose to the 2nd highest CCD in 2010 and maintained its position until 2022 (with a value of $0.6791$). Prior to 2010, the province’s dynastic powerhouses, the Ampatuans, Sunsuats, and Midtimbangs, were seemingly disconnected, with no apparent inter-clan ties connecting them, yielding a very low overall CCD.

\begin{figure}[H]
        \centerline{\includegraphics[width = 0.70\textwidth]{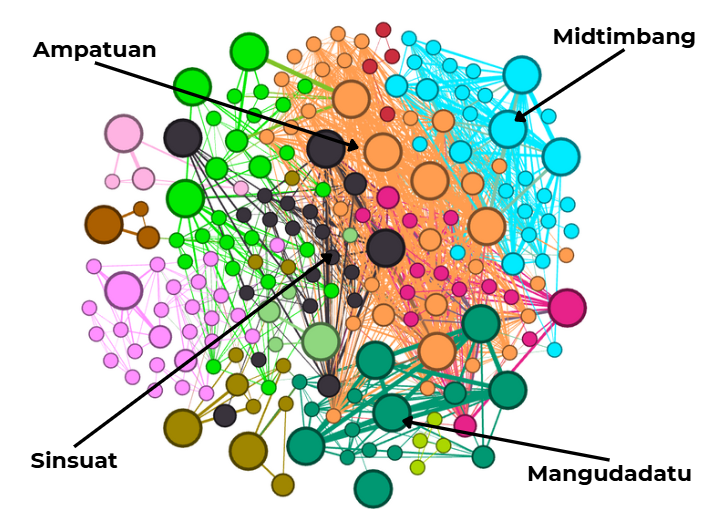}}
        \caption{Giant Component of the Network in Maguindanao (2022)}
        \label{Maguindanao-2022}
\end{figure}

By 2022, however, these families had coalesced into a single ``giant component”, encompassing $56.07$\% of the winning politicians in the province. Figure \ref{Maguindanao-2022} provides a visual analogue of the giant component of Maguindanao in 2022—a large cluster where diverse clans share at least one linkage. Zooming-in to Maguindanao's case, Datu Zaldy U. Ampatuan—son of patriarch Andal Sr.---married Johaira “Bongbong” Midtimbang, daughter of former Guindulungan mayor Midpantao Midtimbang, thereby fusing the Ampatuan and Midtimbang sub-networks. Similarly, Andal Sr.’s sons Hoffer and Sajid Islam Ampatuan married into the Sinsuat family —respectively, Ingrid and Zandria Sinsuat — further knitting these clans together \cite{Lucero_2016}. Although much work could be done in mapping every genealogical tie, the rapid densification observable in the network strongly suggests inter-clan marriages are a strong driving force, echoing Mendoza et al.’s (2020) finding that marriage is a common strategy for power consolidation among Samar dynasties. Such matrimonial linkages create durable bridges between once-discrete family networks, amplifying their collective influence. 

Crucially, this case of Maguindanao highlights yet another case of latent alliances being exposed, one that conventional surname-matching would miss. By considering inter-community connections, CCD highlights how dynasties can form de facto majority blocs without any single family holding an outright seat majority, and as such, merits deeper qualitative and quantitative investigation to uncover more hidden links.

\subsection{Average Community Connectivity (ACC)}

\begin{figure}[H]
        \centerline{\includegraphics[width = 0.75\textwidth]{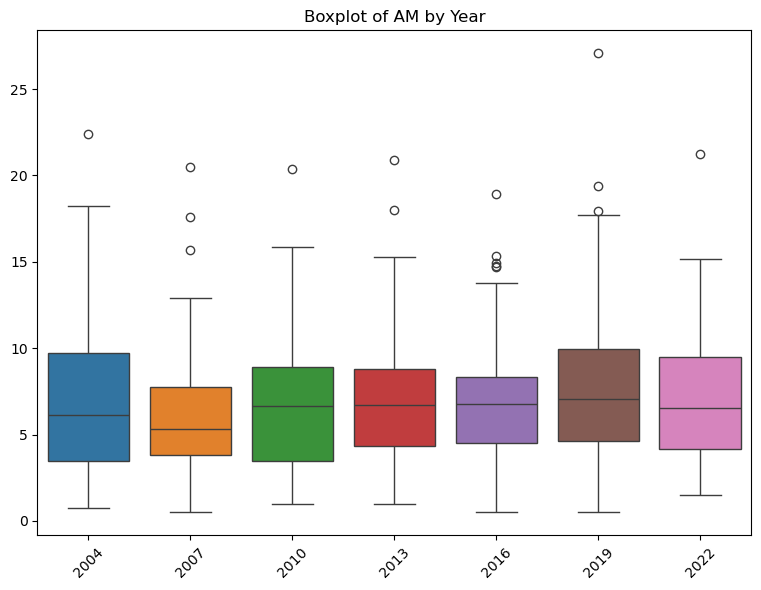}}
        \caption{Boxplot of Average Community Connectivity (ACC)}
        \label{ACC-boxplot}
\end{figure}

Figure \ref{ACC-boxplot} shows the Boxplot of the ACC from $2004$ to $2022$. Unlike with the first three indicators, there are no clear trends in the evolution of ACC. We confirmed this using Linear Regression Trend Analysis (see Appendix \ref{Trend-ACC}) which yielded a p-value of $0.2838 > 0.05$, meaning that there is no significant trend. 

Where CCD gauges inter-connectivity between clans, ACC provides a measure of how strongly intra-connected or tight-knit clans are on average within a given province. A higher ACC indicates that on average clans are tighter-knit, with multiple redundant connections between members. The degree of clan intra-connectivity relates to a variety of key political phenomena, such as the potential for infighting or splintering (i.e. instances where members of the same clan openly compete against each other to contest the same seat races). The lack of a significant trend in ACC suggests that dynastic structures have retained roughly the same degree of closeness throughout the years. 

ACC is less susceptible to the``dilution” problem of Political HHI, whereby small provinces can exhibit extreme HHI values from modest seat gains, making  cross-province comparisons misleading. ACC instead captures the density of ties within each dynastic community irrespective of the total number of seats. 

\begin{figure}[H]
    \centering
    \begin{subfigure}{0.49\textwidth} 
        \includegraphics[width=\linewidth]{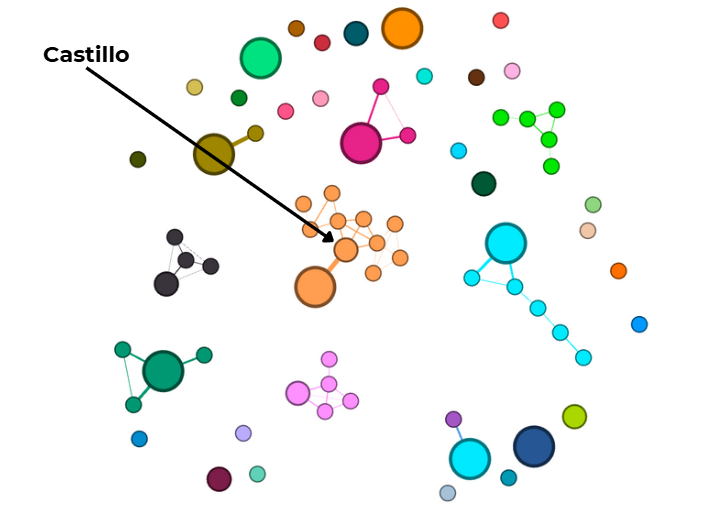}
        \caption{Batanes in 2016}
        \label{fig:first}
    \end{subfigure}
    \hfill
    \begin{subfigure}{0.49\textwidth}
        \includegraphics[width=\linewidth]{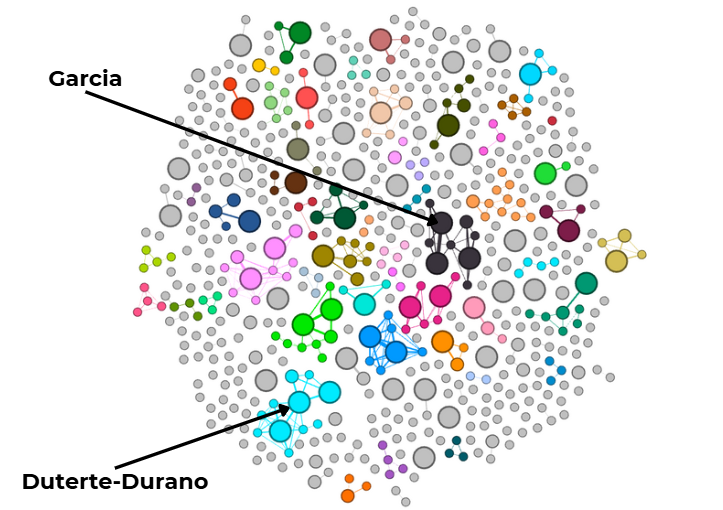}
        \caption{Cebu in 2019}
        \label{fig:second}
    \end{subfigure}
    \caption{Networks with the Lowest and Highest Average Community Connectivity (ACC)}
    \label{fig:network-ACC}
\end{figure}

Figure \ref{fig:network-ACC} shows provinces which yielded the highest and lowest ACC. Multiple provinces shared the lowest recorded ACC value of $0.05$, one of which is Batanes ($2016$). The highest recorded ACC was in Cebu ($2019$) with a value of $27.0738$. It is important to mention that ACC penalizes line- or tree-like clan structures---networks in which nodes have only one or two connections--— because their sparse topology yields low average connectivity scores. This is more illustrative in the case of Batanes in 2016. Since Batanes is a relatively small province and is dominated by a handful of major clans, it had a relatively high Political HHI. However, each of these clans is fairly loose-knit. Consider the Castillo Clan (orange) which is the largest and most central clan in the province. Although Raul Morada De Sagon, the then Mayor of Batanes, belongs to this clan, he is connected to it by only one link through his spouse Sabas Castillo De Sagon. He is only ``loosely" part of the clan. If Sabas (the connecting node) were to suddenly exit the network, the Castillos would lose a large degree of their influence (at least in terms of the network) as they would no longer be connected to a major Mayor. The same could be said for provinces which logged a low ACC such as Marinduque, Siquijor, Kalinga, and Mountain Province.

In contrast, tightly knit communities where members are richly interconnected, score highly on ACC, making it a sensitive gauge of true cohesion rather than mere seat share. Consider Cebu in 2019, a large province with a complex political landscape, dominated by dynasties, albeit fragmented ones situated in their own baluartes. This makes it easily dilutable in terms of the HHI. However, the Garcia family—one of the province’s most prominent political clans, with figures like Governor Gwendolyn Garcia and her relatives holding multiple local offices---forms a dense subgraph in which each member is linked to several others. The same could be said as the Duterte-Durano clan. This is equally as important as a tight-knit community means that the clan as a whole is theoretically more resilient to sudden shocks, such as one member losing a race, or a key figure passing away. Other provinces which registered a high ACC include Iloilo, Quezon, Bohol, and Isabela (see Appendix \ref{ACC-RANK}), all of which notably have histories of multi-generational dynasties that have kept their foothold on local power.

\section{Dynasties and Party Hopping}

\begin{figure}[H]
        \centerline{\includegraphics[width = 0.90\textwidth]{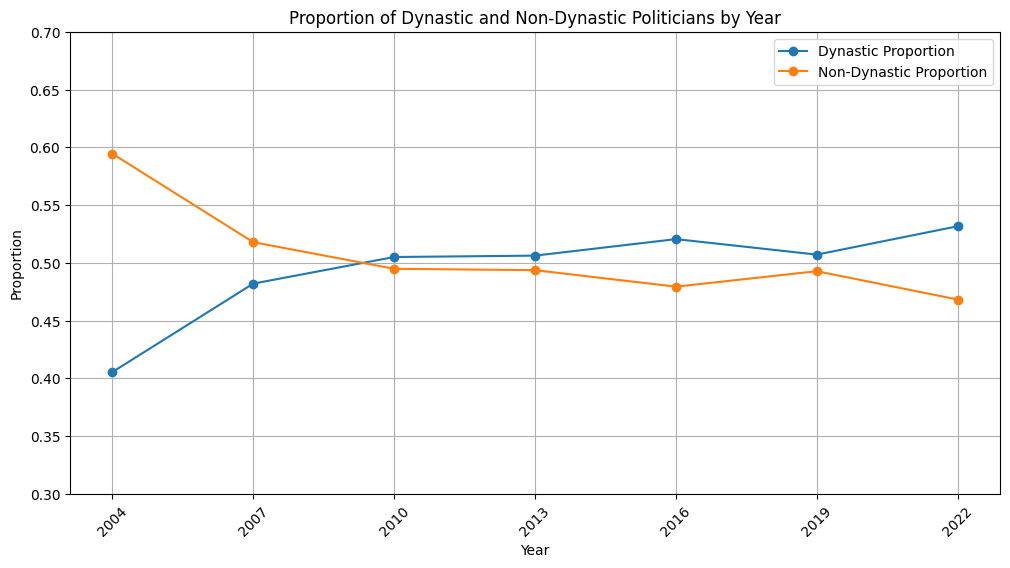}}
        \caption{Proportion of Dynastic and Non-Dynastic Politicians by Electoral Year}
        \label{Dynasty-Count-Line}
\end{figure}

This section discusses our analysis of the dynamics between dynasties and political parties. Recall that we first identified all ``dynastic" candidates in each province and election cycle (where a candidate is classified as dynastic if they take office with another politician in the same community). Note that this captures dynastic fatness rather than thinness. We restrict our analysis here to members of fat dynasties as we primarily want to investigate how dynasties act as a unit of political organization in a comparable sense to political parties, which involves having multiple members of the same unit occupying different seats. Figure \ref{Dynasty-Count-Line} shows the proportion of dynastic politicians from $2004$ to $2022$. 

At a glance, we see that the percentage share of dynastic politicians have increased from $40.73\%$ in $2004$ to $53.22\%$ in $2022$. This further supports the results presented earlier, suggesting that dynasties have not only remained the dominant form of political organization in the country but they have also grown more prevalent over the years. Recall that the term-limit hypothesis (discussed in \hyperref[Chapter 3]{Chapter 3}) posits that dynasties first arose primarily as a way for term-limited incumbents to extend their influence by ruling through the proxy of their relatives \cite{querubin_political_2012}. A crucial ``missing link" for this hypothesis however, is explaining how and why dynasties (rather than parties) remain the preferred way of preserving political influence past one's term limit. 

\begin{figure}[H]
        \centerline{\includegraphics[width = 0.9\textwidth]{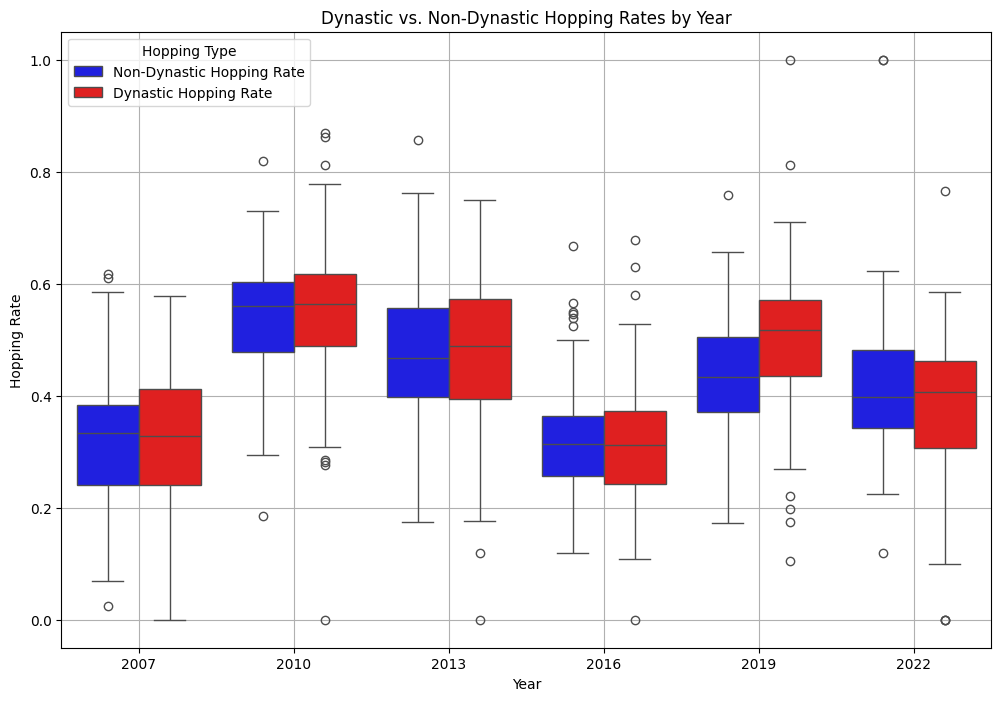}}
        \caption{Boxplot of Party Hopping Rate per Dynastic Status}
        \label{Hopping-Boxplot}
\end{figure}

We first analyze variations in party hopping behavior between dynastic and non-dynastic politicians. Figure \ref{Hopping-Boxplot} depicts the distribution party hopping rates among dynastic and non-dynastic incumbents across each province and election cycle. We see that dynastic politicians appear to have a higher party hopping rate than their non-dynastic counterparts. To substantiate this, we perform a Wilcoxon-Signed Rank Test to determine if there is a significant difference between the two rates. The Wilcoxon-Signed Rank Test is used as the data is non-parametric (as confirmed by a Shapiro-Wilk test for normality) and the samples of interest are paired (we compare hopping rates between dynastic and non-dynastic politicians from the same province and election year). With a test statistic of $56691.5$ and a p-value of $0.0039 < 0.05$, we have sufficient evidence to conclude that the hopping rates of dynastic politicians is significantly higher than those of non-dynastic politicians. 

These results could imply that dynastic politicians, who can rely on their families for resources and political clout, have more leeway when deciding to switch parties precisely because they are less reliant on the party itself for electoral support. Non-dynastic incumbents, on the other hand, are forced to stick with their parties more often as they lack an alternative political network. These findings provide empirical evidence for the ``vicious cycle" described by Querubin \cite{querubin_political_2012} whereby the rise of dynasties stymies the development of mature political parties, which in turn further promotes the need for reliance on kinship networks. Even when non-dynastic candidates do consolidate into parties, they must eventually curry favor with dynastic candidates. As these dynastic candidates can party hop with virtually no consequence, parties are unable to develop a high degree of internal institutional trust, and thus never mature into a viable alternative for political organization.

\begin{figure}[H]
        \centerline{\includegraphics[width = 0.9\textwidth]{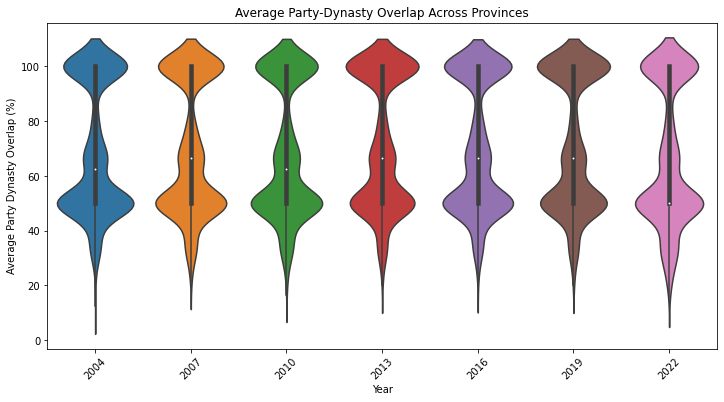}}
        \caption{Violin Plot of Average Party-Dynasty Overlap across election cycles}
        \label{Party-Dynasty Violin}
\end{figure}

Do members of the same clan tend to affiliate with the same political party? Interestingly, as depicted in Figure \ref{Party-Dynasty Violin}, there seems to be a bimodal distribution in the average party-dynasty overlap across provinces. The violin plot depicts the distribution of provinces based on Party-Dynasty overlap. The areas where the plot is widest are the modes of the distribution. In many provinces, as might be expected, members of the same clan all tend to belong to the same political party. However, in many other provinces, the average party-dynasty overlap is only around 50\% (as evidenced by the second peak on lower portion of the plots for each year). This suggests that many dynasties may purposely choose not to affiliate with the same political party. Notably, this pattern is consistent across all election cycles. 

These findings suggest that there may be two dominant ways in which dynasties choose to interact with parties. One is to join parties as a ``block". This strategy could be beneficial for the family as, especially for larger clans, it gives them a higher degree of leverage within the party itself. Another is to ``hedge bets" by diversifying party membership. Families may be choosing to ally with different parties to maximize their sphere of influence in an attempt to ensure that no matter which party wins, the family retains its political influence. 
\begin{table}[H]
    \centering
    \setlength{\tabcolsep}{6pt}      
    \renewcommand{\arraystretch}{1.2} 
    \caption{Number of Politicians per Party per Year and their Dynastic Status for the Four Winning (Presidential) Parties in the National Elections}
    \begin{tabular}{
        l 
        S[table-format=4.0] S[table-format=4.0] S[table-format=4.0] 
        S[table-format=4.0] S[table-format=4.0] S[table-format=4.0]
        S[table-format=4.0]
    }
        \toprule
        {Party} & {2004} & {2007} & {2010} & {2013} & {2016} & {2019} & {2022} \\
        \midrule
        \textbf{LKS-CMD/LKS-KAM} & & & & & & & \\  
        Dynastic & 3021 & 2944 & 3706 & 0 & 0 & 0 & 0 \\
        Non-Dynastic & 4339 & 3171 & 3706 & 0 & 0 & 0 & 0 \\
        \midrule
        \textbf{LP} & & & & & & & \\  
        Dynastic & 545 & 514 & 1258 & 3231 & 3859 & 328 & 204 \\
        Non-Dynastic & 857 & 604 & 1281 & 3088 & 3451 & 338 & 180 \\
        \midrule
        \textbf{PDPLBN} & & & & & & & \\  
        Dynastic & 44 & 86 & 26 & 9 & 125 & 3043 & 1721 \\
        Non-Dynastic & 63 & 79 & 30 & 9 & 131 & 2812 & 1562 \\
        \midrule
        \textbf{PFP} & & & & & & & \\  
        Dynastic & 0 & 0 & 0 & 0 & 0 & 264 & 153 \\
        Non-Dynastic & 0 & 0 & 0 & 0 & 0 & 273 & 154 \\
        \bottomrule
    \end{tabular}
    \label{tab:HoppingCount}
\end{table}

To further analyze this dynamic, we examine the bandwagoning behavior of dynastic versus non-dynastic candidates. Table \ref{tab:HoppingCount} shows the party membership demographics of each Presidential ruling party per election. The Lakas–Christian Muslim Democrats (LKS-CMD), which later merged with Kampi (LKS-KMP) in $2008$,  was the party of Former President Gloria Macapagal Arroyo when she took office in $2004$. The Liberal Party (LP) gained power when Former President Benigno Aquino III won the $2010$ elections. PDP Laban (PDPLBN) rose to power in $2016$ with Former President Rodrigo Roa Duterte, and Partido Federal ng Pilipinas (PFP) is the party of Current President Ferdinand Marcos Jr., who took office in $2022$. 

At the aggregate scale, a clear pattern of bandwagoning emerges. Ruling party membership tends to spike during the midterm elections. Notice the sharp increases in party membership for the Liberal Party in $2013$ and $2016$, and similarly sharp increases in $2019$ to $2022$ for PDPLBN. The case of LKS-CMD/LKS-KAM deviates somewhat from this norm as Former President Gloria Macapagal Arroyo was functionally in office from $2001$ upon the impeachment of Former President Joseph Estrada. Still, after GMA left office in 2010, LKS-CMD split into majority and minority blocks (hence the party membership here is reflected as 0), demonstrating the fragility of party ties once bandwagon effects no longer draw the party together. 

Here, however, there seems to be no discernible distinction between the bandwagoning behavior exhibited by dynastic and non-dynastic politicians. Parties tend to be roughly equally composed of dynastic and non-dynastic members, more or less matching the distribution of dynastic and non-dynastic candidates in the election cycle as a whole. We hypothesize that this may be because the pull of bandwagon effects tend to dominate any differences that may exist in their respective party-switching behaviors. Both dynasts and non-dynasts associate with the ruling party not out of loyalty or ideological alignment, but because they perceive the party as providing them with the most secure route to victory \cite{mendoza_political_2014}. 

\section{Linear Mixed Methods}
This section presents the results of the Linear Mixed Methods regression models developed in this study. Subsection $5.3.1$ discusses how the key traits of dynastic clans captured by our indicators impact local development---comparing our findings to those of previous literature on dynastic concentration and development. Meanwhile, Subsection $5.3.2$ examines the reverse causal association---asking how various developmental conditions impact dynastic power and structure. All regressions performed in this paper satisfied the minimum assumptions formally defined in the Methods and are summarized in the Appendix, specifically in Table \ref{tab:VIF-table} and Figures \ref{fig:D1-QQ}, \ref{fig:D2-QQ-ACC}, \ref{fig:D2-QQ-CCD}, \ref{fig:D2-QQ-CGC}, and \ref{fig:D2-QQ-HHI}.

\subsection{\textbf{How do dynasties impact local development?}}
By viewing dynasties as communities within a network, we are able to gauge key structural features of these clans beyond merely the level of dynastic power concentration. These features are crucial in establishing a more precise understanding of how dynasties impact local development, and which structural traits of ruling dynasties tend towards better or worse outcomes. We first investigated the interaction between each of our dynastic indicators and provincial Human Development Index (HDI).

\subsubsection{Provincial Human Development Index (HDI)}

HDI is an aggregate measure of three essential dimensions of development: \textit{life expectancy}, \textit{education}, and \textit{standard of living}. HDI is commonly used as a gauge of multi-dimensional development, capturing the expansion of individuals' capabilities and opportunities in health, education, and welfare, which are not usually emphasized in the province's level of economic growth. HDI thus provides a more holistic view of the actual quality of life in a given constituency.

Several studies argue on the potential interplay of local politics and HDI by analyzing disparities of human development investments. Lantion et al. \cite{hdi_gov_relationship} argued that an increased government prioritization in development investments such as an increased public spending towards improving the quality of education and health could deliver a positive impact towards the life expectancy and welfare of its constituents. This, then, provides a justification for the need to assess the ``status" and ``quality" of local politics as it interferes with an individual's own experience on development. 

\begin{table}[H]
    \centering
    \setlength{\tabcolsep}{6pt}      
    \renewcommand{\arraystretch}{1.2} 
    \caption{Summary of Pertinent LMM Regression Results for Provincial Human Development Index (HDI)}
    \begin{tabular}{
        l 
        S[table-format=1.6]
        S[table-format=1.6]
        S[table-format=1.6]
        S[table-format=1.6]
    }
        \toprule
        \multicolumn{5}{c}{Dependent Variable: \textit{Provincial Human Poverty Index (HDI)}} \\
        \midrule
        {\textbf{Dynastic Metrics}} & {\textbf{ACC}} & {\textbf{CCD}} & {\textbf{CGC}} & {\textbf{HHI}} \\
        \textbf{p-value} & \num{0.493} & \num{0.041}* & \num{0.000}*** & \num{0.239} \\
        \textbf{Coefficient} & \num{0.001} & \num{-0.170} & \num{-0.336} & \num{0.022} \\
        \bottomrule
        \multicolumn{4}{l}{${}^{*}p<0.05; {}^{**}p<0.01; {}^{***}p<0.001$}
    \end{tabular}
    \label{tab:HDI-LMM}
\end{table}

Table \ref{tab:HDI-LMM} presents the results of an LMM Model that takes provincial HDI as the main dependent variable, while allowing for random effects at the provincial level (i.e. provinces are assumed to have varying random intercepts reflective of their ``base" HDI). We find that the level of inequality between clan members (CGC) and the degree of inter-clan connection (CCD) are significant, negative association with HDI $p<0.05$, while dynastic power concentration (HHI) and intra-clan cohesion (ACC) are not significant. Notably, the estimated coefficients of association between CCD, CGC, and HDI are all \textit{negative} and relatively high compared to the possible range of values for each of the indicators.

\begin{table}[H]
    \centering
    \setlength{\tabcolsep}{6pt}      
    \renewcommand{\arraystretch}{1.2} 
    \caption{Several Comparison Metrics for Three Regression Models with HDI as Dependent Variable}
    \begin{tabular}{
        l 
        S[table-format=1.6]
        S[table-format=1.6]
        S[table-format=1.6]
        S[table-format=1.6]
    }
        \toprule
        \multicolumn{4}{c}{Dependent Variable: \textit{Provincial Human Poverty Index (HDI)}} \\
        \midrule
        {\textbf{Regression Model}} & {\textbf{$R^2$}} & {\textbf{Log-Likelihood}} & {\textbf{AIC}}\\
        \midrule
        \textbf{OLS} & {\num{0.093}} & {\num{316.18}} & \num{-622.3634}\\ 
        \textbf{FE} & {\num{0.0683}} & {\num{335.56}} & \num{-633.1188}\\ 
        \textbf{LMM} & {\num{0.8352}$^+$} & {\num{568.6912}} & \num{-1121.3824} \\ 
        \bottomrule
        \multicolumn{4}{l}{${}^{+}\text{Conditional $R^2$ is presented for LMM}$}
    \end{tabular}
    \label{tab:HDI-SummaryMetrics}
\end{table}

Compared to both a fixed-effects model and a simple OLS base regression, the LMM model yielded a significantly better fit. As seen in Table \ref{tab:HDI-SummaryMetrics}, we found that both OLS and Fixed-Effects models failed to provide a meaningful fit, with low $R^2$ values (OLS: $R^2=0.093$; FE: $R^2 = 0.0683$). However, when incorporating provincial level random-effects through our LMM regression, we obtained a much lower AIC value compared to the two former models, suggesting a better fit. This suggests that there are notable provincial level differences in the relationship between HDI and the various dynastic traits investigated. 

At a $95\%$ level of significance, our LMM model provides strong empirical evidence that the worsening status of the power distribution in the provincial dynastic network (CGC) and increasing cohesion between political clans (CCD) could lead to significant declines in the quality of provincial-level development. Crucially, dynastic power concentration alone does not meaningfully capture the negative impacts dynasties have on development. A possible interpretation for this is that the presence of dominant clans alone does not lead to deteriorating developmental conditions. Rather, when political clans can collude and the political landscape becomes dominated by central clan-heads who amass disproportionate influence over other candidates, they begin to disrupt democratic checks and balances set in place to ensure the provision of better policy. 

The concentration of high political positions such as mayors or governors toward an expanding political clan could lead to a monopoly in provincial governance. This endangers not only how the province thrives economically and socially, but also directly affects how the local government prioritizes essential services such as public spending. Additionally, the dilution of political competition in local sphere enables thriving dynasties to be spared from electoral accountability, providing them a free reign to rule without creating policies that meaningfully improve quality of life of their constituents. 

A similar argument was provided in a study by Villanueva in \cite{villanueva_political_2022} where he regressed human development expenditures with metrics derived from political networks in the Rizal Province. He argued that fat dynastic mayors cause lower human development investments for their constituents compared to their non-dynastic counterparts. Although our study does not explicitly examine fatness and thinness, our alternative indicators in CGC and CCD still capture instances where multiple candidates under the same family name runs simultaneously. 

Additionally, our finding that political power concentration as measured by HHI is insignificantly related to HDI aligns with findings of \cite{mendoza_fat_2019} which similarly found an insignificant relationship between the seat share of fat dynasties in a given province with local poverty incidence. Considering only the share of political dynasties among elected officials may not be enough to capture the potential effects of dynastic rule. Thus, the inherent structure of dynastic clans may be a better indicator of the direction of provincial development than political power concentration alone.

\subsubsection{Provincial Poverty Incidence (POV)}

Presented in Table \ref{tab:POV-LMM} are the results of the LMM regression with the incidence of provincial poverty (POV) as our dependent variable and a random effect imposed at the provincial level. The summary table shows that all dynastic metrics are insignificant in explaining the variations for POV at any desirable level of $\alpha$. 

\begin{table}[H]
    \centering
    \setlength{\tabcolsep}{6pt}      
    \renewcommand{\arraystretch}{1.2} 
    \caption{Summary of Pertinent LMM Regression Results for Provincial Poverty Incidence (POV)}
    \begin{tabular}{
        l 
        S[table-format=1.6]
        S[table-format=1.6]
        S[table-format=1.6]
        S[table-format=1.6]
    }
        \toprule
        \multicolumn{5}{c}{Dependent Variable: \textit{Provincial Poverty Incidence (POV)}} \\
        \midrule
        {\textbf{Dynastic Metrics}} & {\textbf{ACC}} & {\textbf{CCD}} & {\textbf{CGC}} & {\textbf{HHI}} \\
        \textbf{p-value} & \num{0.249} & \num{0.911} & \num{0.975} & \num{0.583} \\
        \textbf{Coefficient} & \num{0.003} & \num{-0.014} & \num{0.005} & \num{-0.017} \\
        \bottomrule
        \multicolumn{4}{l}{${}^{*}p<0.05; {}^{**}p<0.01; {}^{***}p<0.001$}
    \end{tabular}
    \label{tab:POV-LMM}
\end{table}

This result is rather interesting, as examining the LMM's AIC in Table \ref{tab:POV-SummaryMetrics} suggests that it still provides a better fit than either the fixed-effects model or the base OLS. Thus, even when accounting for provincial-level random effects, it seems the relationship between dynastic rule and poverty incidence is statistically negligible.

\begin{table}[H]
    \centering
    \setlength{\tabcolsep}{6pt}      
    \renewcommand{\arraystretch}{1.2} 
    \caption{Several Comparison Metrics for Three Regression Models with POV as Dependent Variable}
    \begin{tabular}{
        l 
        S[table-format=1.6]
        S[table-format=1.6]
        S[table-format=1.6]
        S[table-format=1.6]
    }
        \toprule
        \multicolumn{4}{c}{Dependent Variable: \textit{Provincial Poverty Incidence (POV)}} \\
        \midrule
        {\textbf{Regression Model}} & {\textbf{$R^2$}} & {\textbf{Log-Likelihood}} & {\textbf{AIC}}\\
        \midrule
        \textbf{OLS} & {\num{0.074}} & {\num{217.41}} & \num{-424.8162}\\ 
        \textbf{FE} & {\num{ 0.0706}} & {\num{227.85}} & \num{-447.7058}\\ 
        \textbf{LMM} & {\num{0.6677}$^+$} & {\num{357.20}} & \num{-698.4100} \\ 
        \bottomrule
        \multicolumn{4}{l}{${}^{+}\text{LMM only has conditional $R^2$}$}
    \end{tabular}
    \label{tab:POV-SummaryMetrics}
\end{table}

However, our findings here are unsurprising. Studies have consistently found that dynastic prevalence may not be a good explanatory variable for poverty incidence. e, each using different approaches, also found no strong relationship between dynastic prevalence and poverty incidence. Thus, it seems the existence and dominance of political clans alone neither reduces nor increases poverty in a given constituency.

However, we must be wary of how such results should be interpreted. We offer three possible perspectives, drawn from literature and the data at hand, to make sense of these findings.

First, non-dynastic candidates may be as guilty of failing to meaningfully alleviate poverty as their dynastic rivals. This is the view taken by Mendoza et al. \cite{mendoza_political_2013} and Beja et al. \cite{beja_jr_inequality_2012}. They suggested that non-dynastic candidates fail to deliver genuine reform that results in poverty alleviation. We would note, however, that there is limited empirical evidence regarding the actual differences in policies enacted by dynastic versus non-dynastic incumbents, and thus, much remains to be investigated in support of this hypothesis.

\begin{figure}[H]
        \centerline{\includegraphics[width = 1\textwidth]{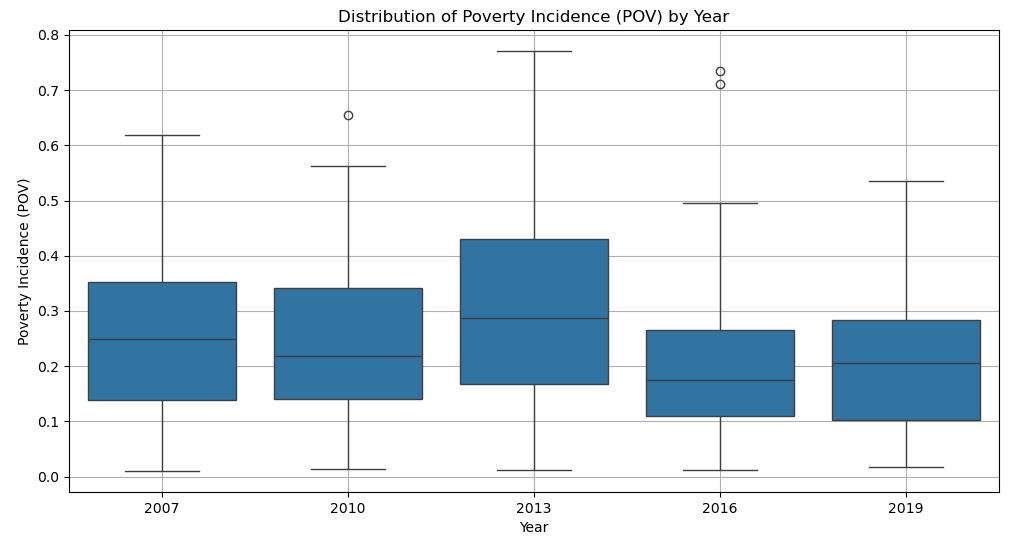}}
        \caption{Distribution of Poverty Incidences (POV) per electoral year}
        \label{PLOT-POV}
\end{figure}

Second, it may also be the case that dynastic prevalence alone is too simple to account for the complex causes of poverty incidence. The effect of broader macroeconomic trends may overpower the influence of individual ruling clans. As shown in the boxplot in Figure \ref{PLOT-POV}, not only has the average poverty incidence decreased (except for in $2013$), the distribution of provincial poverty incidence has also grown narrower, with disparities in poverty between provinces decreasing. It may be that national-level trends in economic development simply have a much greater impact on poverty incidence than local policy. 

Lastly, it may simply be that more sophisticated methods are necessary to ascertain the precise causal effect of dynastic rule on poverty. Such methods would likely require identifying a valid natural experiment that would allow for the use of regression discontinuity or instrumental-variable based methods.
In the subsequent section, we investigate the possibility of a reverse causality between poverty and dynastic prevalence. That is, perhaps poverty incidence may be an explanatory variable for the continued dominance of dynastic clans. 

\subsection{\textbf{How do local development indicators influence persistence of dynasties?}}

We now reverse our line of inquiry and ask how the level of development present at the provincial level could explain the continued persistence of political dynasties. In this direction, we want to explore the possible influences of the same macroeconomic variables---poverty incidence (POV) and human development index (HDI)---on the structure and features of that locale's dynastic clans. 

We treat each of the dynastic indicators (ACC, CCD, CGC, HHI) as dependent variables and regress against POV and HDI while still considering time-based fixed effects and provincial-level random effects. Additionally, we consider the \textit{lagged} values of each macroeconomic variable on every regression performed. This is to account for the possible lagged effects of development caused by the previous administration in shaping the current ruling dynasty.  

Presented in Table \ref{tab:METRICS-LMM} are the summarized results of the LMM regressions performed for this section. From the table, there are only three instances where the LMM returns a significant association at $\alpha \leq 0.05$ level of significance. First, the poverty incidence (and its 3-year lag value) appears to be associated with intra-dynastic strength metric measured by ACC. These associations are \textit{positive}, with the poverty incidence having a high coefficient compared to ACC. Additionally, HDI has a strong significant, negative relationship with CGC.

\begin{table}[H]
    \centering
    \setlength{\tabcolsep}{0pt}      
    \renewcommand{\arraystretch}{1.2} 
    \caption{Summary of Individual LMM Regression for each Development Indicators}
    \begin{tabular}{
        l 
        S[table-format=1.6] S[table-format=1.6] S[table-format=1.6] 
        S[table-format=1.6] 
    }
        \toprule
        \multicolumn{5}{c}{Dependent Variable: \textit{Dynastic Metrics}} \\
        \midrule
        \textbf{Development Indicators} & \textbf{HDI} & \textbf{HDI\_lag\_3year} & \textbf{POV} & \textbf{POV\_lag\_3year}\\
        \midrule 
        \textbf{ACC} \\
        \textbf{p-value} & \num{0.324} & \num{0.338} & \num{0.039}* &  \num{0.092} \\ 
        \textbf{coefficient} & \num{1.735} & \num{-1.675} & \num{2.169} &  \num{-1.813} \\ 
        \midrule 
        \textbf{CCD} \\
        \textbf{p-value} & \num{0.497} & \num{0.105} & \num{0.928} &  \num{0.017}* \\ 
        \textbf{coefficient} & \num{0.029} & \num{-0.069} & \num{0.002} &  \num{0.061} \\ 
        \midrule 
        \textbf{CGC} \\
        \textbf{p-value} & \num{0.001}*** & \num{0.879} & \num{0.548} &  \num{0.510} \\ 
        \textbf{coefficient} & \num{-0.113} & \num{-0.005} & \num{-0.013} &  \num{0.015} \\ 
        \midrule 
        \textbf{HHI} \\
        \textbf{p-value} & \num{0.900} & \num{0.763} & \num{0.326} &  \num{0.088} \\ 
        \textbf{coefficient} & \num{0.021} & \num{-0.051} & \num{-0.097} &  \num{0.173} \\ 
       \bottomrule
    \end{tabular}
    \label{tab:METRICS-LMM}
    ${}^{*}p<0.05; {}^{**}p<0.01; {}^{***}p<0.001$
\end{table}

As with the previous section, each of the LMM models developed performed significantly better than either the equivalent fixed effects or simple OLS. This is evidenced by the low AIC values attained across all the regression instances in Tables \ref{tab:AAC-SummaryMetrics}, \ref{tab:CCD-SummaryMetrics}, \ref{tab:CGC-SummaryMetrics}, and \ref{tab:HHI-SummaryMetrics}. Additionally, both OLS and FE return low values of $R^2$, implying that these models fail to capture a good fit of the possible relationship between these variables. 

\begin{table}[H]
    \centering
    \setlength{\tabcolsep}{6pt}      
    \renewcommand{\arraystretch}{1.2} 
    \caption{Several Comparison Metrics for Three Regression Models with ACC as Dependent Variable}
    \begin{tabular}{
        l 
        S[table-format=1.6]
        S[table-format=1.6]
        S[table-format=1.6]
        S[table-format=1.6]
    }
        \toprule
        \multicolumn{4}{c}{Dependent Variable: \textit{Total Community Connectivity (ACC)}} \\
        \midrule
        {\textbf{Regression Model}} & {\textbf{$R^2$}} & {\textbf{Log-Likelihood}} & {\textbf{AIC}}\\
        \midrule
        \multicolumn{4}{c}{Independent Variable: \textit{Provincial HDI}} \\
        \midrule
        \textbf{OLS} & {\num{0.003}} & {\num{-1155.3}} & \num{2317}\\ 
        \textbf{FE} & {\num{0.0021}} & {\num{-1151.5}} & \num{2307.0369}\\ 
        \textbf{LMM} & {$\num{0.8214}^+$} & {\num{-901.8918}} & \num{1815.7835} \\ 
        \midrule
        \multicolumn{4}{c}{Independent Variable: \textit{Provincial POV}} \\
        \midrule
        \textbf{OLS} & {\num{0.004}} & {\num{-1155.2}} & \num{2316}\\ 
        \textbf{FE} & {\num{0.0057}} & {\num{-1150.8}} & \num{2305.5469}\\ 
        \textbf{LMM} & {$\num{0.8237}^+$} & {\num{-899.7065}} & \num{1811.4130} \\ 
        \bottomrule
        \multicolumn{4}{l}{${}^{+}\text{LMM only has conditional $R^2$}$}
    \end{tabular}
    \label{tab:AAC-SummaryMetrics}
\end{table}

\begin{table}[H]
    \centering
    \setlength{\tabcolsep}{6pt}      
    \renewcommand{\arraystretch}{1.2} 
    \caption{Several Comparison Metrics for Three Regression Models with CCD as Dependent Variable}
    \begin{tabular}{
        l 
        S[table-format=1.6]
        S[table-format=1.6]
        S[table-format=1.6]
        S[table-format=1.6]
    }
        \toprule
        \multicolumn{4}{c}{Dependent Variable: \textit{Connected Component Density (CCD)}} \\
        \midrule
        {\textbf{Regression Model}} & {\textbf{$R^2$}} & {\textbf{Log-Likelihood}} & {\textbf{AIC}}\\
        \midrule
        \multicolumn{4}{c}{Independent Variable: \textit{Provincial HDI}} \\
        \midrule
        \textbf{OLS} & {\num{0.006}} & {\num{252.10}} & \num{-498.2064}\\ 
        \textbf{FE} & {\num{0.0051}} & {\num{253.76}} & \num{-503.5280}\\ 
        \textbf{LMM} & {$\num{0.9091}^+$} & {\num{611.9558}} & \num{-1211.9115} \\ 
        \midrule
        \multicolumn{4}{c}{Independent Variable: \textit{Provincial POV}} \\
        \midrule
        \textbf{OLS} & {\num{0.009}} & {\num{252.72}} & \num{-499.4}\\ 
        \textbf{FE} & {\num{0.0089}} & {\num{254.54}} & \num{-505.0785}\\ 
        \textbf{LMM} & {$\num{0.9100}^+$} & {\num{613.2909}} & \num{-1214.5817} \\ 
        \bottomrule
        \multicolumn{4}{l}{${}^{+}\text{LMM only has conditional $R^2$}$}
    \end{tabular}
    \label{tab:CCD-SummaryMetrics}
\end{table}

\begin{table}[H]
    \centering
    \setlength{\tabcolsep}{6pt}      
    \renewcommand{\arraystretch}{1.2} 
    \caption{Several Comparison Metrics for Three Regression Models with CGC as Dependent Variable}
    \begin{tabular}{
        l 
        S[table-format=1.6]
        S[table-format=1.6]
        S[table-format=1.6]
        S[table-format=1.6]
    }
        \toprule
        \multicolumn{4}{c}{Dependent Variable: \textit{Centrality Gini Coefficient (CGC)}} \\
        \midrule
        {\textbf{Regression Model}} & {\textbf{$R^2$}} & {\textbf{Log-Likelihood}} & {\textbf{AIC}}\\
        \midrule
        \multicolumn{4}{c}{Independent Variable: \textit{Provincial HDI}} \\
        \midrule
        \textbf{OLS} & {\num{0.042}} & {\num{252.10}} & \num{-1001.9912}\\ 
        \textbf{FE} & {\num{0.0416}} & {\num{504.78}} & \num{-1005.5617}\\ 
        \textbf{LMM} & {$\num{0.7616}^+$} & {\num{701.2082}} & \num{-1390.4164} \\ 
        \midrule
        \multicolumn{4}{c}{Independent Variable: \textit{Provincial POV}} \\
        \midrule
        \textbf{OLS} & {\num{0.032}} & {\num{501.86}} & \num{-997.7236}\\ 
        \textbf{FE} & {\num{0.0295}} & {\num{502.21}} & \num{-1000.4136}\\ 
        \textbf{LMM} & {$\num{0.7671}^+$} & {\num{706.2045}} & \num{-1400.4089} \\ 
        \bottomrule
        \multicolumn{4}{l}{${}^{+}\text{LMM only has conditional $R^2$}$}
    \end{tabular}
    \label{tab:CGC-SummaryMetrics}
\end{table}

\begin{table}[H]
    \centering
    \setlength{\tabcolsep}{6pt}      
    \renewcommand{\arraystretch}{1.2} 
    \caption{Several Comparison Metrics for Three Regression Models with HHI as Dependent Variable}
    \begin{tabular}{
        l 
        S[table-format=1.6]
        S[table-format=1.6]
        S[table-format=1.6]
        S[table-format=1.6]
    }
        \toprule
        \multicolumn{4}{c}{Dependent Variable: \textit{Political Herfindahl-Hirschman Index (HHI)}} \\
        \midrule
        {\textbf{Regression Model}} & {\textbf{$R^2$}} & {\textbf{Log-Likelihood}} & {\textbf{AIC}}\\
        \midrule
        \multicolumn{4}{c}{Independent Variable: \textit{Provincial HDI}} \\
        \midrule
        \textbf{OLS} & {\num{0.025}} & {\num{-270.95}} & \num{547.9090}\\ 
        \textbf{FE} & {\num{0.0232}} & {\num{-270.50}} & \num{544.9931}\\ 
        \textbf{LMM} & {$\num{0.8905}^+$} & {\num{55.4267}} & \num{-98.8534} \\ 
        \midrule
        \multicolumn{4}{c}{Independent Variable: \textit{Provincial POV}} \\
        \midrule
        \textbf{OLS} & {\num{0.032}} & {\num{-269.52}} & \num{545.0476}\\ 
        \textbf{FE} & {\num{0.0321}} & {\num{-268.62}} & \num{541.2452}\\ 
        \textbf{LMM} & {$\num{0.8918}^+$} & {\num{57.1476}} & \num{-102.2952} \\ 
        \bottomrule
        \multicolumn{4}{l}{${}^{+}\text{LMM only has conditional $R^2$}$}
    \end{tabular}
    \label{tab:HHI-SummaryMetrics}
\end{table}

Overall, three main insights can be drawn from the results found in this section. Firstly, increased provincial poverty incidence can contribute to a further consolidation of political clans in the area, essentially making existing dynasties more intact and resilient against any perturbation. Secondly, we must account for how the lagged effects of poverty incidence could affect in the dynamic clan cohesions existing between political clans in the province. Lastly, improvements in human development tend to reflect a decentralization of political power in dynastic clans in the province.

\subsubsection{Poverty and Dynastic Entrenchment}
At an $\alpha = 0.05$ level of significance, it was reflected in Table \ref{tab:METRICS-LMM} that an increase in poverty incidence produced a $2.169$ unit increase in the level of intra-cohesions of political clans in the area (ACC). Likewise, an increase in the 3-year lagged poverty incidence induced a $0.061$ unit increase in inter-clan cohesion in the province (CCD). These values suggest that, in provinces with higher rates of poverty incidence, ruling clans tend to grow both more tight-knit and interconnected. 

Similar results were observed in the study led by Mendoza in 2013 \cite{mendoza_political_2013}. They found that although poverty incidence is not a determinant of political power share among clans, the same variable significantly contributes $(p <0.05)$ to the expansion of the largest and strongest political dynasties. They argued that worsening and unchanging poverty conditions would be beneficial to well-established political dynasties. Even though our paper does not exclusively focus on the most powerful clans in each province \cite{mendoza_political_2013}, our findings corroborate the suggestion that major clans tend to grow more expansive in poorer regions. 

\subsubsection{Diverse Perspectives on Poverty and Dynastic Entrenchment}
The associations between poverty and dynasty generated from the LMM regression can be expounded in several viewpoints drawn from established studies. Although dynasty papers that explored this direction of regression remain limited, our group made an extensive effort to interpret the results we had from our regression.

One viewpoint we can look at is to analyze the position of the poor and marginalized sectors as being the most vulnerable victims of dirty tactics such as vote buying, patronage politics, and political intimidation. A similar case was reported in low-income voters in Metro Manila were targeted with vote-buying instances by election aspirants, either by offering them money, groceries, or a favor \cite{canare_mendoza_votebuying}. 

These acts of political patronage and clientism not only impede fair democractic processes entrusted to candidates during elections, but it also generates an unfair advantage to established political clans who are currently eyeing to extend their term through reelections. For Mendoza in \cite{mendoza_political_2013}, incumbent political clans in the local scene had already established a network of patronage and connections and had amassed enough financial and social capital to mobilize large-scale political machineries to secure votes from economically and politically disadvantaged voters. Therefore, the persistence of patron-client relationship between the poor and dynastic candidates further embeds the familial clan into the local politics. In turn, dynastic clans do not only become more intact and well-structured, but it also provides them an opportunity to consolidate greater power through establishing connections with other dynasts.

We can also perceive the linkage between dynasty and development as a consequence of deliberate attempts by incumbent dynastic administration to impede poverty conditions in their jurisdiction. Here, we look at how poverty alleviation programs are transformed from social reforms to political instruments supervised by the incumbent administration to enhance their clan legitimacy and establish ruling power. In the discussion provided by Wang and Guo in \cite{wang_politics_of_poverty}, they argued that poverty alleviation programs are instrumentalized by reigning political parties and politicians as an electoral tool to hold their position in office by means of either disproportionally allocating poverty-relief goods to targeted groups of constituents to whom the politicians would benefit most (voters vs. nonvoters) or choosing the timing of anti-poverty goods delivery (e.g., near the election period or during electoral cycles). These practices become a reflection of  political favoritism and clientelism that destabilizes true nature of poverty reduction programs. Consequently, it produces a misalignment between the allocation of anti-poverty efforts and what the poor actually needs \cite{wang_politics_of_poverty}. 

Our results from the paper may be applied to the observations from these perspectives. Given the increasing trend of percentage share of dynastic politicians from $40.73\%$ in $2004$ to $53.22\%$ to $2022$ presented in the previous sections, there is now a growing concern that incumbent dynastic clans may politicize the release, allocation, and distribution of political alleviation funds and programs to favor their political party or familial ties. As argued by Swamy \cite{swamy_socialprotection_2016}, social protection and poverty alleviation programs in the country like cash condition transfer programs and \textit{pantawid} initiatives may still be associated with local power brokers (e.g, incumbent local politicians) which, in turn, may still induce sustain clientism in electoral contexts.  

The perspective above can be surmised into the following: \textit{The failure to alleviate poverty in provincial areas may be caused by politicization of incumbent dynasts with social protection programs.} Our LMM results further highlight how the worsening conditions of poverty could, in turn, make currently ruling political clans more connected and robust against shocks due to their clientistic tendencies which establishes their name even stronger in the next elections.

The same perspective can also be applied to analyze the lag effects of poverty extracted from our LMM regression. Recall that the results in Table \ref{tab:METRICS-LMM} reveal that worsening poverty incidence measured from the last administration (i.e, past 3-year) results to an increase in inter-dynastic cohesions (CCD) between dynasties. 

We discussed previously that incumbent dynasts see poverty alleviation programs as a political opportunity for them to continue reigning over their respective jurisdictions. The onset of these political tactics generates an electoral pressure to current political clans of different sizes and structures on how they can cope up with such attribution. We argue that these, in turn, could make several political clans to become allies by either intermarriages between members or alliance between clan heads, thereby consolidating dynastic members into one larger community of dynasty, a measure that is captured by CCD. A specific instance was observed in the focus discussion in \cite{mendoza_dynastic_2020} where the emergence of Stephany Uy-Tan, a member of the Uy dynasty in Western Samar, ensured a stronger connection with the Tan dynasty after her marriage with Stephen Tan.

Another explanation that we could impart here revolves around the natural tendencies of incumbent dynastic clans to maintain their political power by either making the incumbent member seek reelection to the same position or higher or inviting their family members to replace them. Thus, the effects of poverty incidences observed in their previous term remain significant in the next administration as the same dynastic clans occupy the local positions, thereby maintaining, if not further consolidating, the landscape of political networks persisting in the local area.

However, current literature about this ``lagged effect" of poverty and dynast incumbencies remain elusive as related papers only focus on the immediate effects of dynastic measures on socioeconomic statuses. Thus, there could only be speculations to be provided from our end as to why poverty incidences are still significant even at the future administration.

\subsubsection{HDI and Clan Power Asymmetry}

At an $\alpha = 0.001$ level of significance, it was found in Table \ref{tab:METRICS-LMM} that an increase in HDI results in a $0.113$ decrease in CGC, reflecting the inequality in power distribution. This implies that improvements in human development tend to result in a more equal distribution of political ties among incumbents.

This result is to be expected, however, as improvements in quality of life tend to directly result from the more robust checks and balances that arise when power is more equally distributed in a given political landscape. When there is no significant disparity in the level of political clout held by the politicians in a network, these politicians are less capable of leveraging their patronage and family ties to evade accountability \cite{mendoza_fat_2019, querubin_political_2012}. When no one clan has a clear advantage over any other and clans do not collude, even though dynasties may dominate the system, there are sufficient electoral incentives to pass meaningful reforms.

Likewise, given the competitive nature of elections in the Philippines, an evenly distributed political power between dynastic clans and non-dynastic partisans could establish a form of surveillance among themselves to ensure that no form of corruption would make one clan or party more powerful than the other. Thus, local incumbents would be disincentivized to perform any actions that would undermine the checks and balances of the system, which, when caught by other clans, could cause their family's ultimate demise.


\chapter{Conclusion and Recommendations}
% ...

This chapter presents the primary conclusions to each of the research questions pursued in this study. We also outline a number of directions for future research building on the methods and results laid out here.

\subsection{Evolution of Dynastic Networks}

This study's findings suggest that, over the period from 2004 to 2022, political dynasties have grown stronger and more interconnected. Across all provinces, dynastic candidates have taken up an increasing share of seats versus their non-dynastic counterparts–supporting the theoretical consensus that the prevalence of dynasties has in fact hampered electoral competitiveness. In a majority of provinces, formerly competing dynastic clans have formed strategic inter-marriage alliances with each other, resulting in more deeply interconnected dynastic networks. This suggests that rather than the powerful political clans``check and balancing" each other, their persistence in provincial political landscape may only provide pathways for stronger collusion.

Moreover, we draw several novel insights regarding the structure of these dynastic clans. We find that dynastic networks tend to be organized around central figures who hold a disproportionate level of influence. Clans tend to be structured around a dominant patriarch or matriarch who anchors the dynastic network. We note however that over time, this pattern seems to be weakening, which could suggest that power in dynasties is becoming more diffuse among their members. Finally, we note that political clans have remained tightly-knit, with members often having multiple overlapping connections that keep them close to their respective dynasties.

\subsection{Party Hopping}

We uncover new evidence surrounding the dynamics between dynasties and political parties. We find that, in all dynastic families surveyed across the $80$ provinces and throughout $5$ election years, are significantly more likely to hop parties than their non-dynastic counterparts. Dynasts can rely on their clans as an independent safety net for resources and political clout, voiding the need to form strong political bonds with other candidates as part of a party. This is further supported by our finding that, while a large proportion of clans choose to run under a unified party banner, a significant number also choose to run under fractured party allegiances. A political party taking in dynastic candidates would have to constantly question whether its dynastic members' loyalty to their own families would not supersede their loyalty to the party. This analysis fills a key gap in the dynastic literature with regards to explaining the persistence of dynasties. Dynasties have remained the dominant form of political organization because dynastic candidates undermine the formation of institutional trust within parties. 

In a legislative climate where an outright dynastic ban seems unlikely to manifest in law, these findings may provide an alternative perspective on how to tackle the dynastic problem. In contexts where strong parties exist that can ensure the continuity of policy and satisfy term-limited incumbents' desire to hold onto power, candidates may feel less need to proliferate their own dynasties. Much investigation remains surrounding the relationship between parties and dynasties. Policymakers seeking to combat the dynastic problem while avoiding the major legislative roadblock of an outright dynastic ban could benefit from taking the perspective that dynasties persist primarily as a form of political organization in a context where no appealing alternatives exist.

Admittedly, our results fall short of providing a precise causal claim regarding the relationship between dynastic membership and party hopping. As dynastic membership is an immutable characteristic, it is difficult to establish the precise causal effect of dynastic status on party loyalty. Our findings here control for province and election year, but with more expansive data (such as information on candidates' social class or political alignments) it may be possible to extract even closer causal approximations. 

\subsection{Developmental Impacts of Dynastic Rule}
Our findings concur with previous studies that dynastic prevalence has a strong relationship with the level of human development attained in the province. However, we found that the dynastic structure and connectedness between and within political clans, and not the distribution and inequality of political power, affects the quality of life attained by their constituents. 

Additionally, our study supports the notion from related literature that dynastic prevalence fails to provide valuable understanding with the provincial poverty index. However, when viewed in reverse causal direction, we discovered that worsening poverty conditions can further consolidate inter- and intra-dynastic ties between political clans. 

Our findings suggest that dynastic prevalence alone may not lead to deteriorating socioeconomic and development outcomes as the non-dynastic candidates that \textit{do} win seats equally fail to enact beneficial policy. However, this paper provides a leading evidence that invites us to perceive the dynasty issue in an alternative perspective. That is, the spread of dynasties in the political sphere may be attributed to worsening development indicators which were ultimately caused by their own corrupt practices. Our results support the idea that incumbent dynastic clans may politicize poverty alleviation procedures for their own benefit, causing their clan to become more powerful and concentrated at the expense of further disefranchisement of the poor and marginalized sectors. 

Overall, our study provides relevant insights and evidence towards the understanding that reigning political clans are marginally electorally accountable and thus can continue to reign without passing beneficial policy. Dynastic structure, rather than dynastic power concentration, may be the determinant of dynasties' true effect on local development. 

\subsection{Recommendations}

This study lays a rich foundation for future work in dynastic research. The indicators we develop, or modifications thereof, open the possibility of analyzing dynasties not just from the perspective of the power they hold, but also in terms of their internal network structure. There are several areas in particular that we believe would prove worthwhile for further investigation.

This study, as with much dynastic literature, has focused exclusively on incumbents–-the candidates that win their respective electoral races. This is largely due to limitations in the available data rather than a result of a conscious choice in research design. Much richer insights regarding the electoral impacts of dynasties could be drawn from a more complete view of each election cycle that considers both winning and losing candidates. Do dynastic candidates receive a higher share of votes than their nondynastic counterparts? Do ``winning" dynasties tend to be more close-knit than ``losing" dynasties? Do members of more central dynasties receive a higher share of votes than those of less-connected dynasties? Such questions and many others can only be answered with more comprehensive election data that at the time of writing, is not easily accessible.

More robust causal inference strategies could also be deployed on the dataset to determine more specifically both the developmental impacts of dynasties and the factors that contribute to dynastic persistence. With richer datasets, it may be worthwhile to focus on developmental outcomes that avoid the problem of endogeneity with dynastic prevalence to minimize the impact of confounding variables. With poverty and HDI as our focused indicators, it is difficult to establish the exact role dynasties play. Future research may want to examine potential instrumental variables that could allow for more robust causal inference, such as measures of corruption, fiscal spending for essential services such as health and education, cases of political violence and terrorism, and strength of media, to name a few. Additionally, we note that the case of BARMM, which recently enacted a strong anti-dynasty law, could become the basis for regression discontinuity strategies in the future for isolating the causal effect of dynastic prevalence. 

Finally, more sophisticated methods of analyzing dynastic network structure could be used in the future to extend on the work done here. Emerging techniques such as Topological Data Analysis may yet reveal unique insight surrounding the structure of dynastic networks and their evolution over time. Future researcher may want to investigate the higher dimensional properties of dynastic graphs, which may perhaps reveal patterns in the formation of political clans or even new taxonomies of political dynasties. 


\newpage

\makeatletter
\def\@makeschapterhead#1{%
  \vspace*{-40pt}%
  {\parindent \z@ \raggedright
    \normalfont \centering
    \interlinepenalty\@M
    \Large \bfseries  #1\par\nobreak
    \vskip0.5truecm
  }}
\makeatother

\renewcommand{\bibname}{References}
\addcontentsline{toc}{chapter}{References}

\printbibliography

@misc{sparsityGINI,
      title={Sparsity Measure of a Network Graph: Gini Index}, 
      author={Swati Goswami and C. A. Murthy and Asit K. Das},
      year={2016},
      eprint={1612.07074},
      archivePrefix={arXiv},
      primaryClass={cs.DM},
      url={https://arxiv.org/abs/1612.07074}, 
}

@book{HHI-Paper,
  title     = "National power and the structure of foreign trade",
  author    = "Hirschman, Albert O",
  publisher = "University of California Press",
  series    = "The Politics of the international economy",
  month     =  feb,
  year      =  1981,
  address   = "Berkeley, CA"
}

@article{GINI-Paper,
    author = {},
    title = {Variability and Mutability},
    journal = {Journal of the Royal Statistical Society},
    volume = {76},
    number = {6},
    pages = {619-622},
    year = {2018},
    month = {12},
    issn = {0952-8385},
    doi = {10.1111/j.2397-2335.1913.tb03065.x},
    url = {https://doi.org/10.1111/j.2397-2335.1913.tb03065.x},
    eprint = {https://academic.oup.com/jrsssa/article-pdf/76/6/619/49683089/jrsssa\_76\_6\_619.pdf},
}

@book{Zweig-2014,
  title     = "Network analysis literacy",
  author    = "Zweig, Katharina A",
  publisher = "Springer",
  series    = "Lecture Notes in Social Networks",
  edition   =  1,
  month     =  aug,
  year      =  2014,
  address   = "Vienna, Austria",
  language  = "en"
}

@book{Bickle-2020,
  title     = "Fundamentals of graph theory",
  author    = "Bickle, Allan",
  publisher = "American Mathematical Society",
  series    = "Pure and Applied Undergraduate Texts",
  month     =  may,
  year      =  2020,
  address   = "Providence, RI",
  language  = "en"
}

@article{mendoza_political_2022,
	title = {Political Dynasties and Terrorism: An Empirical Analysis Using Data on the Philippines},
	volume = {10},
	issn = {22882693, 22882707},
	url = {https://kiss.kstudy.com/Detail/Ar?key=3985795},
	doi = {10.18588/202210.00a266},
	shorttitle = {Political Dynasties and Terrorism},
	pages = {435--459},
	number = {2},
	journaltitle = {Asian Journal of Peacebuilding},
	shortjournal = {{AJP}},
	author = {Mendoza, Ronald U. and Yap, Jurel K. and Mendoza, Gabrielle Ann S. and Pizzaro, Angelika Lourdes J. and Engelbrecht, Georgi},
	urldate = {2024-11-17},
	date = {2022-11-30},
	langid = {english},
}

@misc{mendoza_dynastic_2020,
	location = {Rochester, {NY}},
	title = {Dynastic Political Networks: Insights from the Case of Western Samar, Philippines},
	url = {https://papers.ssrn.com/abstract=3652960},
	doi = {10.2139/ssrn.3652960},
	shorttitle = {Dynastic Political Networks},
	number = {3652960},
	publisher = {Social Science Research Network},
	author = {Mendoza, Ronald U. and Hiwatig, Joshua and Banaag, Miann},
	urldate = {2024-11-17},
	date = {2020-07-16},
	langid = {english},
}

@misc{mendoza_fat_2019,
	location = {Rochester, {NY}},
	title = {From Fat to Obese: Political Dynasties after the 2019 Midterm Elections},
	url = {https://papers.ssrn.com/abstract=3449201},
	doi = {10.2139/ssrn.3449201},
	shorttitle = {From Fat to Obese},
	number = {3449201},
	publisher = {Social Science Research Network},
	author = {Mendoza, Ronald U. and Jaminola, Leonardo M. and Yap, Jurel},
	urldate = {2025-03-11},
	date = {2019-09-01},
	langid = {english},
}

@article{balanquit_measuring_2017,
	title = {Measuring political dynasties in Metro Manila},
	volume = {54},
	rights = {Copyright (c) 2025 Philippine Review of Economics (Online {ISSN} 2984-8156)},
	issn = {2984-8156},
	url = {https://pre.econ.upd.edu.ph/index.php/pre/article/view/952},
	pages = {120--137},
	number = {1},
	journaltitle = {Philippine Review of Economics (Online {ISSN} 2984-8156)},
	author = {Balanquit, Romeo T. and Coronel, Lianca O. and Yambao, Jose Y.},
	urldate = {2025-03-10},
	date = {2017-11-02},
	langid = {english},
	note = {Number: 1},
}

@article{davis_corruption_2024,
	title = {Corruption risk and political dynasties: exploring the links using public procurement data in the Philippines},
	volume = {25},
	issn = {1435-8131},
	url = {https://doi.org/10.1007/s10101-023-00306-4},
	doi = {10.1007/s10101-023-00306-4},
	shorttitle = {Corruption risk and political dynasties},
	pages = {81--109},
	number = {1},
	journaltitle = {Economics of Governance},
	shortjournal = {Econ Gov},
	author = {Davis, Daniel Bruno and Mendoza, Ronald U. and Yap, Jurel K.},
	urldate = {2024-11-17},
	date = {2024-03-01},
	langid = {english},
}

@article{mendoza_political_2013,
	title = {Political Dynasties and Poverty: Resolving the 'Chicken or the Egg' Question},
	issn = {1556-5068},
	url = {http://www.ssrn.com/abstract=2292277},
	doi = {10.2139/ssrn.2292277},
	shorttitle = {Political Dynasties and Poverty},
	journaltitle = {{SSRN} Electronic Journal},
	shortjournal = {{SSRN} Journal},
	author = {Mendoza, Ronald U. and Beja, Edsel L. and Venida, Victor Soriano and Yap Ii, David Barua},
	urldate = {2024-11-17},
	date = {2013},
	langid = {english},
}

@article{labonne_political_2021,
	title = {Political dynasties, term limits and female political representation: Evidence from the Philippines},
	volume = {182},
	issn = {0167-2681},
	url = {https://www.sciencedirect.com/science/article/pii/S0167268120304480},
	doi = {10.1016/j.jebo.2020.12.001},
	shorttitle = {Political dynasties, term limits and female political representation},
	pages = {212--228},
	journaltitle = {Journal of Economic Behavior \& Organization},
	shortjournal = {Journal of Economic Behavior \& Organization},
	author = {Labonne, Julien and Parsa, Sahar and Querubin, Pablo},
	urldate = {2025-04-10},
	date = {2021-02-01},
}

@misc{mendoza_term_2019,
	location = {Rochester, {NY}},
	title = {Term Limits and Political Dynasties: Unpacking the Links},
	url = {https://papers.ssrn.com/abstract=3356437},
	doi = {10.2139/ssrn.3356437},
	shorttitle = {Term Limits and Political Dynasties},
	number = {3356437},
	publisher = {Social Science Research Network},
	author = {Mendoza, Ronald U. and Banaag, Miann and Yusingco, Michael Henry},
	urldate = {2025-04-10},
	date = {2019-03-20},
	langid = {english},
}

@misc{mendoza_political_2014,
	location = {Rochester, {NY}},
	title = {Political Party Switching: It's More Fun in the Philippines},
	url = {https://papers.ssrn.com/abstract=2492913},
	doi = {10.2139/ssrn.2492913},
	shorttitle = {Political Party Switching},
	number = {2492913},
	publisher = {Social Science Research Network},
	author = {Mendoza, Ronald U. and Cruz, Jan Fredrick and Yap {II}, David Barua},
	urldate = {2025-04-11},
	date = {2014-09-23},
	langid = {english},
}

@incollection{engel_network_2021,
	location = {Cham},
	title = {Network Analysis for Economics and Finance: An Application to Firm Ownership},
	isbn = {978-3-030-66891-4},
	url = {https://doi.org/10.1007/978-3-030-66891-4_14},
	shorttitle = {Network Analysis for Economics and Finance},
	pages = {331--355},
	booktitle = {Data Science for Economics and Finance: Methodologies and Applications},
	publisher = {Springer International Publishing},
	author = {Engel, Janina and Nardo, Michela and Rancan, Michela},
	editor = {Consoli, Sergio and Reforgiato Recupero, Diego and Saisana, Michaela},
	urldate = {2025-02-07},
	date = {2021},
	langid = {english},
	doi = {10.1007/978-3-030-66891-4_14},
}

@misc{pimentel_political_2024,
	title = {Political Parties, Family Dynasties, and the Budget Surplus Trap: Addressing the Public Goods Deficit in the Philippines},
	url = {https://blogs.lse.ac.uk/seac/2024/11/21/political-parties-family-dynasties-and-the-budget-surplus-trap-addressing-the-public-goods-deficit-in-the-philippines/},
	shorttitle = {Political Parties, Family Dynasties, and the Budget Surplus Trap},
	author = {Pimentel, Hannah},
	urldate = {2025-02-07},
	date = {2024-11-21},
}

@online{miranda_8_2024,
	title = {8 in every 10 district reps belong to dynasties. More than half are reelectionists in 2025},
	url = {http://pcij.org/2024/10/26/lower-house-district-representatives-political-dynasties-reelection/},
	titleaddon = {{PCIJ}.org},
	author = {Miranda, Maujerie, Guinevere Latoza},
	urldate = {2025-02-06},
	date = {2024-10-26},
	langid = {american},
}

@online{beja_jr_inequality_2012,
	title = {Inequality in democracy: Insights from an empirical analysis of political dynasties in the 15th Philippine Congress},
	url = {https://mpra.ub.uni-muenchen.de/40104/},
	shorttitle = {Inequality in democracy},
	type = {{MPRA} Paper},
	author = {Beja Jr, Edsel and Mendoza, Ronald U. and Venida, Victor S. and Yap, David B.},
	urldate = {2025-03-06},
	date = {2012-07-15},
	langid = {english},
}

@article{balisacan_going_2004,
	title = {Going beyond Crosscountry Averages: Growth, Inequality and Poverty Reduction in the Philippines},
	volume = {32},
	issn = {0305-750X},
	url = {https://www.sciencedirect.com/science/article/pii/S0305750X04001408},
	doi = {10.1016/j.worlddev.2004.05.010},
	shorttitle = {Going beyond Crosscountry Averages},
	pages = {1891--1907},
	number = {11},
	journaltitle = {World Development},
	shortjournal = {World Development},
	author = {Balisacan, Arsenio M. and Fuwa, Nobuhiko},
	urldate = {2025-03-10},
	date = {2004-11-01},
}

@article{ghosh_understanding_2023,
	title = {Understanding layered dominance of political dynasties in India: A de-hyphenated reading of dynastic representation and dynasty-led parties},
	volume = {8},
	issn = {2057-8911},
	url = {https://doi.org/10.1177/20578911221147657},
	doi = {10.1177/20578911221147657},
	shorttitle = {Understanding layered dominance of political dynasties in India},
	pages = {727--747},
	number = {3},
	journaltitle = {Asian Journal of Comparative Politics},
	author = {Ghosh, Ambar Kumar},
	urldate = {2025-02-07},
	date = {2023-09-01},
	langid = {english},
	note = {Publisher: {SAGE} Publications},
}

@article{villanueva_political_2022,
	title = {Political Dynasties and Human Development Investments: Evidence of Linkage from Rizal Province, Philippines},
	volume = {64},
	rights = {Copyright (c)},
	issn = {3028-1296},
	url = {https://journals.upd.edu.ph/index.php/pjpa/article/view/8628},
	shorttitle = {Political Dynasties and Human Development Investments},
	pages = {90--126},
	number = {2},
	journaltitle = {Philippine Journal of Public Administration},
	author = {Villanueva, John Emmanuel B.},
	urldate = {2025-02-06},
	date = {2022-03-14},
	langid = {english},
	note = {Number: 2},
}

@article{mendoza_political_2022-1,
	title = {Political dynasties, business, and poverty in the Philippines},
	volume = {7},
	issn = {2667-3193},
	url = {https://www.sciencedirect.com/science/article/pii/S2667319322000222},
	doi = {10.1016/j.jge.2022.100051},
	pages = {100051},
	journaltitle = {Journal of Government and Economics},
	shortjournal = {Journal of Government and Economics},
	author = {Mendoza, Ronald U. and Yap, Jurel K. and Mendoza, Gabrielle Ann S. and Jaminola, Leonardo and Yu, Erica Celine},
	urldate = {2024-11-17},
	date = {2022-09-01}
}

@article{Traag2019,
  title = {From Louvain to Leiden: guaranteeing well-connected communities},
  volume = {9},
  ISSN = {2045-2322},
  url = {http://dx.doi.org/10.1038/s41598-019-41695-z},
  DOI = {10.1038/s41598-019-41695-z},
  number = {1},
  journal = {Scientific Reports},
  publisher = {Springer Science and Business Media LLC},
  author = {Traag,  V. A. and Waltman,  L. and van Eck,  N. J.},
  year = {2019},
  month = mar 
}

@article{tadem2016,
author = {Tadem, Teresa and Tadem, Eduardo},
year = {2016},
month = {07},
pages = {},
title = {Political dynasties in the Philippines: Persistent patterns, perennial problems},
volume = {24},
journal = {South East Asia Research},
doi = {10.1177/0967828X16659730}
}

@article{Teehankee2020,
author = {Julio C. Teehankee and Cleo Anne A. Calimbahin},
title = {Mapping the Philippines’ Defective Democracy},
journal = {Asian Affairs: An American Review},
volume = {47},
number = {2},
pages = {97--125},
year = {2020},
publisher = {Routledge},
doi = {10.1080/00927678.2019.1702801},
}

@article{mendoza_political_2016,
	title = {Political dynasties and poverty: measurement and evidence of linkages in the Philippines},
	url = {https://archium.ateneo.edu/asog-pubs/63},
	shorttitle = {Political dynasties and poverty},
	journaltitle = {Ateneo School of Government Publications},
	author = {Mendoza, Ronald and Beja, Edsel and Venida, Victor and Yap, David},
	date = {2016-04-01},
}

@article{cruzAER_2017,
Author = {Cruz, Cesi and Labonne, Julien and Querubín, Pablo},
Title = {Politician Family Networks and Electoral Outcomes: Evidence from the Philippines},
Journal = {American Economic Review},
Volume = {107},
Number = {10},
Year = {2017},
Month = {October},
Pages = {3006–37},
DOI = {10.1257/aer.20150343},
URL = {https://www.aeaweb.org/articles?id=10.1257/aer.20150343}
}

@misc{querubin_political_2012,
	location = {Rochester, {NY}},
	title = {Political Reform and Elite Persistence: Term Limits and Political Dynasties in the Philippines},
	url = {https://papers.ssrn.com/abstract=2108036},
	shorttitle = {Political Reform and Elite Persistence},
	number = {2108036},
	publisher = {Social Science Research Network},
	author = {Querubin, Pablo},
	urldate = {2024-11-18},
	year = 2012,
	langid = {english},
}

@article{balindong-sultanate,
author = {Hadji, Sohayle and HADJI ABDUL RACMAN, SOHAYLE and Hassan, D and Shah, Shakeel and Khan, Muhammad},
year = {2020},
month = {10},
pages = {2018},
title = {The Lanao Sultanate in the 17th Century Zakāt System With Special Reference to the Islamic Perspectives of Al-Māwardī}
}

@misc{dynastybill,
  author = {Lacson, Panfilo},
  month = {07},
  number = {30},
  title = {AN ACT PROHIBITING THE ESTABLISHMENT OF POLITICAL DYNASTIES},
  url = {https://legacy.senate.gov.ph/lisdata/3024627075!.pdf},
  year = {2019}
}

@misc{philippine_constitution_1987,
  title = {The 1987 Constitution of the Republic of the Philippines},
  howpublished = {Official text},  % Or "Republic of the Philippines"
  year = {1987} %  The year is now 1987 as you specified.
}

@misc{sk_reform_act_2015,
  title = {Republic Act No. 10755: An Act Strengthening the Sangguniang Kabataan (SK) Reform Act of 2015},
  howpublished = {Official Gazette of the Republic of the Philippines},
  year = {2015}
}

@misc{APC-2016, title = {Ateneo Policy Center Political Dynasties Dataset}, url = {https://www.inclusivedemocracy.ph/data-and-infographics}, author = {Ateneo Policy Center}, year = {2016}}

@misc{Felipe_2011, title={Police nab suspect in ambush-slay of Mt. Province mayor}, url={https://www.philstar.com/nation/2011/02/18/658032/police-nab-suspect-ambush-slay-mt-province-mayor}, abstractNote={Police arrested one of the gunmen in the assassination of Paracelis Mayor Ceasar Balacanao Rafael in Mt. Province on Christmas Day three years ago.}, journal={Philstar.com}, author={Felipe, Bebot Sison Jr, Cecille Suerte}, year={2011}, month=feb }

@misc{HDNPH-Data, title = {Philippine Human Development Report}, author = {Human Development Network PH}, year = {2022}}

@misc{bangsamoro_autonomy_act_35_2022,
  title = {Bangsamoro Autonomy Act No. 35 (Bangsamoro Electoral Code of 2022)},
  howpublished = {Official text, Bangsamoro Parliament}, 
  year = {2022},
}

@article{swamy_socialprotection_2016,
author = {Arun R. Swamy},
title ={Can Social Protection Weaken Clientelism? Considering Conditional Cash Transfers as Political Reform in the Philippines},
journal = {Journal of Current Southeast Asian Affairs},
volume = {35},
number = {1},
pages = {59-90},
year = {2016},
doi = {10.1177/186810341603500103},
URL = { 
    
        https://doi.org/10.1177/186810341603500103
},
eprint = { 
    
        https://doi.org/10.1177/186810341603500103
}
}

@article{wang_politics_of_poverty,
author = {Wang, Zhongyuan and Guo, Sujian},
year = {2022},
month = {06},
pages = {},
title = {Politics of Poverty Governance: an Introduction},
volume = {27},
journal = {Journal of Chinese Political Science},
doi = {10.1007/s11366-022-09804-4}}

@misc{Lucero_2016, title={Ampatuans, Ecleos, Sinsuats, Midtimbangs unrivalled in turfs}, 
url={http://pcij.org/2016/05/04/ampatuans-ecleos-sinsuatsbrmidtimbangs-unrivalled-in-turfs/}, 
abstractNote={AMONG THE 802 unopposed candidates for the 2016 elections, a few stood out not just because their surnames sounded familiar, but also because of the frequency in which these kept popping up.}, journal={PCIJ.org}, 
author={Lucero, Vino}, 
year={2016}, 
month=may, language={en-US} 
}

@article{canare_mendoza_votebuying,
author = {Canare, Tristan and Mendoza, Ronald and Lopez, Mario Antonio},
year = {2018},
month = {02},
pages = {},
title = {An Empirical Analysis of Vote Buying Among the Poor: Evidence from Elections in the Philippines (forthcoming in Southeast Asia Research)}
}

@article{hdi_gov_relationship, title={Assessing the Relationship of Human Development Index (HDI) and Government Expenditure on Health and Education in Selected ASEAN Countries}, volume={4}, url={https://ijosmas.org/index.php/ijosmas/article/view/374}, DOI={10.5555/ijosmas.v4i6.374}, abstractNote={&amp;lt;p&amp;gt;Various studies exhibited the importance of government spending in developing human capital. It can enhance income distribution and employment opportunities, reduce extreme poverty, and increase the consumption of essential services like healthcare and education. The study aims to analyze the relationship between government spending on education and healthcare on the Human Development Index (HDI) of five Southeast Asian countries: the Philippines, Thailand, Malaysia, Indonesia, and Singapore, using the Panel Least Squares method. The study proves that government expenditure on education significantly and positively impacts HDI. In contrast, government spending on health exhibits a positive yet statistically insignificant influence on HDI. These findings imply that directing resources toward education efficiently enhances human capital in the examined countries. However, the same efficiency does not hold true for government allocation to health.&amp;lt;/p&amp;gt;}, number={6}, journal={International Journal of Social and Management Studies}, author={Lantion, Danielle Ann and Musñgi, Gabrielle and Cabauatan, Ronaldo}, year={2023}, month={Nov.}, pages={13–26} }

\begin{appendices}
    \chapter{Appendix}

\subsection{Linear Regression Trend Analysis}

\begin{figure}[H]
        \centerline{\includegraphics[width = 0.9\textwidth]{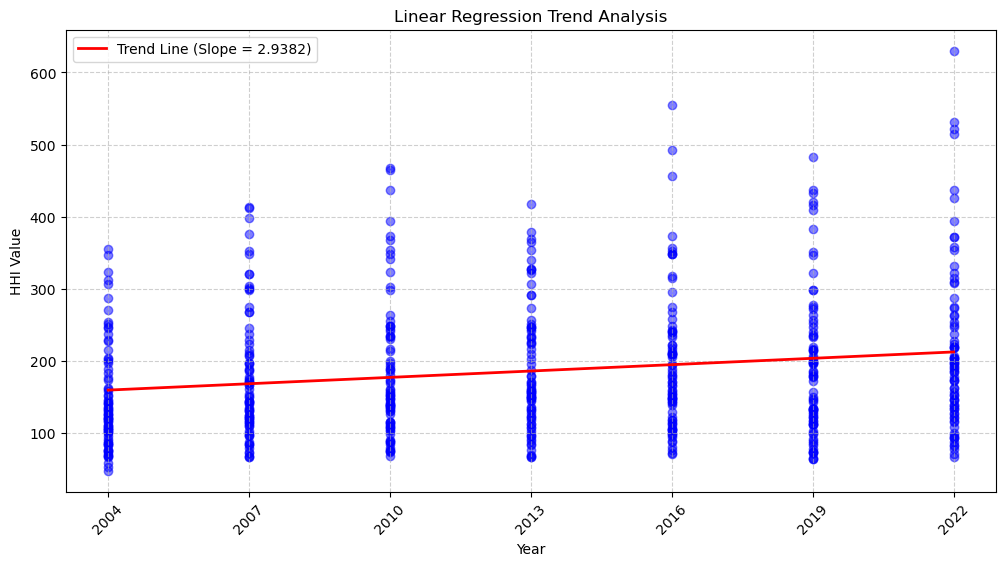}}
        \caption{Trend Analysis of Political Herfindahl-Hirschman Index}
        \label{Trend-HHI}
\end{figure}

\begin{figure}[H]
        \centerline{\includegraphics[width = 0.9\textwidth]{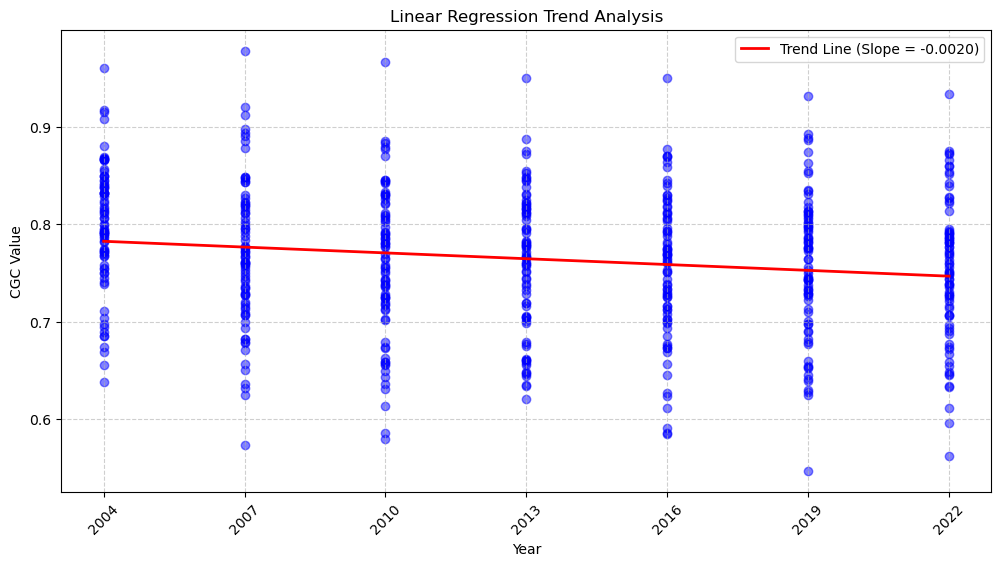}}
        \caption{Trend Analysis of Centrality Gini Coefficient}
        \label{Trend-CGC}
\end{figure}

\begin{figure}[H]
        \centerline{\includegraphics[width = 0.9\textwidth]{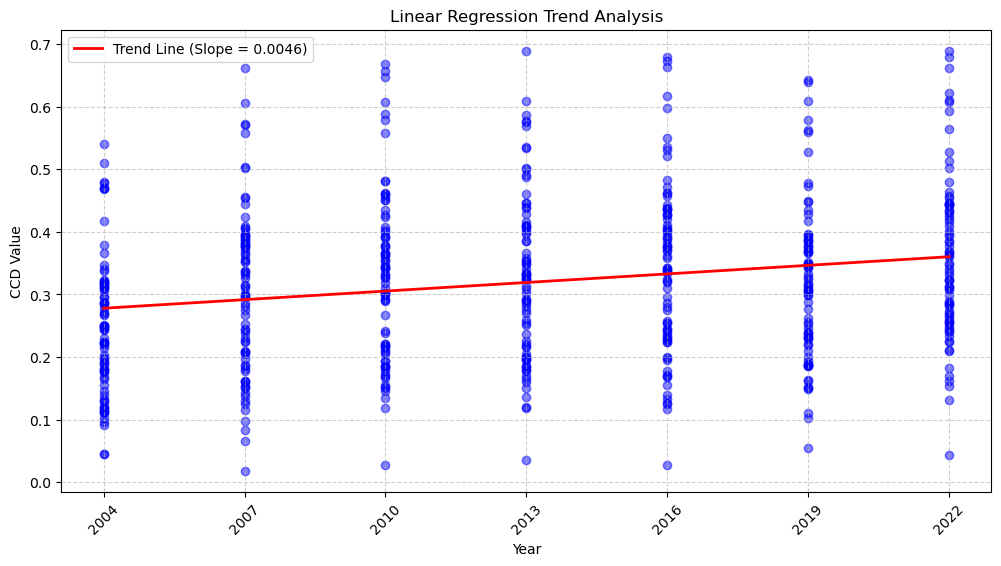}}
        \caption{Trend Analysis of Connected Component Density}
        \label{Trend-CCD}
\end{figure}

\begin{figure}[H]
        \centerline{\includegraphics[width = 0.9\textwidth]{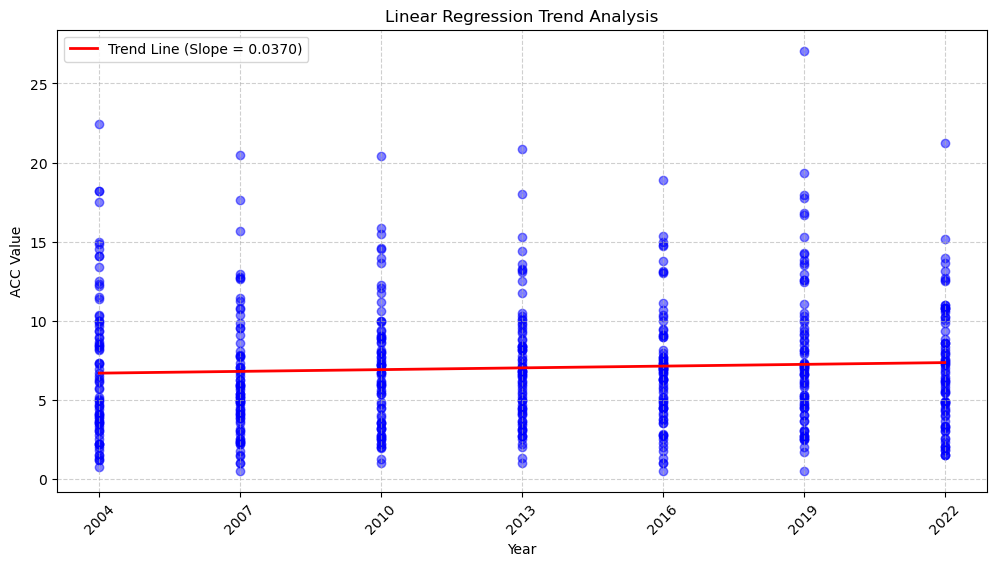}}
        \caption{Trend Analysis of Average Community Connectivity}
        \label{Trend-ACC}
\end{figure}

\subsection{Heatmaps of Metrics}
\begin{landscape}
\begin{figure}[h]
        \centerline{\includegraphics[width = 1.7\textwidth]{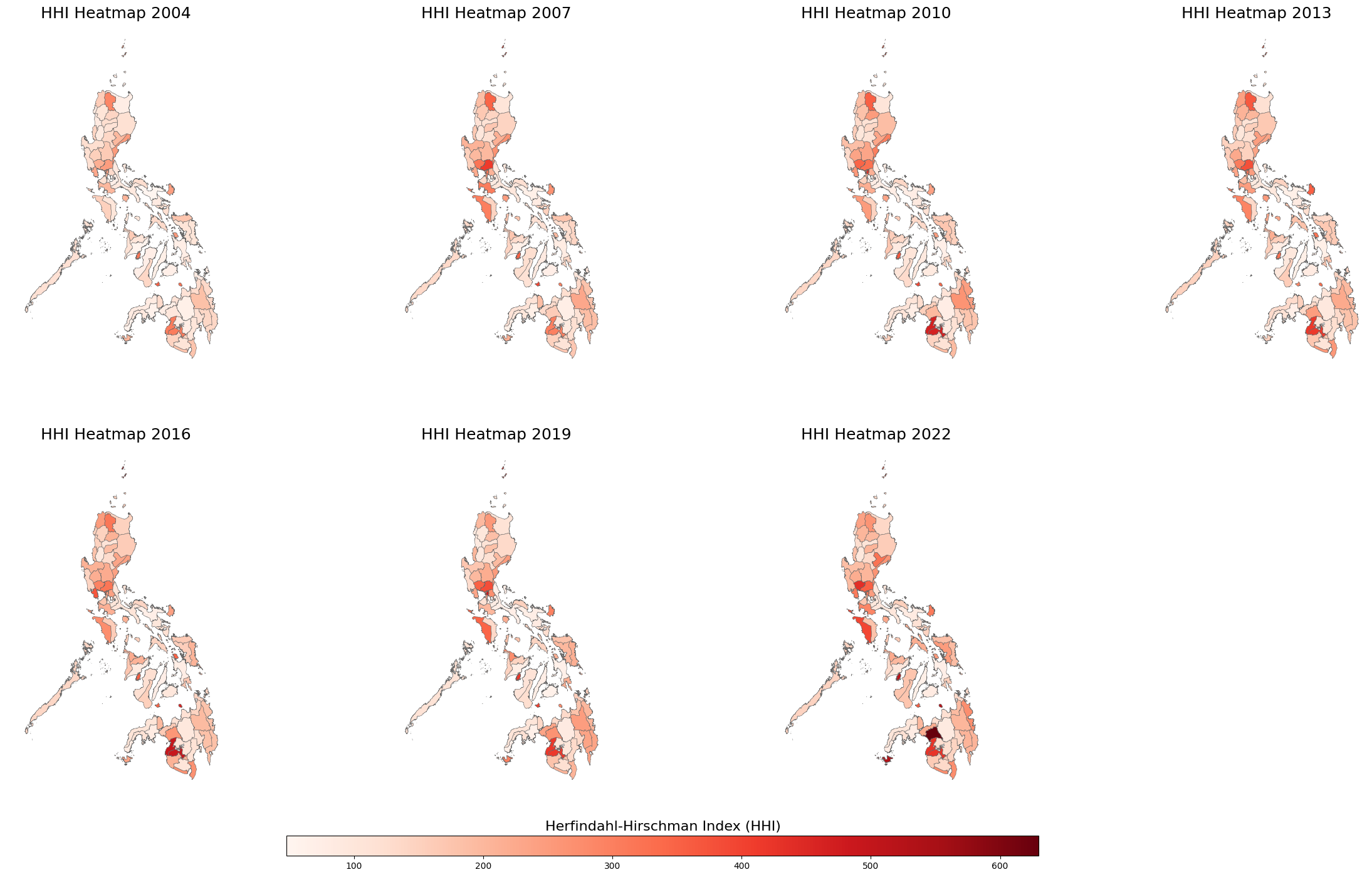}}
        \caption{Heatmap of Political Herfindahl-Hirschman Index}
        \label{Heatmap-HHI}
\end{figure}

\begin{figure}[h]
        \centerline{\includegraphics[width = 1.7\textwidth]{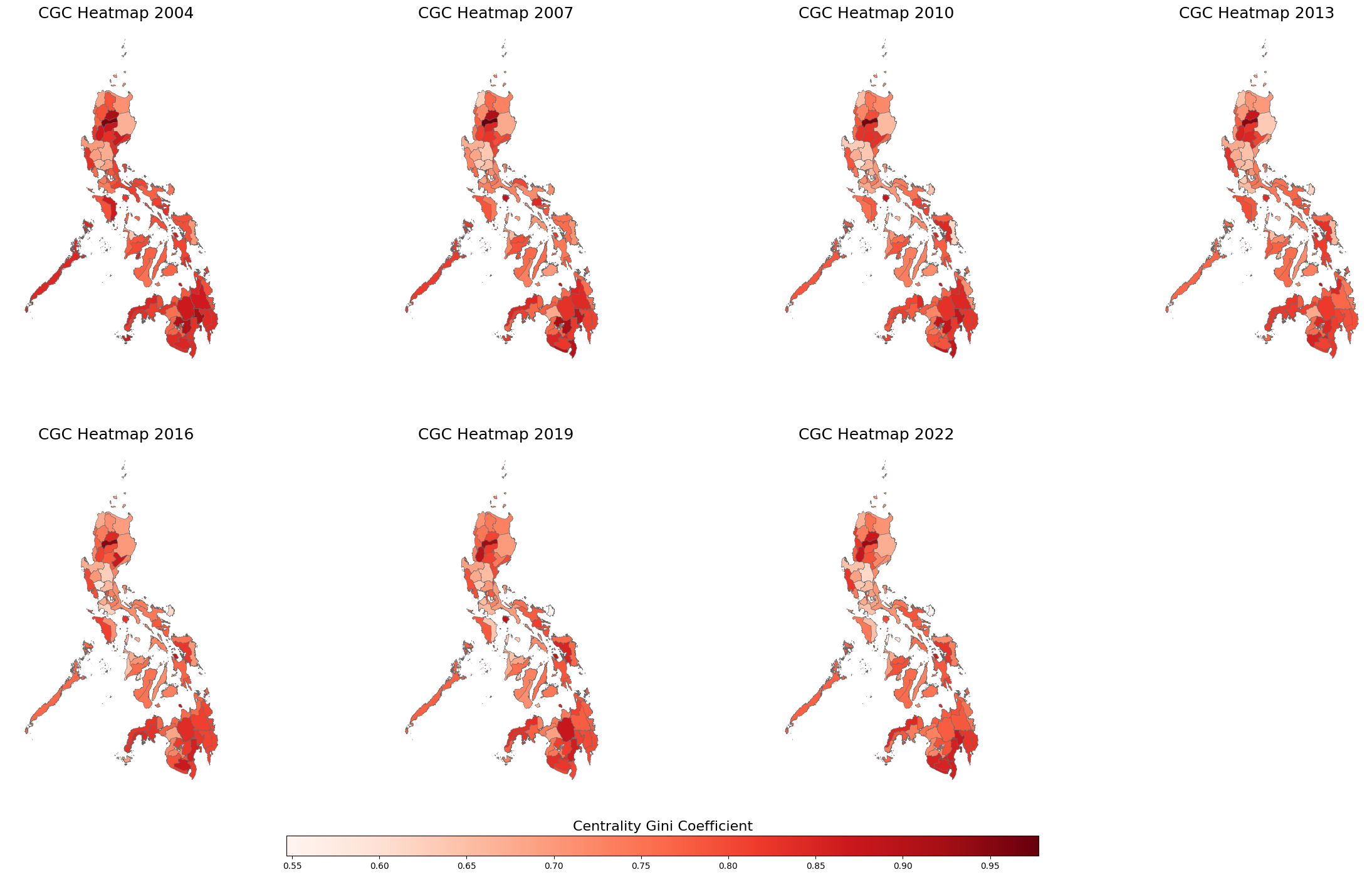}}
        \caption{Heatmap of Centrality Gini Coefficient}
        \label{Heatmap-CGC}
\end{figure}

\begin{figure}[h]
        \centerline{\includegraphics[width = 1.7\textwidth]{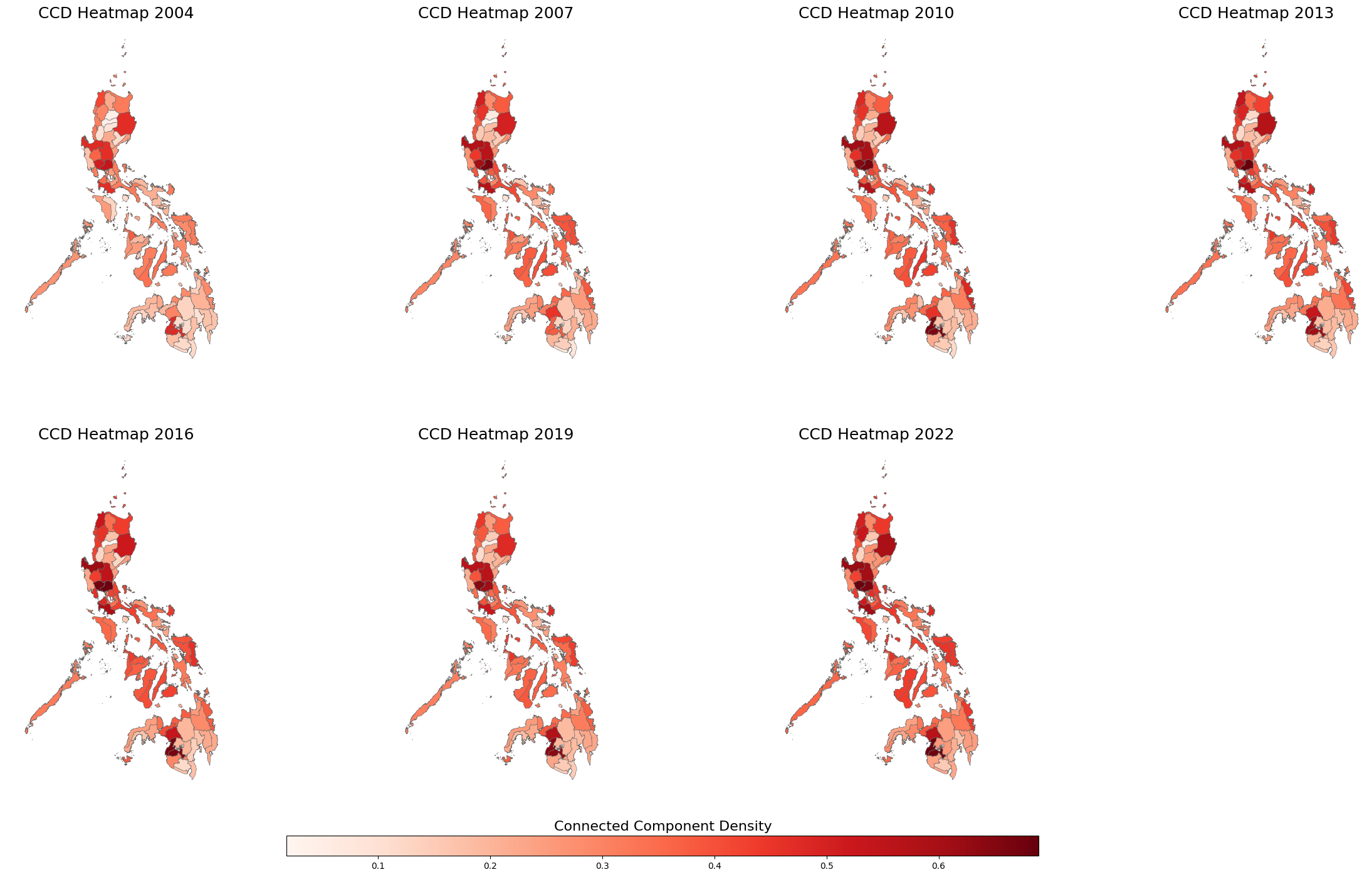}}
        \caption{Heatmap of Connected Component Density}
        \label{Heatmap-CCD}
\end{figure}

\begin{figure}[h]
        \centerline{\includegraphics[width = 1.7\textwidth]{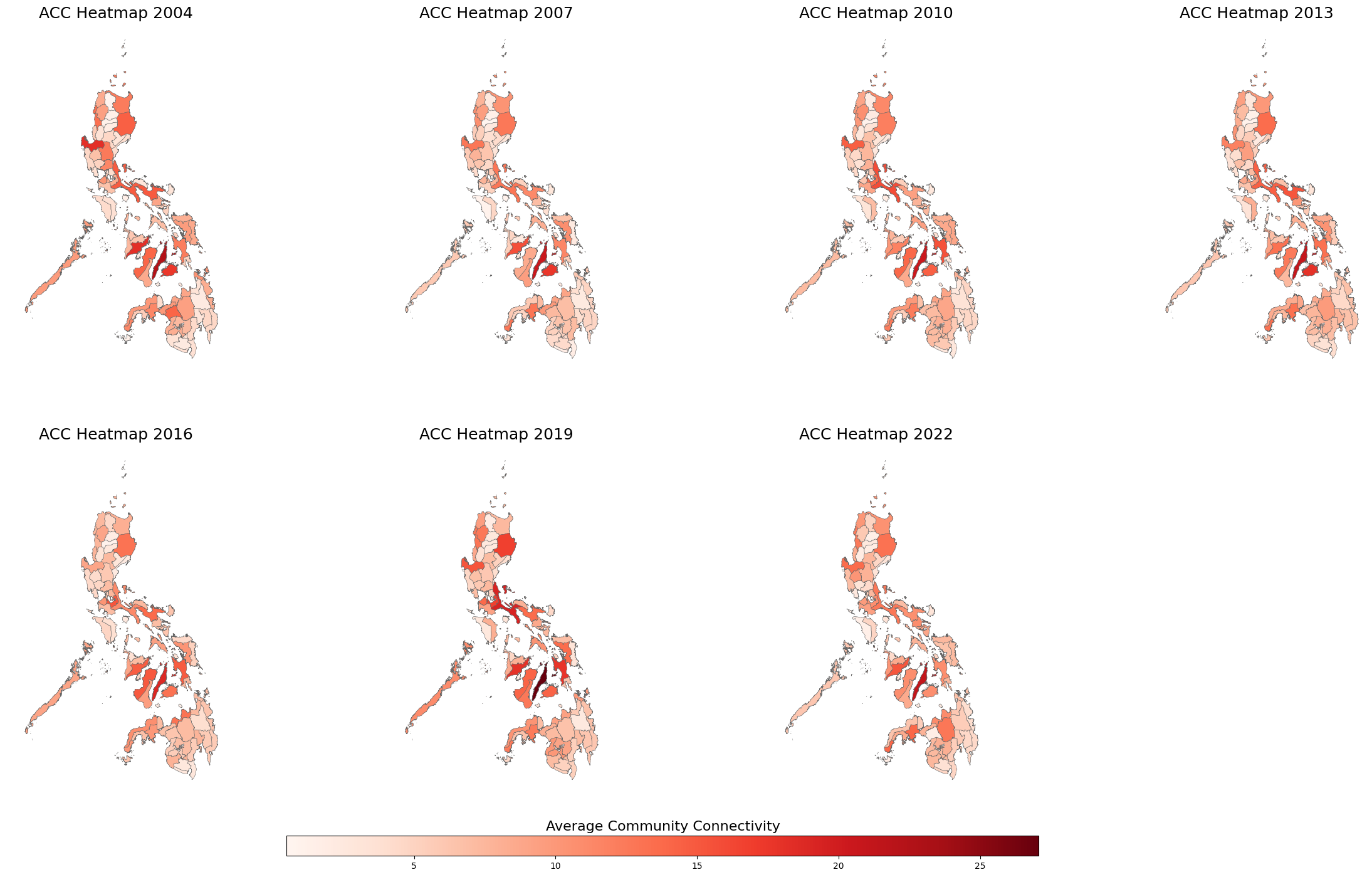}}
        \caption{Heatmap of Average Community Connectivity}
        \label{Heatmap-ACC}
\end{figure}

\end{landscape}

\clearpage

\begin{landscape}
\subsection{Political Herfindahl-Hirschman Index}
\setlength\LTleft{0pt}
\setlength\LTright{0pt}
\setlength{\LTpre}{0pt}   
\setlength{\LTpost}{0pt}  
\centering
\begin{longtable}{lrrrrrrr}
\caption{Political Herfindahl-Hirschman Index per Province for $2004$ to $2022$} \\
\toprule
Province & 2004 & 2007 & 2010 & 2013 & 2016 & 2019 & 2022 \\
\midrule
\endfirsthead

\multicolumn{8}{c}{\textit{(Continued from previous page)}} \\
\toprule
Province & 2004 & 2007 & 2010 & 2013 & 2016 & 2019 & 2022 \\
\midrule
\endhead

\midrule \multicolumn{8}{r}{\textit{Continued on next page}} \\
\endfoot

\bottomrule
\endlastfoot

ABRA & 106.936514 & 166.540432 & 180.850263 & 209.113507 & 149.917827 & 98.563023 & 203.376579 \\
AGUSAN DEL NORTE & 127.796646 & 135.389777 & 143.315843 & 159.256995 & 143.771626 & 178.790396 & 194.869615 \\
AGUSAN DEL SUR & 183.385845 & 228.661474 & 263.411046 & 223.093029 & 194.171981 & 246.801462 & 203.853554 \\
AKLAN & 162.600830 & 171.227691 & 156.028663 & 216.906722 & 221.868150 & 274.400656 & 224.079217 \\
ALBAY & 85.972450 & 82.928726 & 89.092588 & 85.291305 & 99.271887 & 83.058094 & 115.109573 \\
ANTIQUE & 111.610695 & 141.558669 & 170.159889 & 151.124739 & 171.016845 & 147.749466 & 127.485891 \\
APAYAO & 287.304491 & 347.823182 & 353.500000 & 365.500000 & 314.675032 & 256.564349 & 262.512961 \\
AURORA & 251.520757 & 268.392556 & 302.419753 & 235.142809 & 240.591784 & 252.218219 & 263.703704 \\
\end{longtable}

\end{landscape}

\begin{landscape}
\setlength\LTleft{0pt}
\setlength\LTright{0pt}
\setlength{\LTpre}{0pt}   
\setlength{\LTpost}{0pt}  
\centering
\tiny
\begin{longtable}{l@{}r@{}r@{}r@{}r@{}r@{}r@{}r}
\caption{Rank per Province based on Political Herfindahl-Hirschman Index from $2004$ to $2022$} \\
\toprule
Rank & 2004 & 2007 & 2010 & 2013 & 2016 & 2019 & 2022 \\
\midrule
\endfirsthead

\multicolumn{8}{c}{\textit{(Continued from previous page)}} \\
\toprule
Rank & 2004 & 2007 & 2010 & 2013 & 2016 & 2019 & 2022 \\
\midrule
\endhead

\midrule \multicolumn{8}{r}{\textit{Continued on next page}} \\
\endfoot

\bottomrule
\endlastfoot

1 & SIQUIJOR & BULACAN & NCR, FIRST DISTRICT & MAGUINDANAO & BATANES & NCR, SECOND DISTRICT & LANAO DEL SUR \\
2 & NCR, FIRST DISTRICT & BATANES & MAGUINDANAO & BULACAN & MAGUINDANAO & NCR, FIRST DISTRICT & BASILAN \\
3 & CAMIGUIN & SIQUIJOR & NCR, SECOND DISTRICT & BATANES & CAMIGUIN & BATANES & CAMIGUIN \\
4 & GUIMARAS & NCR, FIRST DISTRICT & SIQUIJOR & APAYAO & BATAAN & GUIMARAS & GUIMARAS \\
5 & MAGUINDANAO & GUIMARAS & BATANES & CATANDUANES & GUIMARAS & MAGUINDANAO & PAMPANGA \\
6 & APAYAO & APAYAO & GUIMARAS & NCR, FIRST DISTRICT & NCR, FIRST DISTRICT & SIQUIJOR & MAGUINDANAO \\
7 & BATANES & PAMPANGA & APAYAO & CAMIGUIN & BILIRAN & BULACAN & OCCIDENTAL MINDORO \\
8 & BILIRAN & NCR, SECOND DISTRICT & PAMPANGA & NCR, SECOND DISTRICT & SIQUIJOR & PAMPANGA & NCR, FIRST DISTRICT \\
\end{longtable}

\end{landscape}

\begin{landscape}
\subsection{Centrality Gini Coefficient}
\setlength\LTleft{0pt}
\setlength\LTright{0pt}
\setlength{\LTpre}{0pt}   
\setlength{\LTpost}{0pt}  
\centering
\begin{longtable}{lrrrrrrr}
\caption{Centrality Gini Coefficient per Province for $2004$ to $2022$} \\
\toprule
Province & 2004 & 2007 & 2010 & 2013 & 2016 & 2019 & 2022 \\
\midrule
\endfirsthead

\multicolumn{8}{c}{\textit{(Continued from previous page)}} \\
\toprule
Province & 2004 & 2007 & 2010 & 2013 & 2016 & 2019 & 2022 \\
\midrule
\endhead

\midrule \multicolumn{8}{r}{\textit{Continued on next page}} \\
\endfoot

\bottomrule
\endlastfoot

ABRA & 0.790900 & 0.720400 & 0.723900 & 0.704900 & 0.738300 & 0.745000 & 0.737100 \\
AGUSAN DEL NORTE & 0.849600 & 0.848500 & 0.842400 & 0.845700 & 0.761600 & 0.809500 & 0.791700 \\
AGUSAN DEL SUR & 0.865700 & 0.842800 & 0.845100 & 0.765100 & 0.812900 & 0.776600 & 0.782300 \\
AKLAN & 0.638000 & 0.656500 & 0.703100 & 0.647200 & 0.674700 & 0.645300 & 0.674300 \\
ALBAY & 0.810500 & 0.842900 & 0.809000 & 0.790600 & 0.782000 & 0.813500 & 0.770000 \\
ANTIQUE & 0.775000 & 0.726900 & 0.724800 & 0.781200 & 0.645400 & 0.676700 & 0.696100 \\
APAYAO & 0.787900 & 0.766100 & 0.744100 & 0.706100 & 0.712400 & 0.743300 & 0.749700 \\
AURORA & 0.837800 & 0.805800 & 0.765400 & 0.700100 & 0.725500 & 0.799000 & 0.713900 \\
\end{longtable}

\end{landscape}

\begin{landscape}
\setlength\LTleft{0pt}
\setlength\LTright{0pt}
\setlength{\LTpre}{0pt}   
\setlength{\LTpost}{0pt}  
\centering
\tiny
\begin{longtable}{l@{}r@{}r@{}r@{}r@{}r@{}r@{}r}
\caption{Rank per Province based on Centrality Gini Coefficient from $2004$ to $2022$} \\
\toprule
Rank & 2004 & 2007 & 2010 & 2013 & 2016 & 2019 & 2022 \\
\midrule
\endfirsthead

\multicolumn{8}{c}{\textit{(Continued from previous page)}} \\
\toprule
Rank & 2004 & 2007 & 2010 & 2013 & 2016 & 2019 & 2022 \\
\midrule
\endhead

\midrule \multicolumn{8}{r}{\textit{Continued on next page}} \\
\endfoot

\bottomrule
\endlastfoot

1 & MOUNTAIN PROVINCE & MOUNTAIN PROVINCE & MOUNTAIN PROVINCE & MOUNTAIN PROVINCE & MOUNTAIN PROVINCE & MOUNTAIN PROVINCE & MOUNTAIN PROVINCE \\
2 & DAVAO DEL NORTE & COTABATO & DAVAO DEL NORTE & CAMIGUIN & CAMIGUIN & BENGUET & BENGUET \\
3 & KALINGA & KALINGA & COTABATO & KALINGA & QUIRINO & MARINDUQUE & DAVAO DEL NORTE \\
4 & COTABATO & SARANGANI & SARANGANI & NCR, THIRD DISTRICT & SOUTH COTABATO & BILIRAN & KALINGA \\
5 & IFUGAO & DAVAO DEL NORTE & MARINDUQUE & SULTAN KUDARAT & BILIRAN & BUKIDNON & BILIRAN \\
6 & CAMIGUIN & NCR, THIRD DISTRICT & CAMIGUIN & BENGUET & NCR, FIRST DISTRICT & DAVAO DEL SUR & DAVAO DEL SUR \\
7 & QUIRINO & MARINDUQUE & BILIRAN & BILIRAN & DAVAO DEL SUR & CAMIGUIN & SARANGANI \\
8 & BUKIDNON & CAMIGUIN & AGUSAN DEL SUR & NUEVA VIZCAYA & ZAMBOANGA SIBUGAY & SAMAR & SULTAN KUDARAT \\
\end{longtable}

\end{landscape}

\begin{landscape}
\subsection{Connected Component Density}
\setlength\LTleft{0pt}
\setlength\LTright{0pt}
\setlength{\LTpre}{0pt}   
\setlength{\LTpost}{0pt}  
\centering
\begin{longtable}{lrrrrrrr}
\caption{Connected Component Density per Province for $2004$ to $2022$} \\
\toprule
Province & 2004 & 2007 & 2010 & 2013 & 2016 & 2019 & 2022 \\
\midrule
\endfirsthead

\multicolumn{8}{c}{\textit{(Continued from previous page)}} \\
\toprule
Province & 2004 & 2007 & 2010 & 2013 & 2016 & 2019 & 2022 \\
\midrule
\endhead

\midrule \multicolumn{8}{r}{\textit{Continued on next page}} \\
\endfoot

\bottomrule
\endlastfoot

ABRA & 0.295620 & 0.454545 & 0.456446 & 0.491289 & 0.458042 & 0.387931 & 0.527778 \\
AGUSAN DEL NORTE & 0.138686 & 0.138211 & 0.182482 & 0.196850 & 0.253623 & 0.240602 & 0.255474 \\
AGUSAN DEL SUR & 0.202614 & 0.252033 & 0.309677 & 0.329032 & 0.285714 & 0.313725 & 0.322581 \\
AKLAN & 0.379679 & 0.393258 & 0.365169 & 0.446328 & 0.436170 & 0.450000 & 0.457447 \\
ALBAY & 0.179104 & 0.178218 & 0.194030 & 0.195000 & 0.238579 & 0.185714 & 0.285024 \\
ANTIQUE & 0.271795 & 0.337143 & 0.355670 & 0.325153 & 0.401042 & 0.298969 & 0.323383 \\
APAYAO & 0.222222 & 0.281690 & 0.292683 & 0.353659 & 0.341463 & 0.247059 & 0.285714 \\
AURORA & 0.250000 & 0.274725 & 0.344444 & 0.292683 & 0.244681 & 0.237624 & 0.307692 \\
\end{longtable}

\end{landscape}

\begin{landscape}
\setlength\LTleft{0pt}
\setlength\LTright{0pt}
\setlength{\LTpre}{0pt}   
\setlength{\LTpost}{0pt}  
\centering
\tiny
\begin{longtable}{l@{}r@{}r@{}r@{}r@{}r@{}r@{}r}
\caption{Rank per Province based on Connected Component Density from $2004$ to $2022$} \\
\toprule
Rank & 2004 & 2007 & 2010 & 2013 & 2016 & 2019 & 2022 \\
\midrule
\endfirsthead

\multicolumn{8}{c}{\textit{(Continued from previous page)}} \\
\toprule
Rank & 2004 & 2007 & 2010 & 2013 & 2016 & 2019 & 2022 \\
\midrule
\endhead

\midrule \multicolumn{8}{r}{\textit{Continued on next page}} \\
\endfoot

\bottomrule
\endlastfoot

1 & BULACAN & BULACAN & BULACAN & BULACAN & BULACAN & PAMPANGA & PAMPANGA \\
2 & PAMPANGA & PAMPANGA & MAGUINDANAO & MAGUINDANAO & PAMPANGA & MAGUINDANAO & MAGUINDANAO \\
3 & MAGUINDANAO & PANGASINAN & PAMPANGA & PAMPANGA & MAGUINDANAO & BULACAN & BULACAN \\
4 & PANGASINAN & BATANGAS & PANGASINAN & PANGASINAN & PANGASINAN & LANAO DEL SUR & PANGASINAN \\
5 & ISABELA & NUEVA ECIJA & NUEVA ECIJA & BATANGAS & BATANGAS & PANGASINAN & BATANGAS \\
6 & BATANGAS & ILOCOS NORTE & BATANGAS & ISABELA & NUEVA ECIJA & NUEVA ECIJA & NUEVA ECIJA \\
7 & NUEVA ECIJA & ISABELA & ISABELA & ILOCOS NORTE & LANAO DEL SUR & BATANGAS & ISABELA \\
8 & ILOCOS NORTE & LANAO DEL SUR & SURIGAO DEL SUR & LANAO DEL SUR & ILOCOS NORTE & ISABELA & LANAO DEL SUR \\
\end{longtable}

\end{landscape}

\begin{landscape}
\subsection{Average Community Connectivity}
\setlength\LTleft{0pt}
\setlength\LTright{0pt}
\setlength{\LTpre}{0pt}   
\setlength{\LTpost}{0pt}  
\centering
\begin{longtable}{lrrrrrrr}
\caption{Average Community Connectivity per Province for $2004$ to $2022$} \\
\toprule
Province & 2004 & 2007 & 2010 & 2013 & 2016 & 2019 & 2022 \\
\midrule
\endfirsthead

\multicolumn{8}{c}{\textit{(Continued from previous page)}} \\
\toprule
Province & 2004 & 2007 & 2010 & 2013 & 2016 & 2019 & 2022 \\
\midrule
\endhead

\midrule \multicolumn{8}{r}{\textit{Continued on next page}} \\
\endfoot

\bottomrule
\endlastfoot

ABRA & 9.333333 & 9.500000 & 10.583333 & 7.083333 & 8.942857 & 12.583333 & 6.000000 \\
AGUSAN DEL NORTE & 3.500000 & 4.000000 & 2.666667 & 5.166667 & 7.333333 & 5.800000 & 6.083333 \\
AGUSAN DEL SUR & 2.166667 & 2.333333 & 3.500000 & 4.500000 & 4.000000 & 2.416667 & 5.583333 \\
AKLAN & 6.083333 & 4.833333 & 7.666667 & 6.833333 & 3.694444 & 4.666667 & 4.333333 \\
ALBAY & 8.916667 & 7.000000 & 7.666667 & 10.083333 & 6.916667 & 8.700000 & 10.966667 \\
ANTIQUE & 7.333333 & 5.333333 & 10.000000 & 9.216667 & 9.083333 & 8.466667 & 10.250000 \\
APAYAO & 0.750000 & 2.400000 & 2.000000 & 2.666667 & 4.500000 & 2.000000 & 4.916667 \\
AURORA & 4.000000 & 3.000000 & 2.500000 & 5.333333 & 2.250000 & 4.000000 & 3.250000 \\
\end{longtable}

\end{landscape}

\begin{landscape}
\setlength\LTleft{0pt}
\setlength\LTright{0pt}
\setlength{\LTpre}{0pt}   
\setlength{\LTpost}{0pt}  
\centering
\tiny
\begin{longtable}{l@{}r@{}r@{}r@{}r@{}r@{}r@{}r}
\caption{Rank per Province based on Average Community Connectivity from $2004$ to $2022$} \\
\toprule
Rank & 2004 & 2007 & 2010 & 2013 & 2016 & 2019 & 2022 \\
\midrule
\endfirsthead

\multicolumn{8}{c}{\textit{(Continued from previous page)}} \\
\toprule
Rank & 2004 & 2007 & 2010 & 2013 & 2016 & 2019 & 2022 \\
\midrule
\endhead

\midrule \multicolumn{8}{r}{\textit{Continued on next page}} \\
\endfoot

\bottomrule
\endlastfoot

1 & CEBU & CEBU & CEBU & CEBU & CEBU & CEBU & CEBU \\
2 & PANGASINAN & BOHOL & QUEZON & BOHOL & LAGUNA & QUEZON & ILOILO \\
3 & ILOILO & ILOILO & LEYTE & CAMARINES SUR & NEGROS OCCIDENTAL & ILOILO & ZAMBOANGA DEL SUR \\
4 & BOHOL & ZAMBOANGA DEL SUR & PANGASINAN & QUEZON & LEYTE & LEYTE & PANGASINAN \\
5 & CAMARINES SUR & QUEZON & BOHOL & ISABELA & ILOILO & LAGUNA & ISABELA \\
6 & QUEZON & PANGASINAN & NEGROS OCCIDENTAL & ZAMBOANGA DEL SUR & CAMARINES SUR & ISABELA & BUKIDNON \\
7 & ISABELA & ISABELA & LAGUNA & LEYTE & BOHOL & PANGASINAN & QUEZON \\
8 & LANAO DEL SUR & LEYTE & ZAMBOANGA DEL SUR & ILOILO & MISAMIS ORIENTAL & BOHOL & ILOCOS SUR \\
\end{longtable}

\end{landscape}

\subsection{LMM Assumptions}

\begin{table}[H]
    \centering
    \setlength{\tabcolsep}{6pt}      
    \renewcommand{\arraystretch}{1.2} 
    \caption{Summary of VIF values for LMM Regression}
    \begin{tabular}{
        l 
        S[table-format=1.6]
        S[table-format=1.6]
        S[table-format=1.6]
        S[table-format=1.6]
    }
        \toprule
        {\textbf{Variable}} & {\textbf{VIF}} \\
        \midrule
        POV & \num{3.011988} \\
        HDI & \num{4.881710} \\
        ACC & \num{3.363808} \\
        CCD & \num{4.247681} \\
        HHI & \num{3.243068} \\
        CGC & \num{3.086175} \\
        POV\_lag\_3year & \num{3.661974} \\
        HDI\_lag\_3year & \num{4.206692} \\
        \bottomrule
    \end{tabular}
    \label{tab:VIF-table}
\end{table}

\begin{figure}[h!]
    \caption{Q-Q Residual Plots of LMM Regression under Direction 1}
    \begin{subfigure}{0.48\textwidth} 
        \includegraphics[width=\textwidth]{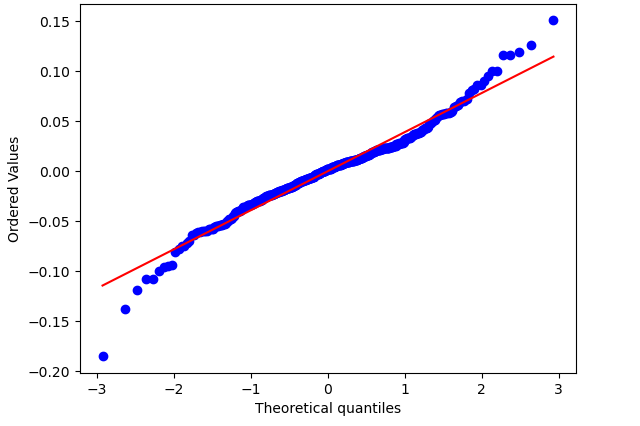} 
        \caption{HDI as Dependent Variable}
        \label{fig:D1-QQHDI}
    \end{subfigure}%
    \hfill 
    \begin{subfigure}{0.48\textwidth} 
        \includegraphics[width=\textwidth]{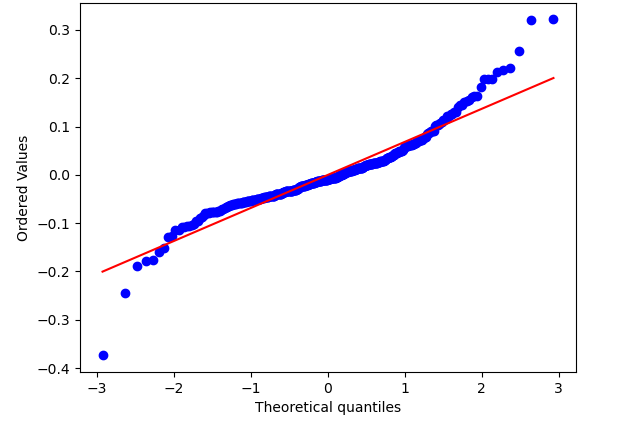}
        \caption{POV as Dependent Variable}
        \label{fig:D1-QQPOV}
    \end{subfigure}
    \label{fig:D1-QQ}
\end{figure}

\begin{figure}[h!]
    \caption{Q-Q Residual Plots of LMM Regression under Direction 2 with ACC as Independent Variable}
    \begin{subfigure}{0.48\textwidth} 
        \includegraphics[width=\textwidth]{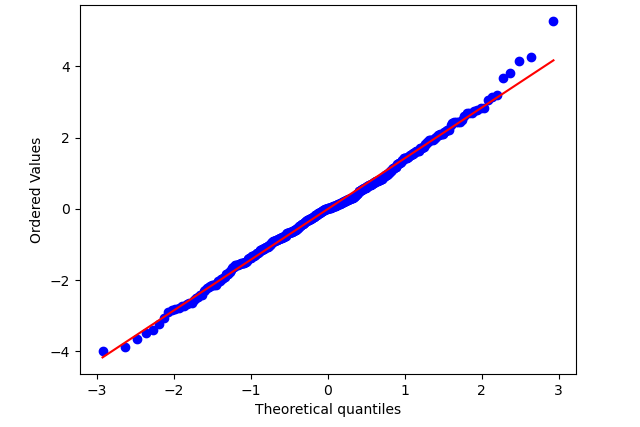} 
        \caption{HDI as Independent Variable}
        \label{fig:D2-QQ-ACC-HDI}
    \end{subfigure}%
    \hfill 
    \begin{subfigure}{0.48\textwidth} 
        \includegraphics[width=\textwidth]{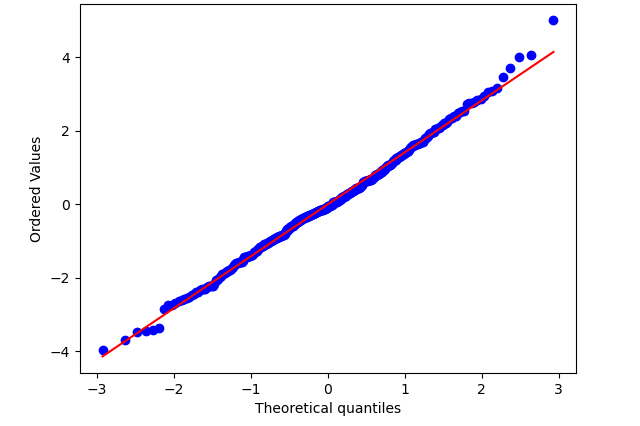}
        \caption{POV as Independent Variable}
        \label{fig:D2-QQ-ACC-POV}
    \end{subfigure}
    \label{fig:D2-QQ-ACC}
\end{figure}

\begin{figure}[h!]
    \caption{Q-Q Residual Plots of LMM Regression under Direction 2 with CCD as Independent Variable}
    \begin{subfigure}{0.48\textwidth} 
        \includegraphics[width=\textwidth]{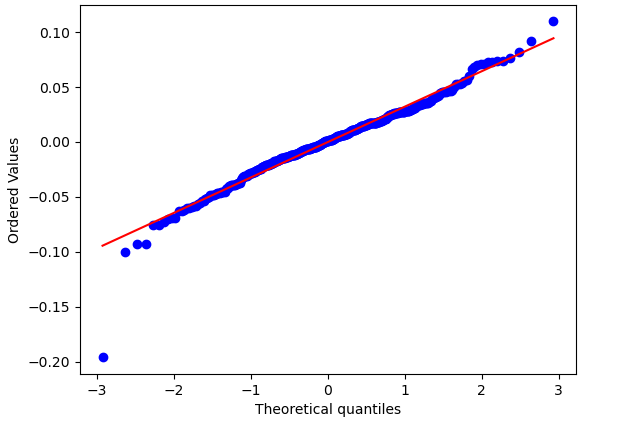} 
        \caption{HDI as Independent Variable}
        \label{fig:D2-QQ-CCD-HDI}
    \end{subfigure}%
    \hfill 
    \begin{subfigure}{0.48\textwidth} 
        \includegraphics[width=\textwidth]{3Manuscript/images/D2-POV_ACC.png}
        \caption{POV as Independent Variable}
        \label{fig:D2-QQ-CCD-POV}
    \end{subfigure}
    \label{fig:D2-QQ-CCD}
\end{figure}

\begin{figure}[h!]
    \caption{Q-Q Residual Plots of LMM Regression under Direction 2 with CGC as Independent Variable}
    \begin{subfigure}{0.48\textwidth} 
        \includegraphics[width=\textwidth]{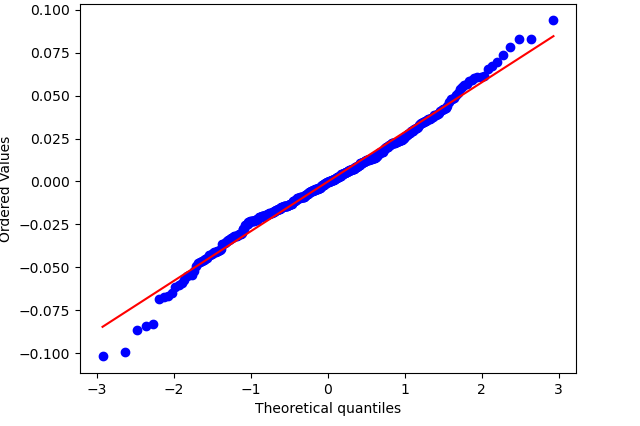} 
        \caption{HDI as Independent Variable}
        \label{fig:D2-QQ-CGC-HDI}
    \end{subfigure}%
    \hfill 
    \begin{subfigure}{0.48\textwidth} 
        \includegraphics[width=\textwidth]{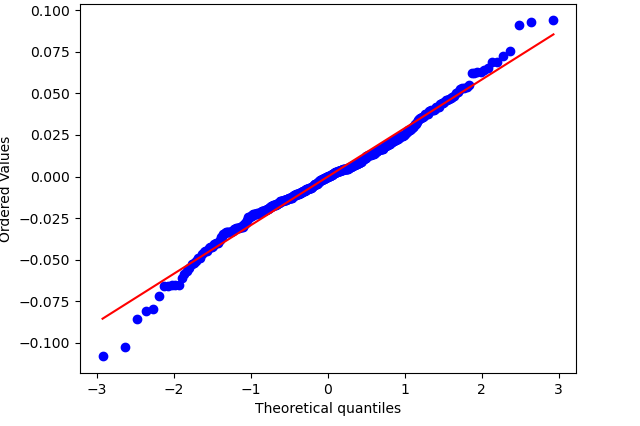}
        \caption{POV as Independent Variable}
        \label{fig:D2-QQ-CGC-POV}
    \end{subfigure}
    \label{fig:D2-QQ-CGC}
\end{figure}

\begin{figure}[h!]
    \caption{Q-Q Residual Plots of LMM Regression under Direction 2 with HHI as Independent Variable}
    \begin{subfigure}{0.48\textwidth} 
        \includegraphics[width=\textwidth]{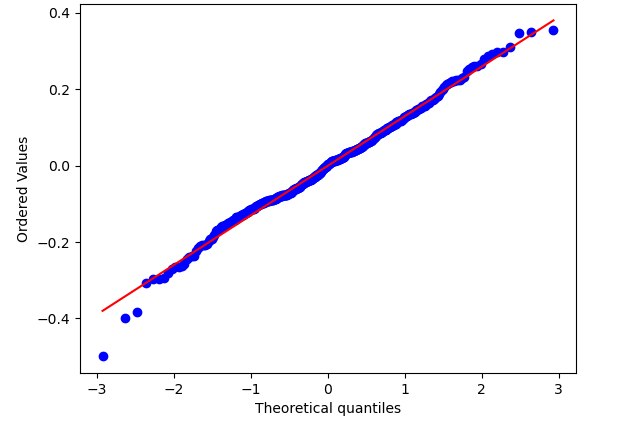} 
        \caption{HDI as Independent Variable}
        \label{fig:D2-QQ-HHI-HDI}
    \end{subfigure}%
    \hfill 
    \begin{subfigure}{0.48\textwidth} 
        \includegraphics[width=\textwidth]{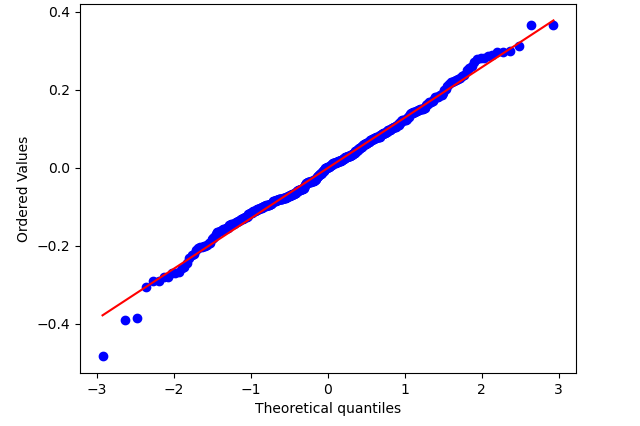}
        \caption{POV as Independent Variable}
        \label{fig:D2-QQ-HHI-POV}
    \end{subfigure}
    \label{fig:D2-QQ-HHI}
    
\end{figure}
\end{appendices}

\end{document}